\documentclass[onecolumn,pre,floats,aps,amsmath,amssymb,nofootinbib]{revtex4-2}
\usepackage{graphicx}
\usepackage{caption}
\usepackage{subfig}
\usepackage{bm}
\usepackage{verbatim}
\usepackage{microtype}
\usepackage{amsmath}
\usepackage{siunitx}
\usepackage{physics}
\usepackage{amssymb}
\usepackage{slashed}
\usepackage{hyperref}
\newcommand{\bea}{\begin{eqnarray}}
\newcommand{\eea}{\end{eqnarray}}
\newcommand{\dfour}[1]{d^{\hspace{1.5pt}4}{#1}}
\newcommand{\dthree}[1]{d^{\hspace{1.5pt}3}{#1}}
\newcommand{\pip}{\pi(\vec{p}\hspace{1.5pt})}
\newcommand{\pirest}{\pi(\vec{0}\hspace{1.5pt})}

\newcommand{\GF}{\frac{G_F}{\sqrt{2}}\,V_{ud}^\ast}
\newcommand{\GFsq}{\frac{G_F^2}{2}\,|V_{ud}|^2}
\newcommand{\pitolnu}{\pi^+\to\ell^+\nu_\ell}
\newcommand{\pitolnug}{\pi^+\to\ell^+\nu_\ell\gamma}

\newcommand{\Elpk}{E_\ell^\prime(\vec{k}\,)}
\newcommand{\Elpmk}{E_\ell^\prime(-\vec{k}\,)}
\newcommand{\Epik}{E_\pi(\vecp{k})}
\newcommand{\vecp}[1]{\vec{#1}\hspace{1.2pt}}
\newcommand{\vecpsq}[1]{\vec{#1}^{\,\,2}}

\newcommand{\Columbia}{Physics Department, Columbia University, New York, NY 10027, USA}
\newcommand{\Peking}{School of Physics, Peking University, Beijing 100871, China}
\newcommand{\QuantumMatter}{Collaborative Innovation Center of Quantum Matter, Beijing 100871, China}
\newcommand{\HEP}{Center for High Energy Physics, Peking University, Beijing 100871, China}
\newcommand{\UConn}{Department of Physics, University of Connecticut, Storrs, CT 06269, USA}
\newcommand{\Soton}{Department of Physics and Astronomy, University of Southampton,\\ Southampton SO17 1BJ, UK}
\newcommand{\BNL}{Computational Science Initiative, Brookhaven National Laboratory,
Upton, NY 11973, USA}

\begin{document}
\title{Radiative corrections to leptonic decays using infinite-volume reconstruction}
\author{Norman H. Christ}\affiliation{\Columbia}
\author{Xu Feng}\affiliation{\Peking}\affiliation{\QuantumMatter}\affiliation{\HEP}
\author{Lu-Chang Jin}\affiliation{\UConn}
\author{Christopher T. Sachrajda}\affiliation{\Soton}
\author{Tianle Wang}\affiliation{\Columbia}\affiliation{\BNL}


\begin{abstract}
Lattice QCD calculations of leptonic decay constants have now reached sub-percent precision so that isospin-breaking corrections, including QED effects, must be included to fully exploit this precision in determining fundamental quantities, in particular the elements of the Cabibbo-Kobayashi-Maskawa (CKM) matrix, from experimental measurements. A number of collaborations have performed, or are performing, such computations.
In this paper we develop a new theoretical framework, based on Infinite-Volume Reconstruction (IVR), for the computation of electromagnetic corrections to leptonic decay widths. 
In this method, the hadronic correlation functions are first processed theoretically in infinite volume, in such a way that the required matrix elements can be determined non-perturbatively from lattice QCD computations with finite-volume uncertainties which are  
exponentially small in the volume.
The cancellation of infrared divergences in this framework is performed fully analytically. 
We also outline how this IVR treatment can be extended to determine the QED effects in semi-leptonic kaon decays with a similar degree of accuracy.
\end{abstract}

\maketitle

 \section{Introduction}

Lattice QCD results for a number of physical quantities have now reached the sub-percent level, e.g. the 2021 review by the Flavour Physics Lattice Averaging Group (FLAG2021)\,\cite{Aoki:2021kgd} quotes the following values for the leptonic decay constants $f_\pi$ and $f_K$\,\footnote{The decay constant $f_\pi$ is frequently used as part of the calibration, including the determination of the lattice spacing, and the value in Eq.\,(\ref{eq:fpifKFLAG}) is obtained from simulations with $N_f=2+1$ light-quark flavours. The values of $f_K$ and $f_K/f_\pi$ are from $N_F=2+1+1$ computations.}:
\begin{equation}\label{eq:fpifKFLAG}
f_\pi=130.2\,(8)\,\mathrm{MeV}, \qquad f_K=155.7\,(3)\,\mathrm{MeV},\qquad\frac{f_K}{f_\pi}=1.1932(21)\,.
\end{equation}
The experimental results for the leptonic decay widths are even more precise.
In order to fully exploit the level of precision in Eq.\,(\ref{eq:fpifKFLAG}) for tests of the Standard Model of particle physics and the 
determination of its parameters, in particular the elements of the Cabibbo-Kobayashi-Maskawa (CKM) matrix,
electromagnetic and strong isospin-breaking corrections need to be included.  
The subject of this paper is the theoretical extension of the Infinite Volume Reconstruction (IVR) method 
for the evaluation of leptonic decay widths of pseudoscalar mesons in lattice QCD computations on a finite Euclidean volume in such a way that i) the cancellation of infrared divergences is explicit and ii) the finite-volume corrections are exponentially small. 
The method is illustrated with the decay of a pion, $\pi^+\to\ell^+\nu_\ell(\gamma)$, where $\ell^+$ is a charged lepton, but applies equally well to the decays of heavier mesons ($K, D, D_s, B$ and $B_c$ mesons).

Infinite volume reconstruction was first proposed in Ref.\,\cite{Feng:2018qpx} to avoid power-like finite-volume uncertainties when computing QED corrections to the hadronic spectrum in a finite volume. It has since been used in studies including: 
long distance contributions to neutrinoless double-$\beta$ decay\,\cite{Tuo:2019bue}, rare kaon decays\,\cite{Christ:2020hwe}; the width for
the decay $K\to \ell\nu(\ell^{\prime\,+}\ell^{\prime\,-})$, where $\ell$ and $\ell^\prime$ represent charged leptons\,\cite{Tuo:2021ewr};
the $\pi^+$\,-\,$\pi^0$ mass splitting\,\cite{Feng:2021zek}; the two-photon exchange contribution to the muonic-hydrogen Lamb shift\,\cite{Fu:2022fgh} and
the contribution from a light sterile neutrino to neutrinoless double-$\beta$ decay\,\cite{Tuo:2022hft}. In these quantities there are no infrared divergences in intermediate stages of the calculation, with the exception of the two-photon exchange contribution to the muonic-hydrogen Lamb shift\,
where the infrared divergence is regulated by the atomic binding energy\,\cite{Fu:2022fgh}. By contract, for leptonic decays the divergences at $O(\alpha_\mathrm{em})$ are cancelled in $\Gamma(\pi^+\to\ell^+\nu_\ell)+\Gamma(\pi^+\to\ell^+\nu_\ell\gamma)$\,\cite{Bloch:1937pw} and an important element of this work is to demonstrate that the width can be computed using IVR after the complete analytic removal of the infrared divergences.

Isospin breaking corrections to leptonic decay widths have been studied in detail in Refs.\,\cite{Carrasco:2015xwa,Lubicz:2016xro,Giusti:2017dwk,DiCarlo:2019thl} in the context of the QED$_\mathrm{L}$ treatment of the photon's zero mode\,\cite{Hayakawa:2008an}. In particular it was shown in Ref.\,\cite{Lubicz:2016xro} that 
the finite-volume dependence of $\Gamma(\pi^+\to\ell^+\nu_\ell)$, the width for the decay $\pi^+\to\ell^+\nu_\ell$, takes the form
\begin{equation}\label{eq:FVexpansion}
\Gamma(\pi^+\to\ell^+\nu_\ell)=c_0(r_\ell)+\tilde{c}_0(r_\ell)\log[m_\pi L]+\frac{c_1(r_\ell)}{m_\pi L}+\cdots
\end{equation}
where $r_\ell=m_\ell/m_\pi$, $m_\pi$ and $m_\ell$ are the masses of the pion and charged lepton respectively and the spatial volume $V=L^3$. The exhibited terms in Eq.\,(\ref{eq:FVexpansion}) are universal, i.e. independent of the structure of the pion, and can therefore be evaluated in perturbation theory treating the pion as an elementary meson. The coefficients $c_0,\,\tilde{c}_0$ and $c_1$ were calculated in Ref.\,\cite{Lubicz:2016xro}, and the corresponding finite-volume effects subtracted from the non-perturbatively computed width in the numerical studies of Ref.\,\cite{Giusti:2017dwk,DiCarlo:2019thl}. The infrared divergence is manifest in the term containing $\log[m_\pi L]$\,\footnote{We have chosen to write the infrared divergent term here as $\log[m_\pi L]$. It can, of course, be written instead as $\log[m_\ell L]$ together with the corresponding redefinition of $c_0(r_\ell)$.}, so that $L$ acts as the infrared regulator.
In the QED$_\mathrm{L}$ formulation the leading finite-volume effects which depend on the structure of the decaying pion are 
therefore of $O(1/(m_\pi L)^2)$ and, together with higher order terms, are represented by the ellipsis in Eq.\,(\ref{eq:FVexpansion}). 
The $O(1/(m_\pi L)^2)$ non-perturbative effects were recently estimated in Ref.\,\cite{DiCarlo:2021apt}, together with a perturbative calculation of the terms of $O(1/(m_\pi L)^3)$ obtained by treating the meson as a point-like particle (see also Ref.\,\cite{Boyle:2022sgz}). It was found that while the structure-dependent terms at $O(1/(m_\pi L)^2)$ are small, the $O(1/(m_\pi L)^3)$ terms corresponding to a point-like pion are significant. The structure-dependent terms at $O(1/(m_\pi L)^3)$ are unknown however, and difficult to estimate without repeating computations at different volumes at the same lattice spacings and quark masses.

The primary aim of the present paper is to develop a framework, based on IVR, in which the finite-volume effects decrease exponentially in the volume and in which the cancellation of infrared divergences is fully controlled. In this approach, in contrast to other implementations of QCD+QED in lattice computations, the decay amplitude is not 
fully computed in a finite volume. 
Instead, as will be discussed in detail below, the infinite-volume amplitude is organised in such a way that effects 
related to the long-distance propagation of the photon are calculated analytically and the
only non-perturbative QCD input which is required is a non-local hadronic matrix element which is obtained with exponentially small finite-volume corrections.

A number of issues which are necessary for the evaluation of leptonic decay widths are generic, and hence are common to the QED$_\mathrm{L}$ and IVR frameworks. These were discussed in Refs.\,\cite{Carrasco:2015xwa,Lubicz:2016xro,Giusti:2017dwk,DiCarlo:2019thl} and we do not add further to that discussion here, beyond briefly recalling the main points. These include:

\noindent 1. \textit{The Effective Lagrangian and determination of the Fermi Constant:}\\[0.1cm]
Lattice calculations are generally performed with an inverse lattice spacing of the order of a few GeV ({\it e.g.} $a^{-1}\simeq 2$\,-\,$4$\,GeV) and, even with techniques such as step-scaling, direct 
computations in the Standard Model, which contain scales of $O(M_W)$, are not possible at present. Instead, weak decay amplitudes are evaluated in an effective theory in which the heavy degrees of freedom, and in particular the $W$ and $Z$ bosons are integrated out. The amplitudes are therefore written in terms of the Fermi constant, $G_F$, which is conventionally determined from the muon lifetime. At $O(\alpha_\mathrm{em})$ and neglecting higher order terms in $m_e^2/m_\mu^2$, the lifetime $\tau_\mu$ is given by the expression\,\cite{Berman:1958ti,Kinoshita:1958ru}:
\begin{equation}\label{eq:muonlifetime}
\frac{1}{\tau_\mu}=\frac{G_F^2m_\mu^5}{192\pi^3}\left[1-\frac{8m_e^2}{m_\mu^2}\right]\left[1+\frac{\alpha_\mathrm{em}}{2\pi}\left(\frac{25}{4}-\pi^2\right)\right]\,,
\end{equation}
leading to the value $G_F=1.16634\times 10^{-5}\,\mathrm{GeV}^{-2}$.
(For an extension of Eq.\,(\ref{eq:muonlifetime}) to $O(\alpha^3)$ and the inclusion of higher powers of $\rho\equiv(m_e/m_\mu)^2$ see Sec.\,10.2.1 of the 2022 edition of the Particle Data Group's review\,\cite{Workman:2022ynf}. 
The authors quote the corresponding value of the Fermi constant to be $G_F=1.1663787(6)\times 10^{-5}\,\mathrm{GeV}^{-2}$.) 

The evaluation of the amplitude for the process $\pi^+\to\ell\nu_\ell$ up to $O(\alpha_{\mathrm{em}})$ can be performed in the effective theory with the effective
Lagrangian\,\cite{Sirlin:1981ie,Braaten:1990ef}
\begin{equation}\label{eq:Leff}
{\cal L}_{\mathrm{eff}}=\frac{G_F}{\sqrt{2}}\,V_{ud}^\ast\left(1+\frac{\alpha_{\mathrm{em}}}{\pi}\,\log\frac{M_Z}{M_W}\right)
\big(\bar{d}\gamma^\mu(1-\gamma^5)u\big)\,\big(\bar{\nu}_\ell\gamma_\mu(1-\gamma^5)\ell)
\end{equation}
and with the Feynman-gauge photon propagator in the W-regularisation\,\cite{Sirlin:1980nh}, i.e with $1/k^2$ replaced by $M_W^2/k^2(M_W^2-k^2)$ where $k$ is the four-momentum of the photon. Since $\alpha_\mathrm{em}/\pi\,\log(M_Z/M_W)\simeq 2.9\times10^{-4}$ we drop this term in the remainder of this paper. It can readily be included if necessary.\\[0.1cm]
\nopagebreak
\noindent 2. \textit{Renormalisation of the lattice operator(s):}\\[0.1cm]
From the previous paragraphs we note that the matrix elements of bare lattice operator(s) determined in a lattice computation need to be converted into the W-regularisation scheme. This is a short-distance issue and, given the large scale $M_W$, in practice this requires some perturbation theory. 
For the Wilson action for both the fermions and gluons, the conversion was performed entirely in perturbation theory at $O(\alpha_\mathrm{em})$ in Ref.\,\cite{Carrasco:2015xwa} (see Eq.\,(10) of this reference). (Note that the lack of chiral symmetry with Wilson fermions implies that the current-current operator in Eq.\,(\ref{eq:Leff}) is a linear combination of 5 four-fermion lattice operators.) The precision of the calculation was subsequently improved from $O(\alpha_\mathrm{em}\,\alpha_s(a))$ to $O(\alpha_\mathrm{em}\,\alpha_s(M_W))$ in Ref.\,\cite{DiCarlo:2019knp}.

The discussion in this paper is independent of the choice of the lattice discretisation of QCD. Whichever choice is made in the computation of the decay width, the bare lattice operators will need to be matched to those in the W-regularisation, either using a combination of non-perturbative renormalisation and perturbation theory or entirely in perturbation theory. \\

\noindent 3. \textit{Quark and meson mass shifts:}\\[0.1cm]
Electromagnetic effects induce a shift in the masses of quarks and hadrons. Computations of hadron masses in the full theory, i.e. including electromagnetic and strong isospin breaking effects, are now performed by a number of groups\,\cite{Blum:2007cy,Blum:2010ym,deDivitiis:2013xla,
Borsanyi:2014jba,Horsley:2015eaa,Giusti:2017dmp,Boyle:2017gzv,MILC:2018ddw,CSSM:2019jmq,Feng:2021zek}. The hadron masses in the full QCD+QED theory are of course unambiguous and the computed quantities reproduce their physical values, up to statistical and systematic uncertainties.
On the other hand,
at $O(\alpha_\mathrm{em})$ computed quantities in QCD (without QED) are convention dependent, i.e. they
depend on the criteria used to determine the input bare quark masses and lattice spacing. For a detailed discussion of this point, 
see section II in Ref.\,\cite{DiCarlo:2019thl}, where a number of possible conventions for the definition of QCD are reviewed. In the present paper we will not discuss strong isospin breaking, since it does not present significant conceptual difficulties, such as the cancellation of infrared divergences and finite-volume effects which are not exponentially small. 
The presentation in this paper does not depend on the convention chosen to define QCD and so we generically label the mass of the charged pion in QCD by $m_\pi^0$ and that in the full theory by $m_\pi=m_\pi^0+\delta m_\pi$. The mass shift $\delta m_\pi$ is obtained from the time behaviour of the correlation functions as explained in Sec.\,\ref{subsec:MA}. Our focus instead, is on the determination of the decay width, which is obtained from the correlation functions after the subtraction of the term proportional to the mass-shift.

While the non-pertubative QCD effects will necessarily be determined from hadronic correlations functions computed on finite Euclidean volumes, the discussion in this paper is presented in an infinite four-dimensional volume. We identify the non-pertubative hadronic elements which need to be calculated and define and process the correlation functions from which they can be determined. We then organise the calculation in such a way that the hadronic matrix elements contributing to the width can subsequently be determined from a finite-volume computation with only exponentially small finite-volume corrections.

The correlation functions studied in Sec.\,\ref{sec:diagrams} all include an interpolating operator to create the pion at rest at time $-t_\pi$ and the hadronic weak current which annihilates the meson at the four-dimensional origin. In an infinite space-time volume $t_\pi$ can be chosen to be arbitrarily large. 
In Euclidean space the Feynman-gauge photon propagator is given in Eq.\,(\ref{eq:SgammaE}):
\begin{equation}
S_\gamma^{\mu\nu}(x,y)=\frac{\delta^{\mu\nu}}{4\pi^2|x-y|^2}\,,
\end{equation}
where $x$ and $y$ are the positions of the two electromagnetic currents in diagrams A, B and C (see Fig.\,{\ref{fig:diagrams}). In the absence of infrared divergences, one can therefore, at arbitrarily large temporal separations, {\it e.g.} $|x^4-y^4|\gtrsim t_\pi$, factor the amplitude, writing it as the product of source, sink and propagation contributions.. Infrared divergences are present however, and without an additional infrared cut-off it is $t_\pi$ which acts as the cut-off, with terms proportional to $\log[m_\pi t_\pi]$ present. Instead, we organise the discussion by implicitly introducing a separate cut-off, e.g. a mass for the photon $m_\gamma$, with $t_\pi m_\gamma\gg 1$, 
so that contributions from $|x^4-y^4|\gtrsim t_\pi$, where the photon propagator joins the source and sink factors, can now legitimately be neglected. The cancellation of the infrared divergences, which are now proportional to 
$\log(m_\pi/m_\gamma)$,  will be handled analytically and IVR will be applied to the finite terms to ensure that the finite-volume corrections are exponentially small. Whilst the logic of the discussion requires us to take the limits in the order $\lim_{m_\gamma\to 0}\lim_{t_\pi}\to\infty$, this limit is taken before the lattice calculations, which are therefore independent of $m_\gamma$ and free of infrared divergences. 

We stress that the cancellation of infrared divergences is performed fully analytically, with no lattice uncertainties. This is different for example, from the computations in QED$_\mathrm{L}$ in which an analytic expression containing the infrared divergence, which is of the form $\log[m_\pi L]$, is subtracted from the amplitude computed numerically. 
\begin{center}
\begin{figure}[t]
\includegraphics[width=0.3\hsize]{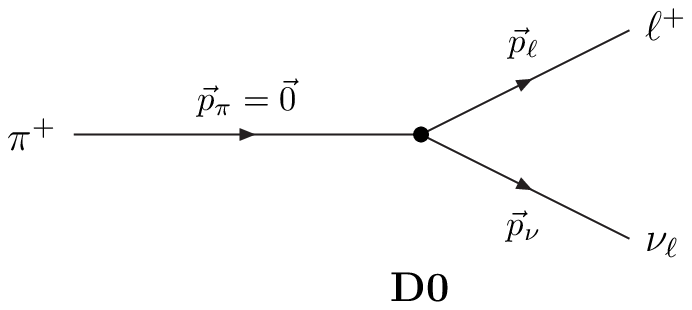}\\[0.25in]
\includegraphics[width=0.3\hsize]{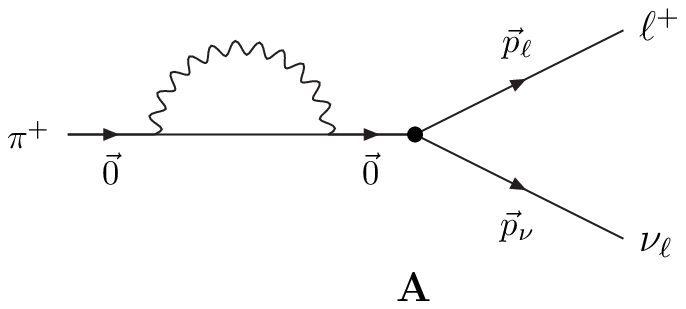}\quad
\includegraphics[width=0.3\hsize]{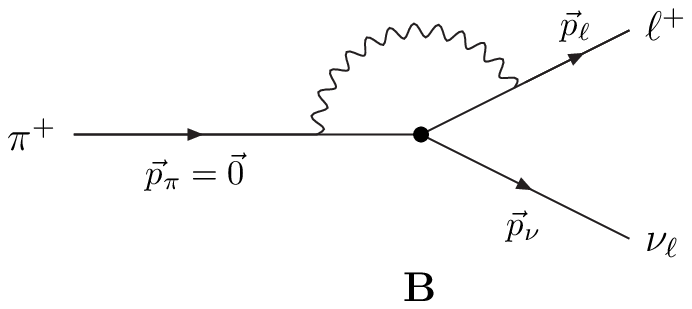}\quad
\includegraphics[width=0.3\hsize]{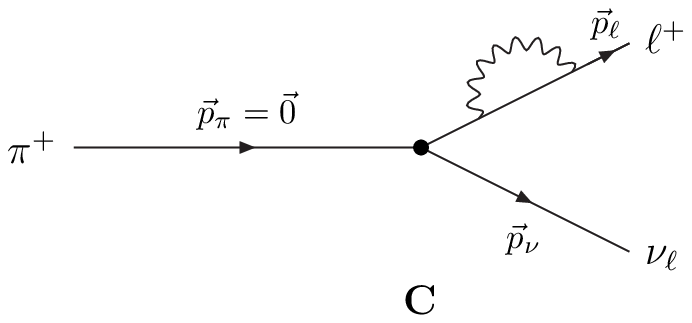}
\\[0.25in]
\includegraphics[width=0.3\hsize]{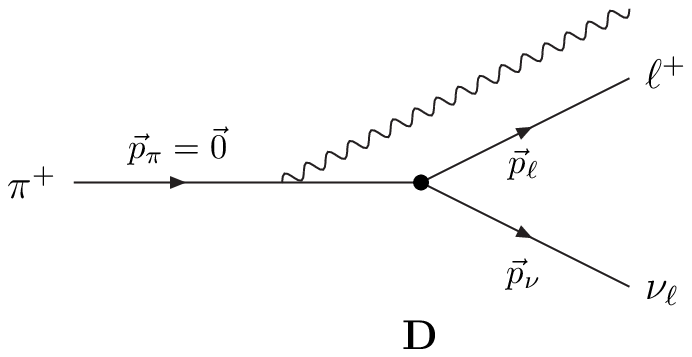}\qquad
\includegraphics[width=0.3\hsize]{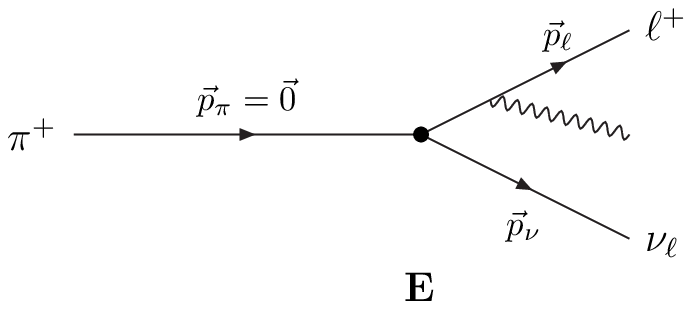}
\caption{Diagram D0 contributes to the amplitude for the decay $\pitolnu$ in the absence of electromagnetism. The remaining 5 connected diagrams contribute to the $O(\alpha_\mathrm{em})$ electromagnetic corrections to the width of the leptonic decay of a pion. Diagrams A-C correspond to the decay $\pi^+\to\ell^+\nu_\ell$ and diagrams D and E to the decay $\pitolnug$. As explained in Sec.\,\ref{sec:diagrams}, each of the five diagrams should be viewed as representing a class of diagrams at the quark and lepton level, without regard for the time ordering suggested by the representatives shown.
\label{fig:diagrams}}
\end{figure}
\end{center}

\vspace{-0.4in} In the following sections we present the implementation of the IVR method in leptonic decays in detail, but we now introduce the main ideas. 
The introduction of radiative corrections, with a photon which can propagate over large distances, results in the presence of both infrared divergences and finite-volume corrections which potentially only decrease slowly with the volume (as inverse powers of $L$, the spacial extent of the volume). 
The fundamental idea of the IVR method is that there is a time interval $t_s\lesssim L$ such that the only hadronic state which contributes significantly to correlation functions when propagating over times greater than $t_s$ is the pion; contributions from states with larger masses are exponentially suppressed. To illustrate the method, consider the hadronic matrix element
\begin{equation}
H(\vec{x},-t)\equiv\langle f| T\,[\,O_2(0)\,O_1(\vec{x},-t)]\,|\pi(\vecp{0})\rangle\,,\label{eq:Hdef}
\end{equation}
where $O_{1,2}$ are local operators, $T$ represents time-ordering and the initial state $|\pi(\vecp{0})\rangle$ is a pion at rest, (i.e. with three-momentum $\vecp{0}$). 
For our specific study of leptonic decays we show in Fig.\,\ref{fig:diagrams} the diagram without electromagnetism and the five diagrams which include electromagnetism and contribute to the $\pi^+\to\ell^+\nu_\ell(\gamma)$ decay amplitude
(we include electromagnetic corrections up to $O(\alpha_\mathrm{em})$ in the decay width).
For diagrams B and D, the final state $|f\rangle=|0\rangle$ and $O_1$ and $O_2$ are electromagnetic and weak currents respectively. For diagram A, 
both $O_{1}$ and $O_2$ are electromagnetic currents and if $-t<0$ and the time at which the weak current is inserted, $t_W$, is sufficiently large and positive so the propagation of states other than the pion
between $O_2(0)$ and the weak current at $t_W$ is suppressed, then $|f\rangle=|\pi(\vecp{0})\rangle$.
In the evaluation of the diagrams, $H(\vec{x},-t)$ is a factor in the integrand of integrals of the genetic form
\begin{equation}\label{eq:integral}
\int\,\dfour x~H(x)\,f(x)\,,
\end{equation}
where $f(x)$ encodes non-hadronic $x$-dependent elements such as the photon and lepton propagators.
We now demonstrate that for $t>t_s$, $H(\vec{x},-t)$ can be determined analytically in terms of $H(\vec{x},-t_s)$. It is therefore unnecessary to perform non-pertubative computations of $H(\vec{x},-t)$ for $t>t_s$.  By the assumption that only pion intermediate states propagate between $(\vec{x},-t_s)$ and the origin we have
\begin{eqnarray}
H(\vec{x},-t_s)&=&\int\frac{\dthree p}{(2\pi)^3}\,\frac1{2E_\pi(\vecp{p})}\,\langle f|  O_2(0)|\pip\rangle
\langle\pip |O_1(\vec{x},-t_s)|\pirest\rangle\nonumber\\ 
&=&\int\frac{\dthree p}{(2\pi)^3}\,\frac1{2E_\pi(\vecp{p})}\,\langle f|  O_2(0)|\pip\rangle
\langle\pip |O_1(0)|\pirest\rangle
e^{-(E_\pi(\vecp{p})-m_\pi)t_s}e^{-i\vec{p}\cdot\vec{x}}
\,,\label{eq:Hts}
\end{eqnarray}
where $E_\pi(\vecp{p})=\sqrt{|\vec{p}\hspace{1.5pt}|^2+m_\pi^2}$\,. Performing the inverse Fourier transform we obtain 
\begin{equation}
\frac1{2E_\pi(\vecp{p})}\,\langle f|  O_2(0)|\pip\rangle
\langle\pip |O_1(0)|\pirest\rangle=
\int\dthree x\,H(\vec{x},-t_s)e^{(E_\pi(\vecp{p})-m_\pi)t_s}e^{i\vec{p}\cdot\vec{x}}\,.
\label{eq:Haux1}\end{equation}
For values of $t>t_s$ following the same steps as in Eq.\,(\,\ref{eq:Hts}) we have
\begin{eqnarray}
H(\vec{x},-t)|_{t>t_s}&=&
\int\frac{\dthree p}{(2\pi)^3}\,\frac1{2E_\pi(\vecp{p})}\,\langle f|  O_2(0)|\pip\rangle
\langle\pip |O_1(0)|\pirest\rangle
e^{-(E_\pi(\vecp{p})-m_\pi)t}e^{-i\vec{p}\cdot\vec{x}}
\label{eq:HL1}\\
&=&
\int\frac{\dthree p}{(2\pi)^3}\,\int\dthree x^\prime~H(\vec{x}^{\,\prime},-t_s)\,
e^{-(E_\pi(\vecp{p})-m_\pi)(t-t_s)}e^{-i\vec{p}\cdot(\vec{x}-\vec{x}^{\,\prime})}\,.
\label{eq:HL2}
\end{eqnarray}
We see therefore that $H(\vec{x},-t)|_{t>t_s}$ can be determined from the knowledge of $H(\vec{x},-t_s)$, which is now the non-perturbative input into the evaluation of the 
decay amplitude. The discussion here is infinite volume, but we envisage that the non-perturbative determination of $H(\vec{x},-t_s)$ will ultimately be performed in a finite-volume lattice QCD computation. The long-distance behaviour of the correlation function in Eq.\,(\ref{eq:HL1})
is of the form $\exp[-m_\pi(\sqrt{|\vec x|^2 + t^2} -t)]$ with prefactors which include negative powers of $\sqrt{|\vec x|^2 + t^2}$. Thus, in the region of large $t$, increasing $|\vecp x|$ has little effect 
until $|\vecp x|^2 m_\pi$ becomes of the order of $t$. 
Furthermore, contributions from regions of large 
$t>\!>|\vecp{x}|^2m_\pi$  are only suppressed by powers of $t$. Consequently, in a finite-volume lattice
calculation, the omission of the large $|\vecp{x}|$ region at large $t$ from the integral in 
Eq.\,(\ref{eq:integral}), where $f(x)$ contains the photon propagator, results in 
power law, finite-volume errors.
However, as
is shown in Eq.\,(\ref{eq:HL2}), IVR allows the contribution from this troublesome region
of large $t$ to be determined analytically in infinite volume from the calculation
of hadronic matrix elements at, for example, a fixed value of $t = t_s$. The exponential
decrease for large $|\vecp{x}|$ at fixed $t_s$ ensures exponentially vanishing corrections
as the volume used in the lattice calculation grows. Moreover, no finite-volume effects are introduced by
the momentum integration on the right-hand side of Eq.\,(\ref{eq:HL2}), since this is always performed in infinite-volume.  In the following sections we exploit the IVR technique illustrated above to develop a complete procedure for the computation of electromagnetic corrections to the leptonic decay widths of pseudoscalar mesons.

In the previous paragraph, in order to illustrate the method, we demonstrated that performing the computations on lattices of increasing volumes with a fixed value of $t_s$ led to exponentially small finite-volume effects. There also exist other possibilities to achieve this, although the rate of decrease of the exponentially falling finite volume effects will be different in each case. For example, we can increase $t_s$ as the volume increases while keeping the ratio $t_s/L$ fixed, with $t_s\lesssim L$ . 
Increasing $t_s$ in this way, enables us to combine the reduction of finite-volume effects with a decrease of any possible contamination from contributions of excited states at $t=t_s$.

The plan for the remainder of the paper is as follows. In the next section we discuss the evaluation of the diagrams.
The terms containing the infrared divergences are separated from the finite terms. The analytic cancellation of the infrared divergences in the width between diagrams with a virtual photon (diagrams A,B and C) and those with a real photon (diagrams D and E) is demonstrated in Sec.\,\ref{sec:ircancellation}. 
We collect all the terms contributing to the final result for the decay width
in Sec.\,\ref{sec:final} and in Sec.\,\ref{sec:concs} we present a brief summary and our conclusions. 

There are three appendices: in Appendix\,\ref{sec:conventions} we present the conventions we use in Minkowski and Euclidean space. 
In the main body of the paper we identify the terms which lead to infrared divergences in the widths. While the cancellation of infrared divergences is manifest, finite terms remain after the addition of the individually divergent terms. 
These residual finite terms are derived in Appendix\,\ref{sec:finiteterms} and only require knowledge of the decay constant and the matrix element of the interpolating operator of the pion. Finally, in Appendix\,\ref{sec:Kl3} we sketch how IVR can be implemented in $K_{\ell 3}$ decays.

\section{Evaluation of the diagrams}\label{sec:diagrams}
 
The five diagrams which contribute to the $\pi^+\to\ell^+\nu_\ell(\gamma)$ decay amplitudes are illustrated in Fig.\,\ref{fig:diagrams}. They indicate whether the photon is attached to the hadronic or leptonic components of the electromagnetic current(s) (see Eq.\,(\ref{eq:Jemdef}) below). Thus, for example, in diagram A the photon is emitted and absorbed on quark propagators, whereas in diagram B it is emitted from a quark propagator and absorbed by the charged lepton. We stress that the diagrams are a representation of QCD+QED, and that their evaluation in lattice computations is to be performed in a discretisation of QCD at the quark and gluon level. The diagrams are not to be interpreted as corresponding to some effective theory.
``Disconnected" diagrams, i.e. those in which 
the photon is emitted and/or absorbed from a closed
quark loop which is connected to the remainder of the diagram only by gluons, are implicitly included in the diagrams of Fig.\,\ref{fig:diagrams}. While it is generally more difficult to compute such disconnected diagrams numerically, they are included in the framework presented in this paper. At $O(\alpha_\mathrm{em})$ there are no diagrams in which a photon is attached to a closed lepton loop. 

Below we discuss the evaluation of the diagrams of Fig.\,\ref{fig:diagrams} which contribute to the $\pi^+\to\ell^+\nu_\ell(\gamma)$ decay amplitudes. 
More precisely, we define the correlation functions corresponding to each of the diagrams and organise them so the hadronic matrix elements which contribute to 
the amplitudes can ultimately be computed on a finite Euclidean lattice with only exponentially small finite-volume effects. 
Infrared divergent contributions are identified and the cancellation of the divergences is performed analytically, so that the matrix elements which need to be computed 
are all individually infrared finite. 
The numerical evaluation of the matrix elements is postponed to a future study.

We start by defining the fundamental ingredients in the computation of the amplitude, and in particular the hadronic matrix elements and leptonic factors. Since the energy-momentum exchanges in this decay are much smaller than the mass of the $W$-boson, the weak vertex is rewritten as a local four-fermion interaction as in Eq.\,(\ref{eq:Leff})
\begin{equation}\label{eq:LW}
{\mathcal L}_\mathrm{eff}=\frac{G_F}{\sqrt{2}}\,V_{ud}^\ast\,g_{\mu\nu}\,J_W^\mu(x)\,\big[\bar{\nu}_\ell(x)\gamma^\nu(1-\gamma^5)\ell(x)\big]\,,
\end{equation}
where the weak hadronic current $J_W^\mu=\bar{d}\gamma^\mu(1-\gamma^5)u$.

Throughout this paper, we take the initial pion to be at rest, denoting the corresponding QCD eigenstate by $|\pi(\vecp{0})\rangle$, and define the hadronic matrix elements as follows:
\begin{eqnarray}
H_0^\mu&=&\langle 0| J_W^\mu(0)|\pi(\vecp{0})\rangle\label{eq:H0def}\\ 
H_1^{\mu\nu}(x)&=&\langle 0| T\big[ J^\nu_\mathrm{em}(x) J_W^\mu(0)\big]|\pi(\vecp{0})\rangle\label{eq:H1def}\\
H_2^{\mu\nu\rho}(x,y)&=&\langle 0| T\big[ J^\nu_\mathrm{em}(x) J^\rho_\mathrm{em}(y) J_W^\mu(0)\big]|\pi(\vecp{0})\rangle
\label{eq:H2def}\\
H_{2s}^{\nu\rho}(z)&=&\langle \pi(\vecp{0}) | T\big[ J^\nu_\mathrm{em}(0)J^\rho_\mathrm{em}(z)\big] |\pi(\vecp{0})\rangle\,.
\label{eq:H2sdef}
\end{eqnarray}
We use the normalization conventions $\langle\pi(\vecp p)|\pi(\vecp p^\prime)\rangle =(2\pi)^3 2E_\pi\delta(\vec p - \vecp p^\prime)$ for the state $|\pi(\vecp p)\rangle$. In the above  equations $J^\mu_W$ is the weak hadronic current and $J^\nu_{\mathrm{em}}$ is the electromagnetic current
\begin{equation}
J^\nu_{\mathrm{em}}=\sum_f Q_f\,\bar{q}_f\gamma^\nu q_f-\sum_{\ell}\bar{\ell}\gamma^\nu\ell~,
\label{eq:Jemdef}
\end{equation}
where the charges $Q_f=+\frac23$ for up-like quarks, $-\frac13$ for down-like ones and $-1$ for the leptons $\ell$. 
The appearance of $H_0$ in diagrams D0, C and E, $H_1$ in diagrams B and D and $H_2$ in diagram A can be readily understood. We will explain the appearance of $H_{2s}$ when discussing the evaluation of diagram A. 

Diagrams A-C contain the propagator of a photon which, in position space with the photon propagating between $x$ and $y$, we denote by $S^\gamma_{\mu\nu}(x,y)$ where $\mu,\nu$ are Lorentz indices. We denote the four-momenta of the final state charged lepton and neutrino by $p_\ell=(E_\ell,\vec{p}_\ell)$ and $p_\nu=(|\vec{p}_\nu|,\vec{p}_\nu)$, and for this two-body $\pitolnu$ decay $E_\ell$ and $|\vec{p}_\ell|$ are fixed by the masses of the pion and lepton. Diagrams D and E have a real photon in the final state and we denote its polarisation vector by $\epsilon_\lambda^{\nu}(k)$, where $k$ is the momentum of the photon and $\lambda$ labels its polarisation. We denote the four-momenta of the final state charged lepton, neutrino and photon in diagrams D and E by $p_\ell=(E_\ell,\vec{p}_\ell)$, $p_\nu=(|\vec{p}_\nu|,\vec{p}_\nu)$ and $k=(|\vecp{k}|,\vec{k}\,)$ respectively. In all the diagrams, the energy of the final state lepton is given by $E_\ell=\sqrt{|\vec{p}_\ell|^2+m_\ell^2}$.

The hadronic matrix elements in Eqs.\,(\ref{eq:H0def})\,-\,(\ref{eq:H2sdef}) are combined with the photon propagator $S^\gamma_{\mu\nu}$ or polarisation vectors $\epsilon_\lambda^{\nu}$ and the corresponding leptonic factors
\begin{eqnarray}
L_0^\mu&=&\bar{u}(p_{\nu_\ell})\gamma^\mu(1-\gamma^5)v(p_\ell) \label{eq:L0def}\\
L_1^{\mu\nu}(x)&=&\bar{u}(p_{\nu_\ell})\gamma^\mu(1-\gamma^5)S_\ell(0,x)\gamma^\nu v(p_\ell) e^{-ip_\ell\cdot x}\label{eq:L1def}\\
L_2^{\mu\nu\rho}(x)&=&\bar{u}(p_{\nu_\ell})\gamma^\mu(1-\gamma^5)S_\ell(0,x)\gamma^\nu 
S_\ell(x,y)\gamma^\rho
v(p_\ell) e^{-ip_\ell\cdot y}\label{eq:L2def}\label{eq:L2}
\end{eqnarray} 
and subsequently integrated over $x$ and $y$ as appropriate.

In the following we start by writing down the contribution from each diagram to the decay amplitude with all quantities, and in particular the $\gamma$-matrices and photon's polarisation vector, in Minkowski space as presented in Appendix\,\ref{subsec:conventionsM}.  
The earlier expressions in this section, from Eq.\,(\ref{eq:LW}) to Eq.\,(\ref{eq:L2}) were all written in terms of these Minkowski-space quantities. 
Since the hadronic matrix elements are eventually to be evaluated in lattice computations in a finite Euclidean volume we rewrite and process these contributions in terms of Euclidean quantities as defined in Appendix\,\ref{subsec:conventionsE}. We stress however, that the expressions in both cases are exactly equivalent. An important point to recall is that the discussion in this section is in infinite volume (both temporal and spatial). We will identify the hadronic elements which need to be calculated and organise them in such a way that they can be computed on a finite lattice with only exponentially small finite-volume corrections.


\subsection{The amplitude in QCD without QED}\label{subsec:M0}
The amplitude for the leptonic decay $\pi^+\to\ell^+\nu_\ell$ in pure QCD, i.e. neglecting electromagnetism, is represented by diagram D0 in Fig.\,\ref{fig:diagrams} and is simply given by 
\begin{equation}
M_0=\frac{G_F}{\sqrt{2}}\,V_{ud}^\ast~H_0^\mu L_0^\nu\, g_{\mu\nu}
=-\frac{G_F}{\sqrt{2}}\,V_{ud}^\ast~H_0^0 L_0^0\,,\label{eq:M0Minkowski}
\end{equation}
where we recall that we are using the metric $g_{\mu\nu}=\mathrm{diag}(-1,1,1,1)$ and that we take the meson in the initial state to be at rest so that only the time component of $H_0^\mu$ is not zero. 
Rewriting this expression in terms of Euclidean $\gamma$-matrices as defined in Sec.\,\ref{subsec:conventionsE} we have
\begin{equation}
M_0=-\GF~H_{0E}^4 L_{0E}^4\,,\label{eq:M0Euclidean}
\end{equation}
where the subscript $E$ is included to indicate that the $\gamma$-matrices are in Euclidean space, distinguishing the expression from Eq.\,(\ref{eq:M0Minkowski}) where they are in Minkowski space. Nevertheless the two expressions are identical of course.  

The matrix element $H_{0E}^4$ is obtained from the following lowest order QCD (without QED) correlation functions with quark masses corresponding to a $\pi^+$ meson with mass $m_\pi^0$:
\begin{eqnarray}
C^0_{\phi\phi}(t_\pi)&=&\int \dthree x\,\langle 0|\phi(0)\,\phi^\dagger(-t_\pi,\vecp{x})\,|0\rangle
= |Z_0|^2\,\frac{e^{-m_\pi^0 t_\pi}}{2m_\pi^0} + \cdots
\label{eq:C0phiphidef}\\
C^{0}_{J_W\phi}(t_\pi)&=&\int \dthree x\,\langle 0|J_W^4(0)\,\phi^\dagger(-t_\pi,\vecp{x})\,|0\rangle
=H^4_{0E}\,Z_0\,\frac{e^{-m_\pi^0 t_\pi}}{2m_\pi^0} + \cdots~,
\label{eq:C0JWphidef}
\end{eqnarray}
where $t_\pi>0$, $\phi^\dagger$ is an interpolating operator which can create the pion from the vacuum and the QCD matrix element $Z_0$ is defined as $Z_0=\langle \pi(\vecp{0})|\phi^\dagger(0)|0\rangle$ evaluated in QCD without QED\,\footnote{Note that our definition of $Z_0$ differs from 
the convention
$\sqrt{Z_0}=\langle \pi(\vecp{0})|\phi^\dagger(0)|0\rangle$ frequently used in the normalisation of quantum fields.}. 
The superscript $0$ on $m_\pi^0$ indicates that this is the pion mass evaluated in QCD, before the shift in quark masses induced by electromagnetic interactions.
It is assumed that $t_\pi$ is sufficiently large that the correlation functions are dominated by the propagation of a single pion at rest and the ellipsis represent the contributions from the excited states. These will be assumed to be negligible and in the following presentation we drop the ellipsis. 
The hadronic matrix element $H^4_{0E}$ is obtained by combining Eqs.\,(\ref{eq:C0phiphidef}) and (\ref{eq:C0JWphidef}):
\begin{equation}
H^4_{0E}=[2m_\pi^0]^\frac12\,\frac{C^{0}_{J_W\phi}(t_\pi)}{[C^0_{\phi\phi}(t_\pi)e^{-m_\pi^0 t_\pi}]^\frac12}\,.\label{eq:H40Eexpression}
\end{equation}

Since throughout the discussion below the initial pion is at rest, in the following we will use the shorthand notation for $\phi^\dagger(-t_\pi)$ (i.e. $\phi^\dagger$ with a single variable):
\begin{equation}
\phi^\dagger(-t_\pi)\equiv\int\dthree x~ \phi^\dagger(-t_\pi,\vecp{x})\,.
\end{equation}
The correlation functions from which the contributions to the decay amplitudes from each of the diagrams in Fig.\,\ref{fig:diagrams} are determined
are defined in the following sections.

The contribution to the decay width in the absence of QED is given by 
\begin{equation}
\Gamma_0(\pi^+\to\ell^+\nu_\ell)=\frac{G_F^2\,|V_{ud}|^2f_\pi^2}{8\pi}\,m_\pi\,m_\ell^2\left(1-\frac{m_\ell^2}{m_\pi^2}\right)^{\!\!2}\,,
\label{eq:Gamma0}
\end{equation}
where the leptonic decay constant $f_\pi$ is obtained from $|H_{0E}^4|^2=m_\pi^2\,f_\pi^2$\,.

\subsection{Contribution to the amplitude from Diagram A}\label{subsec:MA} 

Diagram A contributes at $O(\alpha_\mathrm{em})$ to both the mass of the pion and to the decay amplitude; the latter through the wave function renormalisation of the pion and a correction to the weak interaction vertex. The leptonic factor $L_{0E}^4$ is common to both $M_0$ and the contribution from diagram A and we define the Euclidean correlation function corresponding to diagram A as:
\begin{equation}\label{eq:CJphiA}
C_{J_W\phi}^{A }(t_\pi)=-\frac{e^2}2\int\dfour x\int\dfour y~\langle 0|T\big[J_W^4(0)J_\mathrm{em}^\mu(x)J_\mathrm{em}^\nu(y)\phi^\dagger(-t_\pi)\big]|0\rangle\,S_\gamma^{\mu\nu}(x,y)\,,
\end{equation}
where the $\frac12$ is the standard combinatorial factor and the superscript $A$ indicates the contribution of diagram A. As the discussion in this subsection is presented entirely in Euclidean space, we do not include an explicit subscript $E$ to denote \emph{Euclidean}.
Combining $C_{J_W\phi}^{A }$ with $C_{J_W\phi}^{0}$ (defined in Eq.\,(\ref{eq:C0JWphidef})) gives
\begin{eqnarray}\label{eq:CJWphi2}
C^{0}_{J_W\phi}(t_\pi)+C_{J_W\phi}^{A}(t_\pi)&=&\frac{H^4_\mathrm{full}Z_\mathrm{full}}{2(m_\pi^0+\delta m_\pi)}\,e^{-(m_\pi^0+\delta m_\pi)t_\pi}\,,
\end{eqnarray}
where $\rule[-5pt]{0pt}{15pt}H^4_\mathrm{full}\equiv\langle 0|J_W^4(0)|\pi(\vecp{0})\rangle_\mathrm{full}$, $Z_\mathrm{full}=\langle\pi(\vecp{0})|\phi^\dagger(0)|0\rangle_\mathrm{full}$ and the label {\footnotesize$\mathrm{full}$} implies that the matrix element is defined in QED+QCD up to $\rule[-10pt]{0pt}{20pt}O(\alpha_\mathrm{em})$. At this order we can write
\begin{equation}
C^{0}_{J_W\phi}(t_\pi)+C_{J_W\phi}^{A}(t_\pi)=\left(-\delta m_\pi t_\pi\right)\,\frac{H_0^4\,Z_0}{2m_\pi^0}\,e^{-m_\pi^0t_\pi}
+
\frac{H^4_\mathrm{full}Z_\mathrm{full}}{2(m_\pi^0+\delta m_\pi)}\,e^{-m_\pi^0t_\pi}\,,\label{eq:C0CAJWphi}
\end{equation}
so that the mass-shift $\delta m_\pi$ can be obtained from a study of the $t_\pi$ behaviour of the correlation function.

Similarly, defining 
\begin{equation}\label{eq:phiphiA}
C_{\phi\phi}^{A}(t_\pi)\equiv -\frac{e^2}2\int\dfour x\int\dfour y~\langle 0|T\big[\phi(0)J_\mathrm{em}^\mu(x)J_\mathrm{em}^\nu(y)\phi^\dagger(-t_\pi)\big]|0\rangle\,S_\gamma^{\mu\nu}(x,y)\,,
\end{equation}
and following the same steps we obtain
\begin{eqnarray}\label{eq:Cphiphi2}
C^{0}_{\phi\phi}(t_\pi)+C_{\phi\phi}^{A }(t_\pi)&=&\frac{\left|Z_\mathrm{full}\right|^\mathrm{2}}{2(m_\pi^0+\delta m_\pi)}\,e^{-(m_\pi^0+\delta m_\pi)t_\pi}
\nonumber\\
&=&\left(-\delta m_\pi t_\pi\right)\,\frac{Z_0^2}{2m_\pi^0}\,e^{-m_\pi^0t_\pi}+
\frac{\left|Z_\mathrm{full}\right|^2}{2(m_\pi^0+\delta m_\pi)}\,e^{-m_\pi^0t_\pi}\,.\label{eq:C0CAphiphi}
\end{eqnarray}

The mass-shift $\delta m_\pi$ is obtained from the coefficient of $t_\pi$ in Eqs.\,(\ref{eq:C0CAJWphi}) and (\ref{eq:C0CAphiphi}). In order to simplify the notation in the later discussion we define 
\begin{eqnarray}
\tilde{C}_{J_W\phi}^{A}(t_\pi)&=&C_{J_W\phi}^{A}(t_\pi)-\left(-\delta m_\pi t_\pi\right)\,C^{0}_{J_W\phi}(t_\pi)\label{eq:tildeCAJWphidef}\\ 
\tilde{C}_{\phi\phi}^{A}(t_\pi)&=&C_{\phi\phi}^{A}(t_\pi)-\left(-\delta m_\pi t_\pi\right)\,C^{0}_{\phi\phi}(t_\pi)\,,\label{eq:tildeCAphiphidef}
\end{eqnarray}
so that $\tilde{C}_{J_W\phi}^{A}(t_\pi)$ and $\tilde{C}_{\phi\phi}^{A}(t_\pi)$ are the contributions to the correlation functions after the subtraction of the linear term in $t_\pi$ which is proportional to the mass-shift. It is from these subtracted correlation functions that the contribution to the decay amplitude is obtained.
Thus by studying the $t_\pi$ dependence of $C^{0}_{J_W\phi}(t_\pi)+C_{J_W\phi}^{A}(t_\pi)$ and $C^{0}_{\phi\phi}(t_\pi)+C_{\phi\phi}^{A }(t_\pi)$, the matrix element 
$H_\mathrm{full}^4\equiv\langle 0|J_W^4(0)|\pi(\vec{0})\rangle_\mathrm{full}$
can be determined:
\begin{eqnarray}
H^4_\mathrm{full}&=&2(m_\pi^0+\delta m_\pi)\frac{\big(C^0_{J_W\phi}(t_\pi)+\tilde C^A_{J_W\phi}(t_\pi)\big)e^{m_\pi^0t_\pi}}{Z_\mathrm{full}}\nonumber\\
&=&\big[2(m_\pi^0+\delta m_\pi)\big]^{\frac12}\,\frac{C^0_{J_W\phi}(t_\pi)+\tilde C^A_{J_W\phi}(t_\pi)}{\Big[\big(C^0_{\phi\phi}(t_\pi)+
\tilde C^A_{\phi\phi}(t_\pi)\big)\,e^{-m_\pi^0t_\pi}\Big]^{\frac12}}\nonumber\\
&\simeq&\big[2m_\pi^0\big]^{\frac12}\,
\left[\left(1+\frac{\delta m_\pi}{2m_\pi^0}\right)\frac{C^0_{J_W\phi}(t_\pi)}{[C^0_{\phi\phi}(t_\pi)\,\,e^{-m_\pi^0t_\pi}]^\frac12}
+\frac{\tilde{C}^A_{J_W\phi}(t_\pi)}{[C^0_{\phi\phi}(t_\pi)\,\,e^{-m_\pi^0t_\pi}]^\frac12}
-\frac12\frac{\tilde C^A_{\phi\phi}(t_\pi)}{[C^0_{\phi\phi}(t_\pi)\,\,e^{-m_\pi^0t_\pi}]^\frac12}\,\frac{H_0^4}{Z_0}
\right]\label{eq:H4fullaux1}\\ 
&\equiv& H^4_0+H^4_A\,,\label{eq:H4fullaux2}
\end{eqnarray}
where $H^4_0$ is given in Eq.\,(\ref{eq:H40Eexpression})\,.

The corresponding contribution to the amplitude is the extension of Eq.\,(\ref{eq:M0Euclidean}) to $O(\alpha_\mathrm{em})$:
\begin{equation}\label{eq:MA}
M_A=-\GF~H_{\mathrm{A}}^4 L_{0}^4\,.
\end{equation}

In the evaluation of the correlation functions $C_{J_W\phi}^{A}(t_\pi)$ and $C_{\phi\phi}^{A}(t_\pi)$ we exploit the symmetry under $x\leftrightarrow y$ and consider only the contribution from the region $x_4>y_4$ and introduce a factor of 2. We also divide the time integrations into 4 regions $R_i$, $i=1$\,-\,$4$,
\begin{eqnarray}
R_1\!\!: &\quad x_4>-t_s,~&~y_4>-t_\pi+t_s\nonumber\\ 
R_2\!\!: &\quad x_4<-t_s,~&~y_4>-t_\pi+t_s\nonumber\\
R_3\!\!: &\quad x_4<-t_s,~&~y_4<-t_\pi+t_s\label{eq:regions}\\
R_4\!\!: &\quad x_4>-t_s,~&~y_4<-t_\pi+t_s\,,\nonumber
\end{eqnarray}
where $t_s>0$ is sufficiently large  that the propagation of excited hadronic states can be neglected for time intervals greater than $t_s$. In principle, the limits on $y_4$ defining the four regions in Eq.\,(\ref{eq:regions}) could be $-t_\pi+t_s^\prime$, with $t_s^\prime\neq t_s$ but still sufficiently large for the contribution of excited states to be negligible for time intervals greater than $t_s^\prime$. For the remainder of this paper however, we simply set $t_s^\prime=t_s$ as in Eq.\,(\ref{eq:regions}).

In region $R_4$ the photon propagates for a time interval of at least $t_\pi-2t_s$. In an infinite volume with infinite time extent, the situation being considered in the current discussion, $t_\pi$ can be set to be arbitrarily large and hence the contribution to the correlation functions from this region can be made arbitrarily small. This term can therefore be neglected. 

We now write the expressions for the correlation function $C_{J_W\phi}^{A}(t_\pi)$ and $C_{\phi\phi}^{A}(t_\pi)$ in each of the three remaining regions in a way which will be useful for the discussion below:
\begin{eqnarray}
C_{J_W\phi;R_1}^{A}(t_\pi)&=&-e^2\int^{\infty}_{-t_s} \hspace{-0.05in}dx_4\int_{-t_\pi+t_s}^{x_4} \hspace{-0.2in}dy_4\int\dthree x\int\dthree y~\langle 0|T\big[J_W^4(0)J_\mathrm{em}^\mu(x)J_\mathrm{em}^\nu(y)\big]\phi^\dagger(-t_\pi)|0\rangle\,S_\gamma^{\mu\nu}(x,y)\nonumber\\ 
&=&-e^2\frac{Z_0}{2m_\pi^0}\,e^{-m_\pi^0 t_\pi}\int^{\infty}_{-t_s} \hspace{-0.05in}dx_4\int_{-t_\pi+t_s}^{x_4} \hspace{-0.2in}dy_4\int\dthree x\int\dthree y~\langle 0|T\big[J_W^4(0)J_\mathrm{em}^\mu(x)J_\mathrm{em}^\nu(y)\big]|\pi(\vecp{0})\rangle\,S_\gamma^{\mu\nu}(x,y)\label{eq:CAJphi1}\\[0.15in]
C_{J_W\phi;R_2}^{A}(t_\pi)&=&-e^2\int_{-t_\pi+t_s}^{-t_s} \hspace{-0.05in}dx_4\int_{-t_\pi+t_s}^{x_4} \hspace{-0.2in}dy_4\int\dthree x\int\dthree y~\langle 0|J_W^4(0)J_\mathrm{em}^\mu(x)J_\mathrm{em}^\nu(y)\phi^\dagger(-t_\pi)|0\rangle\,S_\gamma^{\mu\nu}(x,y)\nonumber\\ 
&&\hspace{-0.8in}=-e^2\frac{Z_0H_{0}^4}{(2m_\pi^0)^2}\,e^{-m_\pi^0 t_\pi}\!\!\int_{-t_\pi+t_s}^{-t_s} \hspace{-0.15in}dx_4\int_{-t_\pi+t_s}^{x_4} \hspace{-0.2in}dy_4\int\dthree z~
\langle \pi(\vecp {0})|J_\mathrm{em}^\mu(x_4,\vecp{0})J_\mathrm{em}^\nu(y_4,\vecp{z})|\pi(\vecp{0})\rangle\,S_\gamma^{\mu\nu}\big((x_4,\vecp{0}),(y_4,\vecp{z})\big)
\label{eq:CAJphi2}\\[0.15in]
C_{J_W\phi;R_3}^{A}(t_\pi)&=&-e^2\int_{-\infty}^{-t_\pi+t_s} \hspace{-0.15in}dy_4\int_{y_4}^{-t_s} \hspace{-0.2in}dx_4\int\dthree x\int\dthree y~\langle 0|J_W^4(0)T\big[J_\mathrm{em}^\mu(x)J_\mathrm{em}^\nu(y)\phi^\dagger(-t_\pi)\big]|0\rangle\,S_\gamma^{\mu\nu}(x,y)\nonumber\\ 
&&\hspace{-0.8in}=-e^2\frac{H_{0}^4}{2m_\pi^0}\int_{-\infty}^{-t_\pi+t_s}\hspace{-0.15in} dy_4\int_{y_4}^{-t_s}\hspace{-0.1in} dx_4\int\dthree z
~\langle \pi(\vecp{0})|T\big[J_\mathrm{em}^\mu(x_4,\vecp{0})J_\mathrm{em}^\nu(y_4,\vecp{z})\phi^\dagger(-t_\pi)\big]|0\rangle\,
S_\gamma^{\mu\nu}\big((x_4,\vecp{0}), (y_4,\vecp{z})\big)
\label{eq:CAJphi3}\,.
\end{eqnarray}
The corresponding expressions for the $C_{\phi\phi;R_i}^A(t_\pi)$, $i=1,2,3$, are obtained from those in Eqs.(\ref{eq:CAJphi1})\,-\,(\ref{eq:CAJphi3}) by replacing $H^{4}_0$ by $Z_0$ and $J_W^4(0)$ by $\phi(0)$.

Note that the integrals in Eqs.\,(\ref{eq:CAJphi2}) and (\ref{eq:CAJphi3}) are common to both $C_{J_W\phi;R_i}^A(t_\pi)$ and the corresponding $C_{\phi\phi;R_i}^A(t_\pi)$ ($i=2,3$) so that
\begin{equation}\label{eq:partialcancellation}
\tilde{C}^A_{J_W\phi;R_i}(t_\pi)=\tilde{C}^A_{\phi\phi;R_i}(t_\pi)\,\frac{H_0^4}{Z_0}
\qquad (i=2,3)
\end{equation}
which leads to a partial cancellation in Eq.\,(\ref{eq:H4fullaux1}).
We shall show in the following subsections that Eq.\,(\ref{eq:partialcancellation}) together with the observation that 
$C^A_{\phi\phi:R_1}(t_\pi)=C^A_{\phi\phi:R_3}(t_\pi)$, leads to considerable simplifications, specifically that to obtain $H^4_A$ it is sufficient to compute the $O(\alpha_\mathrm{em})$ correlation functions $\tilde{C}_{J_W\phi;R_1}(t_\pi)$ and $\tilde{C}_{J_W\phi;R_2}(t_\pi)$ (see Eq.\,(\ref{eq:H4Afinal}) below).

The infrared divergent terms in $C^A_{J_W\phi}(t_\pi)$ and $C^A_{\phi\phi}(t_\pi)$, as well as the shift in the pion mass, come from region $R_2$ defined in Eq.\,(\ref{eq:regions}). We therefore start the discussion of these correlation functions by considering the contributions from region $R_2$ in Sec.\ref{subsubsec:R2}. Although the contributions from regions $R_1$ and $R_3$ are infrared finite, they do allow for the propagation of a single pion over large (i.e. $>t_s$) time intervals which in a finite-volume would lead to large, non-exponential, finite-volume effects. These are eliminated by the use of IVR as explained in Secs.\,\ref{subsubsec:R1} and \ref{subsubsec:R3}.  

\subsubsection{Contribution from Region $R_2$}\label{subsubsec:R2}
The infrared divergences, as well as terms which would potentially lead to non-exponential finite-volume effects in a finite-volume computation of the correlation functions, 
come from the propagation of a single pion together with the photon over large time separations $x_4-y_4$. We therefore rewrite $C^A_{J_W\phi;R_2}(t_\pi)$ as the sum of two terms
\begin{equation}
C^A_{J_W\phi;R_2}(t_\pi)=C^{A\,\mathrm{L}}_{J_W\phi;R_2}(t_\pi)+C^{A\,\mathrm{S}}_{J_W\phi;R_2}(t_\pi)\,,
\end{equation}
where the indices L and S represent \textit{Long} and \textit{Short} temporal separations between the insertions of the two currents respectively. Specifically, we define regions L and S as corresponding to $x_4-y_4\ge t_s$ and $x_4-y_4<t_s$ respectively and take $t_s$ to be the same as in   
the definition of the four regions in Eq.\,(\ref{eq:regions}). This is a convenient choice but not a necessary one; all that is required is that the only significant contribution in region L corresponds to a single pion and photon propagating between the two currents. The infrared divergence is contained in $C^{A\,\mathrm{L}}_{J_W\phi;R_2}(t_\pi)$ whereas $C^{A\,\mathrm{S}}_{J_W\phi;R_2}(t_\pi)$ is infrared finite. We now consider these in turn, starting with the contribution from the short temporal separation, $C^{A\,\mathrm{S}}_{J_W\phi;R_2}(t_\pi)$.\\[0.1cm]

The infrared-convergent contribution $C^{A\,\mathrm{S}}_{J_W\phi;R_2}(t_\pi)$ is given by:
\begin{eqnarray}
C_{J_W\phi;R_2}^{A\,\mathrm{S}}(t_\pi)&=&-e^2\frac{Z_0H_{0}^4}{(2m_\pi^0)^2}\,e^{-m_\pi^0 t_\pi}\int_{-t_\pi+t_s}^{-t_s} \hspace{-0.2in}dx_4\int^{x_4}_{\mathrm{max}(x_4-t_s,-t_\pi+t_s)} \hspace{-0.7in}dy_4\hspace{0.5in}\int\dthree z~\langle \pi(\vecp {0})|J_\mathrm{em}^\mu(x_4,\vecp{0})J_\mathrm{em}^\nu(y_4,\vecp{z})|\pi(\vecp{0})\rangle\times\nonumber\\
&&\hspace{2.5in}S_\gamma^{\mu\nu}\big((x_4, \vecp{0}),(y_4,\vecp{z})\big)\,.
\label{eq:CAJphi2con}
\end{eqnarray}
It can be evaluated in lattice computations with exponentially small finite-volume corrections. The term proportional to $t_\pi$ contributes to the mass shift and is subtracted as in Eq.\,(\ref{eq:tildeCAJWphidef}) and the difference is denoted by $\tilde{C}_{J_W\phi;R_2}^{A\,S}(t_\pi)$.

It is instructive to consider $C^{A\,\mathrm{S}}_{J_W\phi;R_2}(t_\pi)$ in a little more detail. Note that the integrand in Eq.\,(\ref{eq:CAJphi2con}) is only a function of the difference $z_4\equiv x_4-y_4$ and $\vec{z}$. Thus one time integration can be eliminated:
\begin{eqnarray}
&&\int_{-t_\pi+t_s}^{-t_s} \hspace{-0.2in}dx_4\int^{x_4}_{\mathrm{max}(x_4-t_s,-t_\pi+t_s)} \hspace{-0.7in}dy_4
\hspace{0.5in}\int\dthree z~
\langle \pi(\vecp {0})|J_\mathrm{em}^\mu(x_4,\vecp{0})J_\mathrm{em}^\nu(y_4,\vecp{z})|\pi(\vecp{0})\rangle\,S_\gamma^{\mu\nu}\big((x_4, \vecp{0}),(y_4,\vecp{z})\big)\nonumber\\
&&\hspace{0.3in}=(t_\pi-2t_s)\int_0^{t_s} dz_4\int\dthree z~\langle \pi(\vecp {0})|J_\mathrm{em}^\mu(z)J_\mathrm{em}^\nu(0)|\pi(\vecp{0})\rangle\,S_\gamma^{\mu\nu}\big(z, 0\big)\nonumber\\
&&\hspace{1.5in} -\int_0^{t_s} dz_4\int\dthree z~z_4\langle \pi(\vecp {0})|J_\mathrm{em}^\mu(z)J_\mathrm{em}^\nu(0)|\pi(\vecp{0})\rangle\,S_\gamma^{\mu\nu}\big(z, 0\big)\,,\label{eq:CAJphi2Saux1}
\end{eqnarray}
where $z=(z_4,\vecp{z})$ and we have used translation invariance. The factor of $t_\pi-2t_s$ in front of the first term on the right-hand side is a reflection of the fact that the temporal range in region $R_2$ for $y_4=x_4$ is $t_\pi-2t_s$. From this region therefore, we would expect a factor of $\delta m_\pi(t_\pi-2t_s)$. Regions $R_1$ and $R_3$ both have temporal extent $t_s$ and hence contributions which cancel the term proportional to $-2t_s$ in the first term on the right-hand side of Eq.\,(\ref{eq:CAJphi2Saux1}). We therefore only subtract the term proportional to $t_\pi$ as in Eq.\,(\ref{eq:tildeCAJWphidef}) so that 
\begin{eqnarray}
\tilde{C}_{J_W\phi;R_2}^{A\,S}(t_\pi)&=&e^2\frac{Z_0H_{0}^4}{(2m_\pi^0)^2}\,e^{-m_\pi^0 t_\pi}
\Bigg\{2t_s\int_0^{t_s} dz_4\int\dthree z~\langle \pi(\vecp {0})|J_\mathrm{em}^\mu(z)J_\mathrm{em}^\nu(0)|\pi(\vecp{0})\rangle\,S_\gamma^{\mu\nu}\big(z, 0\big)\nonumber\\
&&\hspace{0.2in}+\int_0^{t_s} dz_4\int\dthree z~z_4\langle \pi(\vecp {0})|J_\mathrm{em}^\mu(z)J_\mathrm{em}^\nu(0)|\pi(\vecp{0})\rangle\,S_\gamma^{\mu\nu}\big(z, 0\big)\Bigg\}\,.\label{eq:CAJWR2S}
\end{eqnarray}
When the contribution from regions $R_1$, $R_2$ and $R_3$ are combined the $t_s$ dependence is eliminated leaving both $\delta m_\pi$ and the contribution to the amplitude independent of $t_s$.

\vspace{0.3cm}
We now consider the contribution $C^{A\,\mathrm{L}}_{J_W\phi;R_2}(t_\pi)$ which contains the infrared divergence:
\begin{eqnarray}
C_{J_W\phi;R_2}^{A\,\mathrm{L}}(t_\pi)&=&-e^2\frac{Z_0H_{0}^4}{(2m_\pi^0)^2}\,e^{-m_\pi^0 t_\pi}\int_{-t_\pi+2t_s}^{-t_s} \hspace{-0.2in}dx_4\int_{-t_\pi+t_s}^{x_4-t_s} \hspace{-0.2in}dy_4\int\dthree z~
\langle \pi(\vecp{0})|J_\mathrm{em}^\mu(x_4,\vec{0})J_\mathrm{em}^\nu(y_4,\vecp{z})|\pi(\vecp{0})\rangle\times\nonumber\\
&&\hspace{2.5in} S_\gamma^{\mu\nu}\big((x_4,\vecp{0}),(y_4,\vecp{z})\big)\,.\label{eq:CAJphi2div}
\end{eqnarray}

In order to organise the cancellation of infrared divergences we further manipulate $C_{J_W\phi;R_2}^{A\,\mathrm{L}}(t_\pi)$:
\begin{eqnarray}
C_{J_W\phi;R_2}^{A\,\mathrm{L}}(t_\pi)&=&
-e^2\frac{Z_0H_{0}^4}{(2m_\pi^0)^2}\,e^{-m_\pi^0 t_\pi}\int_{-t_\pi+2t_s}^{-t_s} \hspace{-0.2in}dx_4\int_{-t_\pi+t_s}^{x_4-t_s} \hspace{-0.2in}dy_4\int\dthree z~\int\frac{d^3p}{(2\pi)^3}\frac1{2E_\pi(\vecp{p})}\int\frac{d^3k}{(2\pi)^3}\frac1{2E_\gamma}\times\nonumber\\
\rule[-8pt]{0pt}{10pt}&&\hspace{0.75in}\langle \pi(\vecp {0})|J_\mathrm{em}^\mu(x_4,\vecp{0})|\pi(\vecp{p})\rangle\langle\pi(\vecp{p})|J_\mathrm{em}^\mu(y_4,\vecp{z})|\pi(\vecp{0})\rangle\,e^{-E_\gamma(x_4-y_4)}
e^{i\vec{k}\cdot\vec{z}}
\nonumber\\ 
&=&-e^2\frac{Z_0H^{4}_0}{(2m_\pi^0)^2}\,e^{-m_\pi^0 t_\pi}\int_{-t_\pi+2t_s}^{-t_s} \hspace{-0.2in}dx_4\int_{-t_\pi+t_s}^{x_4-t_s} \hspace{-0.2in}dy_4~\int\frac{d^3k}{(2\pi)^3}\frac1{(2E_\gamma)\,(2E_\pi(\vecp{k}))}\times\nonumber\\
&&\hspace{0.7in}\langle \pi(\vecp {0})|J_\mathrm{em}^\mu(0)|\pi(\vecp{k})\rangle\langle\pi(\vecp{k})|J_\mathrm{em}^\mu(0)|\pi(\vecp{0})\rangle\,e^{-(E_\pi(\vecp{k})+E_\gamma-m_\pi^0)(x_4-y_4)},
\end{eqnarray}
where for three momentum $\vec{q}$, $E_\pi(\vecp{q})=\sqrt{\vecpsq{q}+(m_\pi^{0})^2}$ and $E_\gamma=|\vec{k}|$\,\footnote{We write $E_\gamma$ because when regulating the infrared divergences below we could envisage introducing a photon mass $m_\gamma$ so that $E_\gamma=\sqrt{|\vecp{k}|^2+m_\gamma^2}$.}.
Performing the time integrations we obtain
\begin{eqnarray}
C_{J_W\phi;R_2}^{A\,\mathrm{L}}(t_\pi)&=&-e^2\frac{Z_0H_0^4}{(2m_\pi^0)^2}\,e^{-m_\pi^0 t_\pi}
\int\frac{d^3k}{(2\pi)^3}\frac1{2E_\gamma\,2E_\pi(\vecp{k})}\langle \pi(\vecp {0})|J_\mathrm{em}^\mu(0)|\pi(\vecp{k})\rangle\langle\pi(\vecp{k})|J_\mathrm{em}^\mu(0)|\pi(\vecp{0})\rangle\times\nonumber\\
&&\hspace{-0.25in}
\left\{\frac{e^{-(E_\pi(\vecp{k})+E_\gamma-m_\pi^0)t_s}}{E_\pi(\vecp{k})+E_\gamma-m_\pi^0}\,(t_\pi-3t_s)-\frac{e^{-(E_\pi(\vecp{k})+E_\gamma-m_\pi^0)t_s}-e^{-(E_\pi(\vecp{k})+E_\gamma-m_\pi^0)(t_\pi-2t_s)}}{(E_\pi(\vecp{k})+E_\gamma-m_\pi^0)^2}
\right\}\,.\label{eq:CJWAR2div1}
\end{eqnarray}
In Eq.\,(\ref{eq:CJWAR2div1}) the term in braces which is proportional to $t_\pi$ corresponds to a contribution to the pion's electromagnetic mass shift and the remaining terms to a contribution to the amplitude. 
The $t_s$ dependence in the terms proportional to $t_\pi$ in Eq.\,(\ref{eq:CJWAR2div1}) and (\ref{eq:CAJphi2Saux1}) cancel leaving $\delta m_\pi$ independent of $t_s$.
The second contribution in braces can readily be seen to be infrared divergent by noting that for small $|\vec{k}|$, $E_\pi(\vecp{k})+E_\gamma-m_\pi^0=O(|\vec{k}|)$ and so the ($t_\pi$-independent) term in the integrand is $O(1/|\vec{k}|^3)$. 
Note that, as explained in the Introduction, we assume that an infra-red cut-off, such as a mass for the photon, has been introduced so that 
$e^{-(E_\pi(\vecp{k})+E_\gamma-m_\pi^0)\,t_\pi}\to 0$ as $t_\pi\to\infty$ and the term containing this factor in the integrand does not contribute to the amplitude.

Using Eq.\,(\ref{eq:Haux1}) with $f=|\pi(\vecp{0})\rangle$ and with electromagnetic currents for the two operators $O_{1,2}$, i.e. 
\begin{equation}
\frac1{2E_\pi(\vecp{p})}\langle\pi(\vecp{0})|J^\mu_\mathrm{em}(0)|\pi(\vecp{p})\rangle\,\langle\pi(\vecp{p})|J^\nu_\mathrm{em}(0)|\pi(\vecp{0})\rangle=
\int\dthree x \,H_{2s}^{\mu\nu}(\vec{x},-t_s)\,e^{(E_\pi(\vecp{p})-m_\pi)t_s}\,e^{i\vec{p}\cdot\vec{x}}\,,
\label{eq:IVRaux1}\end{equation}
the term contributing to the amplitude in Eq.\,(\ref{eq:CJWAR2div1}) can be rewritten in the form
\begin{eqnarray}
\tilde{C}_{J_W\phi;R_2}^{A\,L}(t_\pi)&=&e^2\frac{Z_0H_0^4}{(2m_\pi^0)^2}\,e^{-m_\pi^0 t_\pi}\!
\int\frac{d^3k}{(2\pi)^3}\frac1{2E_\gamma}\int\dthree z\, H_{2s}^{\mu\mu}(\vec{z},-t_s)\,e^{i\vec{k}\cdot\vec{z}}\times\nonumber\\
&&\hspace{0.75in}\left\{3t_s\frac{e^{-E_\gamma t_s}}{E_\pi(\vecp{k})+E_\gamma-m_\pi^0}+
\,\frac{e^{-E_\gamma t_s}}{(E_\pi(\vecp{k})+E_\gamma-m_\pi^0)^2}\right\}.\label{eq:tildeCJWphiR2}
\end{eqnarray}
The tilde on the left-hand side of Eq.\,(\ref{eq:tildeCJWphiR2}) indicates that the term proportional to $t_\pi$ has been subtracted as explained in the discussion around Eq.\,(\ref{eq:tildeCAJWphidef}).

In order to organise the cancellation of the infrared divergences it is convenient to separate $\tilde{C}_{J_W\phi;R_2}^{A\,L}(t_\pi)$ into a divergent and convergent contribution
\begin{equation}
\tilde{C}_{J_W\phi;R_2}^{A\,L}(t_\pi)=\tilde{C}_{J_W\phi;R_2}^{A\,L\,\mathrm{div}}(t_\pi)+\tilde{C}_{J_W\phi;R_2}^{A\,L\,\mathrm{con}}(t_\pi)\,,
\end{equation}
where
\begin{eqnarray}
\tilde{C}_{J_W\phi;R_2}^{A\,L\,\mathrm{div}}(t_\pi)&=&e^2
Z_0H_0^4\,e^{-m_\pi^0 t_\pi}\int\frac{d^3k}{(2\pi)^3}\frac1{2E_\gamma}\,\frac{e^{-E_\gamma t_s}}{(E_\pi(\vecp{k})+E_\gamma-m_\pi^0)^2}
\label{eq:tildeCAJphi2div}
\\ 
\tilde{C}_{J_W\phi;R_2}^{A\,L\,\mathrm{con}}(t_\pi)&=&e^2
\frac{Z_0H_0^4}{(2m_\pi^0)^2}\,e^{-m_\pi^0 t_\pi}
\int\frac{d^3k}{(2\pi)^3}\frac1{2E_\gamma}\int\!\!\dthree z\, H_{2s}^{\mu\mu}(\vec{z},-t_s)\times\nonumber\\
&&\hspace{0.75in}\left\{\frac{3t_s\,e^{i\vec{k}\cdot\vec{z}}\,e^{-E_\gamma t_s}}{E_\pi(\vecp{k})+E_\gamma-m_\pi^0}
+(e^{i\vec{k}\cdot\vec{z}}-1)
\,\frac{e^{-E_\gamma t_s}}{(E_\pi(\vecp{k})+E_\gamma-m_\pi^0)^2}\right\}\,.\label{eq:tildeCAJphi2con}
\end{eqnarray}
Collecting up all the terms we have
\begin{equation}
\tilde{C}^A_{J_W\phi;R_2}(t_\pi)=\tilde{C}^{A\,L\,\mathrm{div}}_{J_W\phi;R_2}(t_\pi)+\tilde{C}^{A\,L\,\mathrm{con}}_{J_W\phi;R_2}(t_\pi)+
\tilde{C}^{A\,S}_{J_W\phi;R_2}(t_\pi)\,,\label{eq:CJWphiR2tot}
\end{equation}
where $\tilde{C}^{A\,L\,\mathrm{div}}_{J_W\phi;R_2}(t_\pi)$ is given in Eq.\,({\ref{eq:tildeCAJphi2div}), $\tilde{C}^{A\,L\,\mathrm{con}}_{J_W\phi;R_2}(t_\pi)$ in Eq.\,({\ref{eq:tildeCAJphi2con}) and $\tilde{C}^{A\,S}_{J_W\phi;R_2}(t_\pi)$ in Eq.\,(\ref{eq:CAJphi2con}) after subtraction of the term linear in $t_\pi$. \\

From Eq.\,(\ref{eq:CAJphi2}) we see that the expression for the correlation function $C^A_{\phi\phi;R_2}(t_\pi)$ is simply obtained from $C^A_{J_W\phi;R_2}(t_\pi)$
with the replacement of $H^4_0$ by $Z_0$. Therefore, recalling the expression in Eq.\,(\ref{eq:H4fullaux1}), the contribution to $H^4_{\textrm{full}}$ from region $R_2$ is 
\begin{equation}
\big[2m_\pi^0\big]^{\frac12}\,
\left[\frac{\tilde{C}^A_{J_W\phi;R_2}(t_\pi)}{[C^0_{\phi\phi}(t_\pi)\,\,e^{-m_\pi^0t}]^\frac12}
-\frac12\frac{\tilde C^A_{\phi\phi;R_2}(t_\pi)}{C^0_{\phi\phi}(t_\pi)}\,\frac{C^0_{J_W\phi}(t_\pi)}{[C^0_{\phi\phi}(t_\pi)\,\,e^{-m_\pi^0t}]^\frac12}\right]
=\frac12\,\big[2m_\pi^0\big]^{\frac12}\,
\frac{\tilde{C}^A_{J_W\phi;R_2}(t_\pi)}{[C^0_{\phi\phi}(t_\pi)\,\,e^{-m_\pi^0t}]^\frac12}\,.
\end{equation}
The divergent contribution to $H^{4}_\mathrm{full}$ from diagram A is therefore
\begin{equation}
H_A^{4\,\mathrm{div}}=e^2
\frac{H^{4}_0}{2}\int\frac{d^3k}{(2\pi)^3}\frac1{2E_\gamma}\,\frac{e^{-E_\gamma t_s}}{(E_\pi(\vecp{k})+E_\gamma-m_\pi^0)^2}\,.
\label{eq:H4div}\end{equation}
and for later use we define
\begin{equation}
M_A^{\mathrm{div}}=-\GF~H_{\mathrm{A}}^{4\,\mathrm{div}} L_{0}^4\,.\label{eq:MAdiv}
\end{equation}
The factor $e^{-E_\gamma t_s}$ in the integrand ensures that $H_A^{4\,\mathrm{div}}$ is ultra-violet convergent.

\subsubsection{Contribution from Region $R_1$}\label{subsubsec:R1}
 
 The contributions to $M_A$ from region $R_1$ is infrared convergent and in this  subsection we present the corresponding expression.
We start with a discussion of the correlation function in region $R_1$ presented in Eq.\,(\ref{eq:CAJphi1}) which we rewrite here for convenience:  
 \begin{equation}
C_{J_W\phi;R_1}^{A}(t_\pi) 
=-e^2\frac{Z_0}{2m_\pi^0}\,e^{-m_\pi^0 t_\pi}\int^{\infty}_{-t_s} \hspace{-0.05in}dx_4\int_{-t_\pi+t_s}^{x_4} \hspace{-0.2in}dy_4\int\dthree x\int\dthree y~\langle 0|T\big[J_W^4(0)J_\mathrm{em}^\mu(x)J_\mathrm{em}^\nu(y)\big]|\pi(\vecp{0})\rangle\,S_\gamma^{\mu\nu}(x,y)\,.\label{eq:CAJphi1p}
\end{equation}
We now subdivide $R_1$, where $x_4>-t_s$ and $y_4>-t_\pi+t_s$, into subregions in which either the contributions to $C_{J_W\phi;R_1}^{A}(t_\pi)$ can be computed directly or the IVR procedure is implemented:\\[0.05in]

\begin{center}
\vspace{-0.3in}
\begin{tabular}{ccccc}\\ 
$R_{1a}$:&\quad&$x_4<0$,&\quad& $x_4-y_4\ge t_s$\\ 
$R_{1b}$:&\quad& $x_4>0$,&\quad&$y_4\le -t_s$\\
$R_{1c}$:&\quad&$x_4<0$,&\quad& $x_4-y_4<t_s$ \\
$R_{1d}$:&\quad&$x_4>0$,&\quad&$y_4>-t_s$
\end{tabular}
\end{center}
and in each case it is to be implicitly understood that $x_4>-t_s$ and $y_4>-t_\pi+t_s$ with $x_4>y_4$, which is the definition of region $R_1$. In region $R_{1a}$ the temporal separation between the two electromagnetic currents is greater than $t_s$ and hence the correlation function is dominated by the propagation between these currents of states which consist of a single pion and a photon. Similarly in region $R_{1b}$ the temporal separation between the electromagnetic current at $y$ and the weak current is greater than $t_s$ and again the correlation function is dominated by the propagation of a single pion and photon between these currents. 
We envisage that when lattice computations are performed of the contributions from both these regions,
IVR will be implemented to avoid finite-volume corrections which are not exponentially small. In regions $R_{1c}$ and $R_{1d}$ there are no contributions corresponding to the propagation of a single pion and photon over distances greater than $t_s$ and hence the finite-volume effects are exponentially small. The contributions from these two regions can therefore be computed directly in a finite volume.

We start by considering the contribution from region $R_{1a}$:
\begin{eqnarray}
C_{J_W\phi;R_{1a}}^{A}(t_\pi) 
&=&-e^2\frac{Z_0}{2m_\pi^0}\,e^{-m_\pi^0 t_\pi}\int^{0}_{-t_s} \hspace{-0.05in}dx_4\int_{-t_\pi+t_s}^{x_4-t_s} \hspace{-0.2in}dy_4\int\dthree x\int\dthree y~\langle 0|J_W^4(0)J_\mathrm{em}^\mu(x)J_\mathrm{em}^\nu(y)|\pi(\vecp{0})\rangle\,S_\gamma^{\mu\nu}(x,y)\nonumber\\
&&\hspace{-1in}=-e^2\frac{Z_0}{2m_\pi^0}\,e^{-m_\pi^0 t_\pi}\int^{0}_{-t_s} \hspace{-0.05in}\hspace{-0.07in}dx_4\int_{-t_\pi+t_s}^{x_4-t_s} \hspace{-0.25in}dy_4
\int\dthree z\int\dthree y~\langle 0|J_W^4(0)J_\mathrm{em}^\mu(x_4,\vec{z}+\vecp{y})J_\mathrm{em}^\nu(y_4,\vecp{y})|\pi(\vecp{0})\rangle\,S_\gamma^{\mu\nu}\big((x_4-y_4,\vecp{z}),0\big)).\label{eq:CAJWphoRIaaux1}
\end{eqnarray}
In the second line of Eq.\,(\ref{eq:CAJWphoRIaaux1}) we have noted that for any $(x,y)$ the photon propagator only depends on $x-y$ so that 
$S_\gamma^{\mu\nu}(x,y)=S_\gamma^{\mu\nu}(x-y,0)$.
It is now convenient to consider the hadronic component separately and to define
\begin{eqnarray}
H_{R_{1a}}(x_4,y_4,\vecp{z})&\equiv&\int\dthree y~
\langle 0|J_W^4(0)J_\mathrm{em}^\mu(x_4,\vec{z}+\vecp{y})J_\mathrm{em}^\nu(y_4,\vecp{y})|\pi(\vecp{0})\rangle\\
&=&\int\dthree y\int\frac{\dthree p}{(2\pi)^3}\,\frac1{2E_\pi(\vecp{p})}\,\langle 0|J^4_W(0)J^\mu_{\mathrm{em}}(x_4,\vecp{y})|\pi(\vecp{p})\rangle
\,\langle\pi(\vecp{p})|J^\nu_\mathrm{em}(y_4,\vec{y}-\vecp{z})|\pi(\vecp{0})\rangle\nonumber\\
&&\hspace{-1.25in}=\int\dthree y\int\frac{\dthree p}{(2\pi)^3}\,\frac1{2E_\pi(\vecp{p})}\,e^{i\vec{p}\cdot\vec{z}}\,e^{-(E_\pi(\vecp{p})-m_\pi^0)(x_4-y_4-t_s)}
\,\langle 0|J^4_W(0)J^\mu_{\mathrm{em}}(x_4,\vecp{y})|\pi(\vecp{p})\rangle
\,\langle\pi(\vecp{p})|J^\nu_\mathrm{em}(x_4-t_s,\vecp{y})|\pi(\vecp{0})\rangle\,,\label{eq:H21a}
\end{eqnarray}
where we recall that $E_\pi(\vecp{p})=\sqrt{|\vecp{p}|^2+(m_\pi^0)^2}$. Taking the inverse Fourier transform at $y_4=x_4-t_s$ we obtain
\begin{equation}\label{eq:tildeH2IVR}
\int\dthree z^{\prime}\, H_{R_{1a}}(x_4,x_4-t_s,\vec{z}^{\,\prime})\,e^{-i\vec{p}\cdot\vec{z}^{\,\prime}}=\frac1{2E_\pi(\vecp{p})}\,
\int\dthree y~\langle 0|J^4_W(0)J^\mu_{\mathrm{em}}(x_4,\vecp{y})|\pi(\vecp{p})\rangle
\,\langle\pi(\vecp{p})|J^\nu_\mathrm{em}(x_4-t_s,\vecp{y})|\pi(\vecp{0})\rangle\,.
\end{equation}
Inserting Eq.\,(\ref{eq:tildeH2IVR}) into Eq.\,(\ref{eq:H21a}) gives
\begin{equation}\label{eq:R1aux1}
H_{R_{1a}}(x_4,y_4,\vecp{z})=\int\dthree z^\prime~H_{R_{1a}}(x_4,x_4-t_s,\vec{z}^{\,\prime})\,\int\frac{\dthree p}{(2\pi)^3}\,
e^{i\vec{p}\cdot(\vec{z}-\vec{z}^{\,\prime})}\,e^{-(E_\pi(\vecp{p})-m_\pi^0)(x_4-y_4-t_s)}\,.
\end{equation}
Thus for sufficiently large $t_s$ it is enough to compute $H_{R_{1a}}(x_4,y_4,\vecp{z})$ at $x_4-y_4=t_s$ and to use Eq.\,(\ref{eq:R1aux1}) to obtain 
$H_{R_{1a}}(x_4,y_4,\vecp{z})$ at values of $x_4-y_4>t_s$.
It is not necessary to compute $H_{R_{1a}}(x_4,y_4,\vecp{z})$ directly in a finite volume for $x_4-y_4>t_s$.

The calculation in region $R_{1b}$ follows a similar procedure with only a single pion and photon propagating between $J^\nu_\mathrm{em}(y)$ and the weak current. The hadronic matrix element is now
\begin{eqnarray}
H_{R_{1b}}(x_4,y_4,\vecp{z})&\equiv&\int\dthree y~
\langle 0|J_\mathrm{em}^\mu(x_4,\vec{z}+\vecp{y})J_W^4(0)J_\mathrm{em}^\nu(y_4,\vecp{y})|\pi(\vecp{0})\rangle\\
&=&\int\dthree z^\prime~H_{R_{1b}}(x_4,-t_s,\vec{z}^{\,\prime})\,\int\frac{\dthree p}{(2\pi)^3}\,
e^{i\vec{p}\cdot(\vec{z}-\vec{z}^{\,\prime})}\,e^{(E_\pi(\vecp{p})-m_\pi^0)(y_4+t_s)}\,.
\end{eqnarray}
We see that also in this case we don't have to compute directly in a finite volume the matrix element for $-t_\pi+t_s<y_4<-t_s$.

The contributions from regions $R_{1c}$ and $R_{1d}$ do not have on-shell single-pion states propagating over long time separations and hence do not have  non-exponential finite-volume corrections. Collecting up the terms from the four subregions, the correlation function in region $R_1$ can be written as:
\begin{equation}\label{eq:CJWphiR1tot}
C_{J_W\phi;R_1}^{A}(t_\pi) =C_{J_W\phi;R_{1a}}^{A}(t_\pi) +C_{J_W\phi;R_{1b}}^{A}(t_\pi) +C_{J_W\phi;R_{1c}}^{A}(t_\pi) +C_{J_W\phi;R_{1d}}^{A}(t_\pi)\,, 
\end{equation}
where
\begin{eqnarray}
C_{J_W\phi;R_{1a}}^{A}(t_\pi) 
&=&-e^2\frac{Z_0}{2m_\pi^0}\,e^{-m_\pi^0 t_\pi}
\int^{0}_{-t_s} \hspace{-0.05in}dx_4\int_{-t_\pi+t_s}^{x_4-t_s} \hspace{-0.2in}dy_4\int\dthree z
\int\dthree z^\prime\, H_{R_{1a}}(x_4,x_4-t_s,\vec{z}^{\,\prime})\nonumber\\
&&\hspace{0.5in}\times\int\frac{\dthree p}{(2\pi)^3}\,e^{i\vec{p}\cdot(\vec{z}-\vec{z}^{\,\prime})}\,e^{-(E_\pi(\vecp{p})-m_\pi^0)(x_4-y_4-t_s)}S_\gamma^{\mu\nu}(x_4-y_4,\vecp{z})
\label{eq:CJWphiR1a}\\
C_{J_W\phi;R_{1b}}^{A}(t_\pi) 
&=&-e^2\frac{Z_0}{2m_\pi^0}\,e^{-m_\pi^0 t_\pi}
\int_{0}^{\infty} \hspace{-0.05in}dx_4\int_{-t_\pi+t_s}^{-t_s} \hspace{-0.2in}dy_4\int\dthree z
\int\dthree z^\prime\, H_{R_{1b}}(x_4,-t_s,\vec{z}^{\,\prime})\nonumber\\
&&\hspace{0.5in}\times\int\frac{\dthree p}{(2\pi)^3}\,e^{i\vec{p}\cdot(\vec{z}-\vec{z}^{\,\prime})}\,e^{(E_\pi(\vecp{p})-m_\pi^0)(y_4+t_s)}S_\gamma^{\mu\nu}(x_4-y_4,\vecp{z})
\label{eq:CJWphiR1b}\\
C_{J_W\phi;R_{1c}}^{A}(t_\pi) &=&
-e^2\frac{Z_0}{2m_\pi^0}\,e^{-m_\pi^0 t_\pi}\int^{0}_{-t_s} \hspace{-0.05in}dx_4\int^{x_4}_{x_4-t_s} \hspace{-0.2in}dy_4\int\dthree x\int\dthree y~\langle 0|J_W^4(0)J_\mathrm{em}^\mu(x)J_\mathrm{em}^\nu(y)|\pi(\vecp{0})\rangle\,S_\gamma^{\mu\nu}(x,y)
\label{eq:CJWphiR1c}\\
C_{J_W\phi;R_{1d}}^{A}(t_\pi) &=&
-e^2\frac{Z_0}{2m_\pi^0}\,e^{-m_\pi^0 t_\pi}\int_{0}^{\infty} \hspace{-0.05in}dx_4\int^{x_4}_{-t_s} \hspace{-0.05in}dy_4\int\dthree x\int\dthree y~\langle 0|J_\mathrm{em}^\mu(x)T[J_W^4(0)J_\mathrm{em}^\nu(y)]|\pi(\vecp{0})\rangle\,S_\gamma^{\mu\nu}(x,y)\,.
\label{eq:CJWphiR1d}
\end{eqnarray}

The derivation and results for the correlation function $C_{\phi\phi;R_{1}}^{A}(t_\pi)$ follows in precisely the same way with the weak current $J_W^4$ replaced by the annihilation operator $\phi$. However, as we shall see in the following subsection $C_{\phi\phi;R_{1}}^{A}(t_\pi)$ can be combined with $C_{\phi\phi;R_{3}}^{A}(t_\pi)$ to cancel the contribution of $C_{J_W\phi;R_{3}}^{A}(t_\pi)$. The latter therefore does not have to be computed.

\subsubsection{Contribution from Region $R_3$}\label{subsubsec:R3}

The contribution to the correlation function from region $R_3$ is presented in Eq.\,(\ref{eq:CAJphi3}) and we rewrite it here for convenience:  
\begin{equation}
C_{J_W\phi;R_3}^{A}(t_\pi)=-e^2\frac{H_{0}^4}{2m_\pi^0}\int_{-\infty}^{-t_\pi+t_s}\hspace{-0.15in} dy_4\int_{y_4}^{-t_s}\hspace{-0.1in} dx_4\int\dthree z
~\langle \pi(\vecp{0})|T\big[J_\mathrm{em}^\mu(x_4,\vecp{0})J_\mathrm{em}^\nu(y_4,\vecp{z})\phi^\dagger(-t_\pi)\big]|0\rangle\,
S_\gamma^{\mu\nu}\big((x_4-y_4,-\vecp{z}),0\big)\,.\label{eq:CJWphiR3A}
\end{equation}
The evaluation of the correlation function $C_{J_W\phi;R_{3}}^{A}(t_\pi)$ follows in a similar way to that from region $R_1$. 
However $C_{J_W\phi;R_{3}}^{A}(t_\pi)$ is not needed to obtain the result for $H_A^4$ as we now explain. The expression for $C_{\phi\phi;R_{3}}^{A}(t_\pi)$ is the same as for $C_{J_W\phi;R_3}^{A}(t_\pi)$ in Eq.\,(\ref{eq:CJWphiR3A}) with $H_0^4$ replaced by $Z_0$.
We see from Eqs.\,(\ref{eq:H4fullaux1}) and (\ref{eq:H4fullaux2}) that the contribution to $H_A^4$ contains a term proportional to 
\begin{eqnarray}
\frac{1}{[C_{\phi\phi}^0(t_\pi)e^{-m_\pi^0t_\pi}]^\frac12}\,\bigg\{
\tilde{C}^A_{J_W\phi;R_3}(t_\pi)
-\frac12\,\tilde{C}^A_{\phi\phi;R_3}(t_\pi)\,\frac{C^0_{J_W\phi}(t_\pi)}{C^0_{\phi\phi}(t_\pi)}\bigg\}=
\frac{1}{[C_{\phi\phi}^0(t_\pi)e^{-m_\pi^0t_\pi}]^\frac12}\,\frac12\,\tilde{C}^A_{J_W\phi;R_3}(t_\pi)\,,
\end{eqnarray}
i.e. the second term in braces on the left-hand side cancels half of the first term. This can readily be understood as in both cases the electromagnetic currents are well separated in time (by at least $t_s$) from the pion creation operator. Moreover, note also that $C^A_{\phi\phi;R_1}(t_\pi)=C^A_{\phi\phi;R_3}(t_\pi)$ so that 
\begin{eqnarray}
\frac{1}{[C_{\phi\phi}^0(t_\pi)e^{-m_\pi^0t_\pi}]^\frac12}\,\bigg\{
\tilde{C}^A_{J_W\phi;R_3}(t_\pi)
-\frac12\,\Big(\tilde{C}^A_{\phi\phi;R_1}(t_\pi)+\tilde{C}^A_{\phi\phi;R_3}(t_\pi)\Big)\,\frac{C^0_{J_W\phi}(t_\pi)}{C^0_{\phi\phi}(t_\pi)}
\bigg\}=0\,.
\end{eqnarray}
It is therefore not necessary to compute $\tilde{C}^A_{J_W\phi;R_3}$ and 
$\tilde{C}^A_{\phi\phi;R_1}=\tilde{C}^A_{\phi\phi;R_3}$.

\subsubsection{Summary of the contribution to the amplitude from Diagram A}

Since the discussion in this subsection has been lengthy we collect here all the different contributions:
\begin{eqnarray}\label{eq:H4Afinal}
H^4_A=\frac{\delta m_\pi}{2m_\pi^0}H_0^4+\left(\frac{2m_\pi^0}{Z_0e^{-m_\pi^0 t_\pi}}\right)
\,\left(\tilde{C}^A_{J_W\phi;R_1}(t_\pi)+\frac12\tilde{C}^A_{J_W\phi;R_2}(t_\pi)\right)\,,
\end{eqnarray}  
where $\tilde{C}^A_{J_W\phi;R_1}(t_\pi)=C^A_{J_W\phi;R_1}(t_\pi)$ (since $C^A_{J_W\phi;R_1}(t_\pi)$ does not contain terms proportional to $t_\pi$) is given in Eqs.\,(\ref{eq:CJWphiR1tot})\,-\,(\ref{eq:CJWphiR1d}) and $\tilde{C}^A_{J_W\phi;R_2}(t_\pi)$ is presented in Eq.\,(\ref{eq:CJWphiR2tot})
together with Eqs.(\ref{eq:CAJphi2con}), ({\ref{eq:tildeCAJphi2div}) and {\ref{eq:tildeCAJphi2con}) after taking care to subtract the term proportional to $t_\pi$.

\subsection{Contribution to the amplitude from Diagram B}\label{subsec:MB}

The contribution to the amplitude for the decay $\pi^+\to\ell^+\nu_\ell$ from diagram $B$ is 
\begin{equation}
M_B=e^2\,\GF\,g_{\mu\mu^\prime}\int\!\dfour x~ H_1^{\mu\nu}(x)\!\int\!\dfour y~ L_1^{\mu^\prime\nu^\prime}(y)\,S^\gamma_{\nu\nu^\prime}(x,y)\,,\label{eq:MBMink}
\end{equation}
where all quantities are in Minkowski space as in Appendix\,\ref{subsec:conventionsM}. Rewriting $M_B$ in terms of Euclidean quantities defined in Appendix\,\ref{subsec:conventionsE} (including the lepton propagator in $L_{1E}$) we obtain
\begin{equation}
M_B=-e^2\GF\,\int\!\dfour x~ H_{1E}^{\mu\nu}(x)\!\int\!\dfour y~ L_{1E}^{\mu\nu^\prime}(y)\,S^\gamma_{\nu\nu^\prime E}(x,y)\label{eq:MBEucl}\,,
\end{equation}
where the subscript $E$ denotes \emph{Euclidean}.
Since the entire discussion in the remainder of this subsection is presented in terms of Euclidean space quantities, in order to simplify the notation we now drop the subscript $E$.

As anticipated in the Introduction we divide the integration over $x_4$ into two regions, labelled $L$ for \emph{Long}, i.e. $x_4\le -t_s$,  and $S$ for \emph{Short}, i.e. $x_4>-t_s$. 
The hadronic matrix element $H_1^{\mu\nu}(x)$ in region $S$ can be computed directly using lattice methods with only exponentially suppressed finite-volume corrections
\begin{equation}
M_B^S=-e^2\GF\,\int_{-t_s}^\infty\!\! dx_4\int\! \dthree x~ H_{1}^{\mu\nu}(x)\!\int\!\dfour y~ L_{1}^{\mu\nu^\prime}(y)\,S^\gamma_{\nu\nu^\prime}(x,y)\,.\label{eq:MBS}
\end{equation}
For the long-distance contribution we exploit Eq.\,(\ref{eq:HL2}) in order to avoid computing $H_1^{\mu\nu}(x)$ directly at large time separations between the weak and electromagnetic currents and hence introducing finite-volume corrections which decrease only as inverse powers of the volume,
\begin{eqnarray}
M_B^L&=&-e^2\GF\,\int^{-t_s}_{-\infty}\!\! dx_4\int\! \dthree x~ H_{1}^{\mu\nu}(x)\!\int\!\dfour y~ L_{1}^{\mu\nu^\prime}(y)\,S^\gamma_{\nu\nu^\prime}(x,y)
\nonumber\\
&=&-e^2\GF\,\int^{-t_s}_{-\infty}\hspace{-7pt} dx_4\!\int\! \dthree\! x\!\int\!\frac{\dthree p}{(2\pi)^3}
\int\!\dthree x^\prime~H_1^{\mu\nu}(\vec{x}^{\,\prime},-t_s)\,
e^{(E_\pi(\vecp{p})-m_\pi^0)(x_4+t_s)}e^{-i\vec{p}\cdot(\vec{x}-\vec{x}^{\,\prime})}
\!\int\!\dfour y\, L_{1}^{\mu\nu^\prime}(y)\,S^\gamma_{\nu\nu^\prime}(x,y) \nonumber\\
&&\hspace{-0.3in}=-e^2\GF\,\int^{-t_s}_{-\infty}\hspace{-7pt} dx_4\!\int\! \dfour y\, L_{1}^{\mu\nu}(y)\!\int\!\dthree x\,
H_1^{\mu\nu}(\vec{x},-t_s)\!\int\!\!\frac{\dthree k}{(2\pi)^3}\,\frac1{2E_\gamma}\,
e^{(E_\pi(\vecp{k})-m_\pi^0)(x_4+t_s)}\,e^{-E_\gamma |y_4-x_4|}\,e^{-i\vec{k}\cdot(\vec{x}-\vec{y})}\,,
\end{eqnarray}
where we recall that, up to an infrared cut-off, $E_\gamma=|\vec{k}\,|$ and  $E_\pi(\vecp{q})=\sqrt{|\vecp{q}|^2+m_\pi^{0\,2}}$ for any three-momentum $\vec{q}$. We have used Eq.\,(\ref{eq:SgammaE}) for the photon propagator in Euclidean space. 

For notational convenience we rewrite $M_B^L$ in the form
\begin{equation}
M_B^L=-e^2\GF\,\big[\bar{u}(p_{\nu_\ell})\gamma^\mu(1-\gamma^5)\big]_\alpha N_{\alpha\beta}^{\mu\nu}\,\big[\gamma^\nu v(p_\ell)\big]_\beta\,,
\end{equation}
where $\alpha,\beta$ are spinor indices and
\begin{eqnarray}
N^{\mu\nu}_{\alpha\beta}&=&\int^{-t_s}_{-\infty}\hspace{-7pt} dx_4\!\int\! \dfour y\int\frac{\dfour p}{(2\pi)^4} \tilde{S}_{\ell\,\alpha\beta}(p)\,e^{-i(p+p_\ell)\cdot y}\!\int\!\dthree x\,
H_1^{\mu\nu}(\vec{x},-t_s)\times\nonumber\\
&&\hspace{1.5in}\int\!\!\frac{\dthree k}{(2\pi)^3}\,\frac1{2E_\gamma}\,
e^{(E_\pi(\vec{k}\hspace{1.25pt})-m_\pi^0)(x_4+t_s)}\,e^{-E_\gamma |y_4-x_4|}\,e^{-i\vec{k}\cdot(\vec{x}-\vec{y})}\nonumber\\
&=&\int^{-t_s}_{-\infty}\hspace{-7pt} dx_4\int\!\!\frac{\dthree k}{(2\pi)^3}\,\frac1{2E_\gamma}\,\int\! dy_4\int\frac{dp_4}{2\pi} \,
\tilde{S}_{\ell\,\alpha\beta}(p_4,-(\vec{p}_\ell-\vec{k}))\,e^{-ip_4 y_4}\,e^{E_\ell y_4}\!\int\!\dthree x\,
H_1^{\mu\nu}(\vec{x},-t_s)\times\nonumber\\
&&\hspace{2in}
e^{(E_\pi(\vecp{k})-m_\pi^0)(x_4+t_s)}\,e^{-E_\gamma|y_4-x_4|}\,e^{-i\vec{k}\cdot\vec{x}}\,,
\end{eqnarray}
where $\vec{p}_\ell$ is the momentum of the final-state lepton and $E_\ell=\sqrt{\vec{p}_\ell^{\,\,2}+m_\ell^2}$. The infrared divergence arises from the region in which $y_4>0$ and so we start by considering this region:
\begin{eqnarray}
N^{\mu\nu}_{\alpha\beta}\big|_{y_4>0}&=&
\int^{-t_s}_{-\infty}\hspace{-7pt} dx_4\!\int\!\!\frac{\dthree k}{(2\pi)^3}\,\frac{e^{E_\gamma x_4}}{2E_\gamma}\,\int_0^\infty\! dy_4\left(\int\frac{dp_4}{(2\pi)} \,
\tilde{S}_{\ell\,\alpha\beta}(p_4,-(\vec{p}_\ell-\vec{k}))\,e^{-ip_4 y_4}\right)\,e^{(E_\ell-E_\gamma) y_4}\!\int\!\dthree x\,
H_1^{\mu\nu}(\vec{x},-t_s)\nonumber\\
&&\hspace{2in}\times
e^{(E_\pi(\vecp{k})-m_\pi^0)(x_4+t_s)}\,e^{-i\vec{k}\cdot\vec{x}}\nonumber\\[0.01in]
&=&\int^{-t_s}_{-\infty}\hspace{-7pt} dx_4\!\int\!\!\frac{\dthree k}{(2\pi)^3}\,\frac{e^{E_\gamma x_4}}{2E_\gamma}\,\int_0^\infty
\hspace{-5pt}
dy_4\,\frac{(-\Elpmk\gamma_4+i(\vec{p}_\ell-\vec{k})\cdot\vec{\gamma}+m_\ell)_{\alpha\beta}}{2\Elpmk}\,
e^{-(\Elpmk+E_\gamma-E_\ell) y_4}\nonumber\\
&&\hspace{2in}\times\int\!\dthree x\,
H_1^{\mu\nu}(\vec{x},-t_s)
e^{(E_\pi(\vecp{k})-m_\pi^0)(x_4+t_s)}\,e^{-i\vec{k}\cdot\vec{x}}\nonumber\\[0.1in]
&=&\int^{-t_s}_{-\infty}\hspace{-7pt} dx_4\!\int\!\!\frac{\dthree k}{(2\pi)^3}\,\frac{e^{E_\gamma x_4}}{2E_\gamma}\,\frac{(-\Elpmk\gamma_4+i(\vec{p}_\ell-\vec{k})\cdot\vec{\gamma}+m_\ell)_{\alpha\beta}}{2\Elpmk}\,
\frac1{\Elpmk+E_\gamma-E_\ell}\nonumber\\ 
&&\hspace{2in}\times\int\!\dthree x\,
H_1^{\mu\nu}(\vec{x},-t_s)
e^{(E_\pi(\vecp{k})-m_\pi^0)(x_4+t_s)}\,e^{-i\vec{k}\cdot\vec{x}}\,,\nonumber\\
&=&\int\!\!\frac{\dthree k}{(2\pi)^3}\,\frac{e^{-E_\gamma t_s}}{2E_\gamma}\,\frac{(-\Elpmk\gamma_4+i(\vec{p}_\ell-\vec{k})\cdot\vec{\gamma}+m_\ell)_{\alpha\beta}}{2\Elpmk\,(\Elpmk+E_\gamma-E_\ell)(E_\pi(\vecp{k})+E_\gamma-m_\pi^0)}\int\!\dthree x\,
H_1^{\mu\nu}(\vec{x},-t_s)
\,e^{-i\vec{k}\cdot\vec{x}}\,,\label{eq:N1}
\end{eqnarray}
where $\Elpmk=\sqrt{(\vec{p}_\ell-\vecp{k})^2+m_\ell^2}$. As expected the right-hand side of Eq.\,(\ref{eq:N1}) is infrared divergent; each of the factors $E_\gamma$, $\Elpmk+E_\gamma-E_\ell$ and $E_\pi(\vecp{k})+E_\gamma-m_\pi^0$ in the denominator of the integrand is $O(|\vecp{k}|)$ at small $\vec{k}$. In order to obtain an expression for the inclusive decay rate which is free from infrared divergences it is convenient to rewrite the factor $e^{-i\vec{k}\cdot\vec{x}}$ as 
$1+(e^{-i\vec{k}\cdot\vec{x}}-1)$ and to separate $M_B^L$ into divergent and convergent contributions:
\begin{equation}
M_B^{L}=M_B^{L\,\mathrm{div}}+M_B^{L\,\mathrm{con}}\,,
\end{equation}
where
\begin{equation}
M_B^{L\,\mathrm{div}}=-\GF\,H_0^4\int\!\!\frac{\dthree k}{(2\pi)^3}\,\,e^{-E_\gamma t_s}
\left\{\frac{\bar{u}(p_{\nu_\ell})\gamma^4(1-\gamma^5)(-\Elpmk\gamma_4+i(\vec{p}_\ell-\vec{k})\cdot\vec{\gamma}+m_\ell)\gamma^4 v(p_\ell)}{4E_\gamma\Elpmk(\Elpmk+E_\gamma-E_\ell)(E_\pi(\vecp{k})+E_\gamma-m_\pi^0)}
\right\}\label{eq:MBLdiv}
\end{equation}
and
\begin{eqnarray}
M_B^{L\,\mathrm{con}}&=&-\GF\int\!\!\frac{\dthree k}{(2\pi)^3}\,\,e^{-E_\gamma t_s}
\left\{\frac{\bar{u}(p_{\nu_\ell})\gamma^\mu(1-\gamma^5)(-\Elpmk\gamma_4+i(\vec{p}_\ell-\vec{k})\cdot\vec{\gamma}+m_\ell)\gamma^\nu v(p_\ell)}{4E_\gamma\Elpmk(\Elpmk+E_\gamma-E_\ell)(E_\pi(\vecp{k})+E_\gamma-m_\pi^0)}
\right\}\nonumber\\
&&\hspace{2in}\times\int\!\dthree x\,H_1^{\mu\nu}(\vec{x},-t_s)\Big(e^{-i\vec{k}\cdot\vec{x}}-1\Big)
\nonumber\\[0.05in]
&&\hspace{-0.8in}-\GF\int^{-t_s}_{-\infty}\hspace{-7pt} dx_4\!\int\!\!\frac{\dthree k}{(2\pi)^3}\,\frac1{4E_\gamma\Elpmk}\,
\int_{-\infty}^0\hspace{-6pt} dy_4\,e^{(E_\ell+\Elpmk)y_4}\,\Big\{
\bar{u}(p_{\nu_\ell})\gamma^\mu(1-\gamma^5)(\Elpmk\gamma_4+i(\vec{p}_\ell-\vec{k})\cdot\vec{\gamma}+m_\ell)\gamma^\nu v(p_\ell)\Big\}\nonumber\\
&&\hspace{2in}\times\int\!\dthree x\,
H_1^{\mu\nu}(\vec{x},-t_s)
e^{(E_\pi(\vecp{k})-m_\pi^0)(x_4+t_s)}\,e^{-E_\gamma|y_4-x_4|}\,e^{-i\vec{k}\cdot\vec{x}}\,.
\label{eq:MBLcon}
\end{eqnarray}

\subsubsection{Summary of the contribution to the amplitude from Diagram B}
In summary, the contribution to the amplitude from diagram B is 
\begin{equation}
M_B=M_B^S+M_B^{L\,\mathrm{div}}+M_B^{L\,\mathrm{con}}\,,
\end{equation}
where $M_B^S$, $M_B^{L\,\mathrm{div}}$ and $M_B^{L\,\mathrm{con}}$ are given in Eqs.\,(\ref{eq:MBS}), (\ref{eq:MBLdiv}) and (\ref{eq:MBLcon})
respectively.

\subsection{Contribution to the amplitude from Diagram C}

The hadronic element in the contribution of diagram C to the amplitude is simply $H_0^4$ 
and the wavefunction renormalisation of the final state electron can be calculated in QED perturbation theory.  In evaluating the width it is natural to combine the result from the interference of diagrams C and D0 with that of diagram E with itself. The result of this perturbative calculation is reported in Sec.\,\ref{subsec:CEE} below\,\cite{Carrasco:2015xwa}.

\subsection{Contribution to the amplitude from Diagram D}\label{subsec:MD}
The contribution to the amplitude for the decay $\pi^+\to\ell^+\nu_\ell\gamma$ from diagram D, written in terms of Minkowski space quantities is 
\begin{equation}
M_D=ie\,\GF\,g_{\mu\mu^\prime}g_{\nu\nu^\prime}\,\epsilon^{\nu}_\lambda(k)\,L_0^\mu\,\int\dfour x~H_1^{\mu^\prime\nu^\prime}\!(x)\,e^{-ik\cdot x}\,,
\label{eq:MDMink}
\end{equation}
where $k$ is the momentum of the final state photon and $\lambda$ its polarisation with polarisation vector $\epsilon^{\nu}_\lambda(k)$. The charge $e$ is that of the positron. Rewriting the right-hand side in terms of Euclidean space quantities we have
\begin{equation}
M_D=-ie\GF\,\epsilon^{\nu}_{\lambda E}(k)\,L_{0E}^\mu\,\int dx_4\int \dthree x~H_{1E}^{\mu\nu}(x)\,e^{E_\gamma x_4}e^{-i\vec{k}\cdot\vec{x}}\,,
\label{eq:MDEucl}\end{equation}
where again the subscript $E$ reminds us that all $\gamma$-matrices and $\epsilon_\lambda$ are in Euclidean space as defined in Appendix\,\ref{subsec:conventionsE}. Again, since the reminder of this subsection is presented in Euclidean space, for notational convenience we now drop the label $E$.

The corresponding Euclidean hadronic correlation function from which $H_1^{\mu\nu}(x)$ is determined is 
\begin{equation}
C_{H_1}^{\mu\nu}(t_\pi,x)=\langle 0|T\big[J_W^\mu(0)J_{\mathrm{em}}^\nu(x)\phi^\dagger(-t_\pi)\big]|0\rangle
\label{eq:CBdef}\end{equation}
and with the assumption that $t_\pi$ is sufficiently large 
\begin{equation}
C_{H_1}^{\mu\nu}(t_\pi,x)=\frac{Z_0}{2m_\pi^0}\,H_{1}^{\mu\nu}(x)\,e^{-m_\pi^0t_\pi}\label{eq:CDtoH1}\,.
\end{equation}
Thus a computation of $C_{H_1}^{\mu\nu}(t_\pi,x)$ together with the values of $m_\pi^0$ and $Z_0$ obtained from $C^0_{\phi\phi}(t_\pi)$ in Eq.\,(\ref{eq:C0phiphidef}) allows us to determine $H_{1}^{\mu\nu}(x)$. 
As in the previous subsections, we divide the integration over $x_4$ in Eq.\,(\ref{eq:MDEucl}) into two regions, labelled $L$ for \emph{Long}, i.e. $x_4\le -t_s$,  and $S$ for \emph{Short}, i.e. $x_4>-t_s$.

The hadronic matrix element $H_1^{\mu\nu}(x)$ in region $S$ is obtained directly from the correlation function $C_{H_1}^{\mu\nu}(t_\pi,x)$, which can be computed in finite volume with only exponentially suppressed finite volume errors, so that
\begin{equation}
M_D^S=-ie\GF\,\epsilon^{\nu}_{\lambda}(k)\,L_{0}^\mu\,\int_{-t_s}^\infty dx_4\int \dthree x~H_{1}^{\mu\nu}(x)\,e^{E_\gamma x_4}e^{-i\vec{k}\cdot\vec{x}}
\label{eq:MDS}
\end{equation}
is evaluated directly using the computed values of $H_1^{\mu\nu}(x)$. In region $L$ on the other hand we use IVR to write 
\begin{eqnarray}
\int^{-t_s}_{-\infty} \!dx_4\int \dthree x~H_{1}^{\mu\nu}(x)\,e^{E_\gamma x_4}e^{-i\vec{k}\cdot\vec{x}}
&=&\int^{-t_s}_{-\infty} dx_4\int \dthree x\int\frac{\dthree p}{(2\pi)^3}\int\dthree x^\prime 
~H_{1}^{\mu\nu}(\vec{x}^{\,\prime},-t_s)\times\nonumber\\
&&\hspace{0.8in}e^{(E_\pi(\vecp{p})-m_\pi^0)(x_4+t_s)}\,e^{-i\vec{p}\cdot(\vec{x}-\vec{x}^\prime)}e^{E_\gamma x_4}e^{-i\vec{k}\cdot\vec{x}}\nonumber\\
&&\hspace{-1.5in}=e^{(E_\pi(\vecp{k})-m_\pi^0)t_s}\int^{-t_s}_{-\infty} \!\!dx_4\int \dthree x ~H_1^{\mu\nu}(\vec{x},-t_s)\,e^{(E_\pi(\vecp{k})+E_\gamma-m_\pi^0)x_4}\,
e^{-i\vec{k}\cdot\vec{x}}\nonumber\\
&&\hspace{-1.5in}=\frac{e^{-E_\gamma t_s}}{(E_\pi(\vecp{k})+E_\gamma-m_\pi^0)}\int\dthree x~H_1^{\mu\nu}(\vec{x},-t_s)\,e^{-i\vec{k}\cdot\vec{x}}\,.
\end{eqnarray}
We now write $e^{-i\vec{k}\cdot\vec{x}}=1+(e^{-i\vec{k}\cdot\vec{x}}-1)$ and separate the term which leads to the infrared divergence in the rate from the convergent term
\begin{equation}
M_D^{L}=M_D^{L\,\mathrm{div}}+M_D^{L\,\mathrm{con}}\,,
\end{equation}
where
\begin{equation}
M_D^{L\,\mathrm{div}}=-ie\GF\,\epsilon^{4}_{\lambda}(k)\,L_{0}^4\,H_0^4\,
\frac{e^{-E_\gamma t_s}}{(E_\pi(\vecp{k})+E_\gamma -m_\pi^0)}
\label{eq:MDLdiv}
\end{equation}
and
\begin{eqnarray}
M_D^{L\,\mathrm{con}}&=&ie\GF\,\epsilon^{\nu}_{\lambda}(k)\,L_{0}^\mu\,
\frac{e^{-E_\gamma t_s}}{(E_\pi(\vecp{k})+E_\gamma-m_\pi^0)}\int\dthree x~H_1^{\mu\nu}(\vec{x},-t_s)\,\big(1-e^{-i\vec{k}\cdot\vec{x}}\big)\,.
\label{eq:MDLcon}
\end{eqnarray}
Collecting up the three contributions in Eqs.\,(\ref{eq:MDLdiv}), (\ref{eq:MDLcon}) and (\ref{eq:MDS}), the result for this diagram is
\begin{equation}
M_D=M_D^{L\,\mathrm{div}}+M_D^{L\,\mathrm{con}}+M_D^S\,.
\end{equation}

\subsection{Contribution to the amplitude from Diagram E}

The contribution to the amplitude for the decay $\pi^+\to\ell^+\nu_\ell\gamma$ from diagram E is 
\begin{eqnarray}
M_E&=&-ie\GF\,\,g_{\mu\mu^\prime}\,g_{\nu\nu^\prime}\,\epsilon_\lambda^{\nu}(k)\,H_0^\mu\,\int\dfour x~L_{1}^{\mu^\prime\nu^\prime}(x)\,e^{-ik\cdot x}\,.\nonumber\\
&=&-ie\GF\,\,g_{\mu\mu^\prime}\,g_{\nu\nu^\prime}\,\epsilon_\lambda^{\nu}(k)\,H_0^\mu\,
\big\{\bar{u}(p_{\nu_\ell})\gamma^{\mu^\prime}(1-\gamma^5)\big\}_\alpha\int\dfour x\,S_{\alpha\beta}(0,x)
e^{-i(k+p_\ell)\cdot x}\big\{\gamma^{\nu^\prime} v(p_\ell)\big\}_\beta\,,
\label{eq:MEMink}
\end{eqnarray}
where $k$ and $p_\ell$ are the four-momenta of the final state photon and lepton respectively and 
$E_\ell^\prime(\vec{k})=\sqrt{(\vec{p}_\ell+\vecp{k})^2+m_\ell^2}$.
Note that the subscript $E$ denotes diagram $E$ and not \textit{Euclidean}. We start by presenting the discussion in Minkowski space.
The $x$ integration can be performed as follows,
\begin{eqnarray}
\int\dfour x\,S_{\alpha\beta}(0,x)\,e^{-i(k+p_\ell)\cdot x}&=&
i\int\dfour x\,\int\frac{\dfour p}{(2\pi)^4}\,\frac{(p_0\gamma_0-\vec{p}\cdot\vec{\gamma}+m_\ell)_{\alpha\beta}}{p_0^2-\vec{p}^{\,2}-m_\ell^2+i\epsilon}\,e^{-i(p+k+p_\ell)\cdot x}\nonumber\\
&=&i\int dx_0\int\frac{dp_0}{2\pi}\,\frac{(p_0\gamma_0+(\vec{p}_\ell+\vec{k})\cdot\vec\gamma+m_\ell)_{\alpha\beta}}
{p_0^2-\Elpk^2+i\epsilon}\,e^{i(p_0+E_\gamma+E_\ell)x_0}\nonumber\\
&&\hspace{-1.2in}=-i\,\frac{1}{2\Elpk}\,\bigg\{\frac{\big(-\Elpk\gamma_0+(\vec{p}_\ell+\vec{k})\cdot\vec{\gamma}+m_\ell\big)_{\alpha\beta}}{\Elpk-E_\gamma-E_\ell}+\frac{\big(\Elpk\gamma_0+(\vec{p}_\ell+\vec{k})\cdot\vec{\gamma}+m_\ell\big)_{\alpha\beta}}{\Elpk+E_\gamma+E_\ell}\bigg\}\,.\label{eq:MBaux1}
\end{eqnarray}

Note that from the second line of Eq.\,(\ref{eq:MBaux1}) one could have performed the $x_0$ integration, obtaining $\delta(p_0+E_\ell+k)$ so that
\begin{equation}
\int\dfour x\,S_{\alpha\beta}(0,x)\,e^{-i(k+p_\ell)\cdot x}=i\,
\frac{(-(E_\ell+E_\gamma)\gamma_0+(\vec{p}_\ell+\vec{k})\cdot\vec\gamma+m_\ell)_{\alpha\beta}}
{(E_\ell+E_\gamma)^2-\Elpk^2}\,,
\end{equation} 
which is equal to the expression in the third line of Eq.\,(\ref{eq:MBaux1}). However, as will become apparent in Sec.\,\ref{sec:ircancellation}, for the implementation of the IVR framework, it is convenient to write the result in the form of the third line in Eq.\,(\ref{eq:MBaux1}). The first term in braces is the result from the integration over positive values of $x_0$ and the denominator vanishes in the limit of the photon's momentum $\vec{k}\to\vec{0}$. This leads to an infrared divergence in the $\pi^+\to\ell^+\nu_\ell\gamma$ decay rate. The second term is the contribution from the integration over $x_0<0$ and does not lead to an infrared divergence. The two terms will therefore be treated separately. 

Until now the discussion has been presented entirely in Minkowski space. Since the evaluation of the hadronic matrix elements is necessarily performed in Euclidean space, we now rewrite $M_E$ in terms of Euclidean $\gamma$-matrices and polarisation vectors (see Eq.\,(\ref{eq:epsilonE})):
\begin{eqnarray}
M_E&=&M_E^\mathrm{div}+M_E^\mathrm{con}\label{eq:MEEuclidean}
\end{eqnarray}
where
\begin{eqnarray}
M_E^\mathrm{div}&=&ie\,\frac1{2\Elpk}\,\GF\,\epsilon_\lambda^{\nu}(k)\,H_0^4\quad
\frac{\bar{u}(p_{\nu_\ell})\gamma^{4}(1-\gamma^5)
\big(-\Elpk\gamma^4+i(\vec{p}_\ell+\vec{k})\cdot\vec{\gamma}+m_\ell\big)\gamma^{\nu} v(p_\ell)}{\Elpk-E_\gamma-E_\ell}
\label{eq:MEdiv}\\ 
M_E^\mathrm{con}&=&ie\,\frac1{2\Elpk}\,\GF\,\epsilon_\lambda^{\nu}(k)\,H_0^4\quad
\frac{\bar{u}(p_{\nu_\ell})\gamma^{4}(1-\gamma^5)
\big(\Elpk\gamma^4+i(\vec{p}_\ell+\vec{k})\cdot\vec{\gamma}+m_\ell\big)\gamma^{\nu} v(p_\ell)}{\Elpk+E_\gamma+E_\ell}
\,.\label{eq:MEcon}
\end{eqnarray}
In order not to over-complicate the notation we have not included labels to denote explicitly that the $\gamma$-matrices and polarisation vector in Eqs.\,(\ref{eq:MEdiv}) and (\ref{eq:MEcon}) are the Euclidean ones as defined in Sec.\,\ref{subsec:conventionsE}.

\section{Cancellation of infrared divergences}\label{sec:ircancellation}

In Sec.\,\ref{sec:diagrams} we have presented expressions for the contributions from each of the diagrams in Fig.\,\ref{fig:diagrams} to the amplitude for the process $\pi^+\to\ell^+\nu_\ell(\gamma)$ in terms of the hadronic matrix elements $H_i$, the leptonic factors $L_i$  (in both cases i=0,1,2) and the photon propagator $S^\gamma$. 
%
%
In this section we demonstrate how to handle the well-known problem of the cancellation of infrared divergences. At $O(\alpha_\mathrm{em})$ these divergences cancel between the rate for the decay $\pitolnu$ with the propagator of a virtual photon and that for the process $\pitolnug$ with a real photon in the final state\,\cite{Bloch:1937pw}. When calculating the decay rates we perform integrals over the two-body ($\Phi_2$) or three-body ($\Phi_3$) phase-space of the schematic form
\begin{equation}
\int d\Phi_2~\langle\, \pi^+|\,T^\dagger\,|\,\ell^+\nu_\ell\rangle~\langle\ell^+\nu_\ell\, |\,T\,|\,\pi^+\rangle 
\qquad\mathrm{and}\qquad
\int d\Phi_3~\langle\, \pi^+|\,T^\dagger\,|\,\ell^+\nu_\ell\gamma\rangle~\langle\ell^+\nu_\ell\gamma\, |\,T\,|\,\pi^+\rangle\,.\label{eq:rates}
\end{equation}
We will take the virtual photon to be in the Feynman gauge and Eq.\,(\ref{eq:polsum}) for the sum over polarisations of the real photon. 

The cancellation of the infrared divergences occurs between subsets of the diagrams in Fig.\,\ref{fig:diagrams}. The subsets are shown in Tab.\,\ref{tab:subsets} in which the cancellation occurs between the contributions in each of the three rows. Thus the infrared divergence from the two body phase-space integral of the contribution of diagram $A$ to the amplitude $T$ and the lowest order diagram $D0$ to $T^\dagger$ (and vice-versa) cancels that from the three-body phase-space integral in which the contribution from diagram $D$ is taken in both $T$ and $T^\dagger$. Similarly for the remaining two rows. We therefore consider the subsets of diagrams in each of the three rows separately in Subsecs.\,\ref{subsec:ADD}\,-\,\ref{subsec:CEE} respectively.  

\begin{table}[t]
\begin{center}
\begin{tabular}{cc|cc}
\multicolumn{2}{l|}{\footnotesize{$\pitolnu$}}&\multicolumn{2}{c}{\footnotesize{$\pitolnug$}}\\ \hline
\rule[-0.1cm]{0pt}{0.2in}\hspace{0.07in}$T$&\hspace{0.07in}$T^\dagger$&\hspace{0.07in}$T$&\hspace{0.07in}$T^\dagger$\\ 
\hline
\rule[-0.1cm]{0pt}{0.2in}\hspace{0.07in}A &\hspace{0.07in} D0 & \hspace{0.07in}D & \hspace{0.07in}D \\
\hspace{0.07in}B &\hspace{0.07in} D0 & \hspace{0.07in}D & \hspace{0.07in}E \\
\hspace{0.07in}C & \hspace{0.07in} D0 & \hspace{0.07in}E & \hspace{0.07in}E 
\end{tabular}
\end{center}
\caption{The infrared divergences cancel between the phase-space integrals of the contributions to the matrix elements of $T$ and its conjugate $T^\dagger$ from the diagrams in each of the final three rows separately. \label{tab:subsets}}
\end{table}

\subsection{IR cancellation for diagram A and DD}\label{subsec:ADD}

In this subsection we consider the cancellation of infrared divergences between the contribution of the interference of diagrams D0 and A to the decay width of the process $\pitolnu$ and the contribution to the square of diagram D to the width of the decay $\pitolnug$ (see the first of the final three rows of Tab.\,\ref{tab:subsets}).

The $O(\alpha_\mathrm{em})$ infrared divergent contribution to the decay width from the interference of the QCD diagram D0 and diagram A is given by
\begin{equation}\label{eq:Gamma0Adiv0}
\Gamma_{0A}^\mathrm{div}=\frac1{2m_\pi}\,\int d\Phi_2(p_\pi;p_\ell,p_{\nu_\ell})\,2\mathrm{Re}\big[M_A^\mathrm{div}M_0^\dagger\big]\,,
\end{equation}
where $\Phi_2(p_\pi;p_\ell,p_{\nu_\ell})$ is the phase space of the two-body final state consisting of the charged lepton $\ell$ and its neutrino $\nu_\ell$, with $p_\ell+p_{\nu_\ell}=p_\pi=(m_\pi,\vec{0}\,)$ and $M_A^\mathrm{div}$ is defined in Eq.(\ref{eq:MAdiv}). Combining Eqs.\,(\ref{eq:M0Euclidean}), (\ref{eq:MA}) and (\ref{eq:H4div}) we rewrite $\Gamma_{0A}^\mathrm{div}$ as 
\begin{equation}\label{eq:Gamma0Adiv}
\Gamma_{0A}^\mathrm{div}=\frac{e^2}{2m_\pi}\left(\frac{G_F^2|V_{ud}|^2}{2}\right)\big|H_0^4\big|^2\,\int\frac{\dthree k}{(2\pi)^3}\,\frac1{2E_\gamma} \,
\frac{e^{-E_\gamma t_s}}{(E_\pi(\vecp{k})+E_\gamma-m_\pi^0)^2}\,\int d\Phi_2(p_\pi;p_\ell,p_{\nu_\ell})\,\big|L_0^4(p_\ell,p_\nu)\big|^2
\end{equation}
and a sum over the polarisations of the final-state leptons is implicit.

The infrared divergent contribution to the decay width for the process $\pi^+\to\ell^+\bar\nu\gamma$ from the square of diagram D is 
\begin{eqnarray}
\Gamma_{DD}^\mathrm{div}&=&\frac1{2m_\pi}\,\int d\Phi_3(p_\pi;p_\ell,p_{\nu_\ell},k) \big|M_D^\mathrm{div}\big|^2\nonumber\\
&=&-\frac{e^2}{2m_\pi}\left(\frac{G_F^2|V_{ud}|^2}{2}\right)\big|H_0^4\big|^2
\int\frac{\dthree k}{(2\pi)^3}\,\frac1{2E_\gamma} \,
\frac{e^{-2E_\gamma t_s}}{(E_\pi(\vecp{k})+E_\gamma-m_\pi^0)^2}
\int d\Phi_2(p_\pi-k;p_\ell,p_{\nu_\ell})\,\big|L_0^4(p_\ell,p_\nu)\big|^2,
\label{eq:GammadivDD}
\end{eqnarray}
where $\Phi_3(p_\pi;p_\ell,p_{\nu_\ell},k)$ is the phase space of the three-body final state consisting of the charged lepton $\ell$, its neutrino $\nu_\ell$ and a photon, with $p_\ell+p_{\nu_\ell}+k=p_\pi=(m_\pi,\vec{0}\,)$
and $M_D^\mathrm{div}$ is given in Eq.\,(\ref{eq:MDLdiv}). 
In the sum of diagrams D and E only photons with physical polarisations contribute of course. However in order for the infrared divergences to cancel separately in the three rows of Tab.\,\ref{tab:subsets}, we exploit the electromagnetic Ward identity and define the diagrams with a virtual photon (diagrams A, B and C) to be in the Feynman gauge and 
and take for the sum over the photon polarisations 
$\sum_\lambda\epsilon_\lambda^{\mu}(k)\epsilon_\lambda^{\nu\,\ast}(k)=g^{\mu\nu}$ in both diagrams D and E so that 
$\sum_\lambda |\epsilon^4_\lambda(k)|^2=-1$\,. The sum over the lepton polarisations is again implicit in Eq.\,(\ref{eq:GammadivDD}).

As $k\to 0$, the integrands in Eqs.(\ref{eq:Gamma0Adiv}) and (\ref{eq:GammadivDD}) become equal and opposite so that $\Gamma_{0A}^\mathrm{div}+\Gamma_{DD}^\mathrm{div}$ is infrared finite. The finite terms can be determined without any lattice calculations (beyond the evaluation of $H_0^4$ in QCD)
as explained in detail in Appendix\,\ref{subsec:0ADDdiv}. Thus, by using the analytic control of the long-distance portion of the electromagnetic corrections given by IVR, we are able to realize the usual cancellation of infrared divergences before any lattice calculation is undertaken.

\subsection{IR cancellation in diagrams D0, B, D and E}\label{subsec:BDE}

In this subsection we consider the second of the three rows in Tab.\ref{tab:subsets} and demonstrate the cancellation of the infrared divergences between the contributions from the interference of diagrams D0 and B to the decay width for the process $\pi^+\to\ell^+\nu_\ell$ (we denote this contribution by $\Gamma_{0B}$) and the interference of diagrams D and E to the decay width for the process $\pi^+\to\ell^+\nu_\ell\gamma$ (denoted by $\Gamma_{DE}$). 

We start by considering $\Gamma_{0B}$ which can be written in the form
\begin{equation}
\Gamma_{0B}=\frac1{2m_\pi}\,\int d\Phi_2(p_\pi;p_\ell,p_{\nu_\ell})\,2\mathrm{Re}\big[M_BM_0^\dagger\big]\,,
\end{equation}
where $\Phi_2(p_\ell,p_{\nu_\ell})$ is the phase space of the two-body final state consisting of the charged lepton $\ell$ and its neutrino $\nu_\ell$, with $p_\ell+p_{\nu_\ell}=p_\pi=(m_\pi,\vec{0}\,)$. The infrared divergent contribution comes from the component $M_B^{L\,\mathrm{div}}$ in $M_B$, presented in Eq.\,(\ref{eq:MBLdiv}) and here we focus on this contribution:
\begin{eqnarray}
\Gamma_{0B}^{\mathrm{div}}&=&\frac1{2m_\pi}\,\int d\Phi_2(p_\pi;p_\ell,p_{\nu_\ell})\,2\mathrm{Re}\big[M_B^{L\,\mathrm{div}}M_0^\dagger\big]
\nonumber\\
&=&\frac{e^2}{m_\pi}\,\frac{G_F^2}{2}\,|V_{ud}|^2\,\big|H_0^4\big|^2\,\int d\Phi_2(p_\pi;p_\ell,p_{\nu_\ell})L_0^{4\,\dagger}\,\times\nonumber\\
&&\int\frac{\dthree k}{(2\pi)^3}\,
\frac{e^{-E_\gamma t_s}}{2E_\gamma}
\left\{\frac{\bar{u}(p_{\nu_\ell})\gamma^4(1-\gamma^5)(-\Elpmk\gamma_4+i(\vec{p}_\ell-\vec{k})\cdot\vec{\gamma}+m_\ell)\gamma^4 v(p_\ell)}{2\Elpmk(E_\ell^\prime(-\vec{k})+E_\gamma-E_\ell)(E_\pi(\vec{k})+E_\gamma-m_\pi^0)}\right\}\,,\label{eq:Gammadiv0B}
\end{eqnarray}
where $M_B^{L\,\mathrm{div}}$ is given in Eq.\,(\ref{eq:MBLdiv}).

The contribution to the decay width from the interference of diagrams D and E can be written in the form:
\begin{equation}
\Gamma_{DE}=\frac{1}{2m_\pi}\,\int d\Phi_3(p_\pi;p_\ell,p_{\nu_\ell},k)\,2\mathrm{Re}\big[M_D^\dagger M_E\big]\,,
\end{equation}
where $\Phi_3(p_\pi;p_\ell,p_{\nu_\ell},k)$ is the phase space of the three-body final state consisting of the charged lepton $\ell$, the neutrino $\nu_\ell$ and the photon $\gamma$ with $p_\ell+p_{\nu_\ell}+k=p_\pi=(m_\pi,\vec{0}\,)$. The infrared divergent term in the width comes from the $M_D^{L\,\mathrm{div}}$  contribution to $M_D$ (see Eq.\,(\ref{eq:MDLdiv})) and the $M_E^\mathrm{div}$ contribution to $M_E$ (see Eq.\,(\ref{eq:MEdiv})) and is given by
\begin{eqnarray}
\Gamma_{DE}^{\mathrm{div}}&=&\frac{1}{2m_\pi}\,\int d\Phi_3(p_\pi;p_\ell,p_{\nu_\ell},k)\,2\mathrm{Re}\big[M_D^{L\,\mathrm{div}\dagger} M_E^\mathrm{div}\big]\nonumber\\
&=&\frac{e^2}{m_\pi}\GFsq\,\big|H_0^4\big|^2\,
\,\int d\Phi_3(p_\pi;p_\ell,p_{\nu_\ell},k)\,L_0^{4\,\dagger}\,e^{-E_\gamma t_s}\times\nonumber\\
&&\bar{u}(p_{\nu_\ell})\gamma^{4}(1-\gamma^5)\bigg\{\frac{\big(-\Elpk\gamma^4+i(\vec{p}_\ell+\vec{k})\cdot\vec{\gamma}+m_\ell\big)}{2\Elpk\,
(E_\ell^\prime(\vec{k}\,)-E_\gamma-E_\ell)\,(\Epik+E_\gamma-m_\pi^0)}\bigg\}\gamma^{4} v(p_\ell)\,,\label{eq:GammadivDE}
\end{eqnarray}
where the sum of the photon polarisations has been performed. $M_D^{L\,\mathrm{div}}$ and $M_E^{\mathrm{div}}$ are given in  Eqs.\,(\ref{eq:MDLdiv}) and (\ref{eq:MEdiv}) respectively.

Both $\Gamma_{0B}^\mathrm{div}$ and $\Gamma_{DE}^\mathrm{div}$ are infrared divergent, with the integrand over $\vec{k}$ proportional to $1/k^3$ at small $k$. Noting however that at small $k$
\begin{equation}
E_\ell^\prime(\pm\vec{k}\,)\simeq E_\ell\pm\frac{\vec{p}_\ell\cdot\vec{k}}{E_\ell}\
\end{equation}
we see that 
\begin{equation}
E_\ell^\prime(-\vec{k})+E_\gamma-E_\ell\simeq E_\gamma-\frac{\vec{p}_\ell\cdot\vec{k}}{E_\ell}\simeq-(E_\ell^\prime(\vec{k}\,)-E_\gamma-E_\ell)
\end{equation}
so that $\Gamma_{0B}^{\mathrm{div}}+\Gamma_{DE}^{\mathrm{div}}$ is indeed infrared convergent. We explain the evaluation of the finite terms in $\Gamma_{0B}^{\mathrm{div}}+\Gamma_{DE}^{\mathrm{div}}$ in Appendix\,\ref{subsec:0BDEdiv}. Again, this does not require any lattice calculations beyond the determination of $H_0^4$.

\subsection{IR cancellation for diagram C and EE}\label{subsec:CEE}

The hadronic matrix element which contributes to diagrams C and E with the initial pion at rest is $H_0^4$ which is readily obtained 
from two-point correlation functions in Eqs.\,(\ref{eq:C0phiphidef}) and (\ref{eq:C0JWphidef}) with only exponentially small finite-volume corrections. Once $H_0^4$, or equivalently the decay constant $f_\pi$, have been computed, the contributions to the decay width from these diagrams only requires $O(\alpha_{\mathrm{em}})$ calculations within QED. These have been performed in Ref.\,\cite{Carrasco:2015xwa}.

The contribution to the decay width from the interference of diagrams D0 and the wave-function renormalisation of the lepton from diagram C in the Feynman gauge is given by
\begin{equation}\label{eq:D0C}
\Gamma_{0C}=  \frac{\alpha_\mathrm{em}}{4\pi}\,\Gamma_0\,
\left\{\log\left(\frac{m_\ell^2}{M_W^2}\right)  - 2 \,  \log \left( \frac{m_\gamma^2}{m_\ell^2}\right)-  \frac{9}{2}
\right\}\,,
\end{equation}
where we use the $W$-regularisation for the ultra-violet divergences and have introduced a mass $m_\gamma$ for the photon in order to regulate the infrared divergences.

The contribution to the decay width from the square of diagram E, with photon energies integrated up to $\Delta E$ in the pion rest frame is
\begin{eqnarray}\label{eq:EE}
\Gamma_{EE}&=&\frac{\alpha_\mathrm{em}}{4\pi}\,\Gamma_0\,\left( R_{EE1}+ R_{EE2} \right)\,,
\end{eqnarray}
where
\begin{eqnarray}  
R_{EE1}&=&
2 \log\left(\frac{m_\gamma^2}{4\Delta E^2}\right) 
-2\frac{1+r_\ell^2}{1-r_\ell^2} \log(r_\ell^2)\,,
\nonumber \\
R_{EE2}&=& 
\frac{r_E^2-1 +  (4 r_E-6) r_\ell^2}{(1-r_\ell^2)^2}\ \log(1-r_E)
-\frac{r_E(r_E + 4 r_\ell^2)}{(1-r_\ell^2)^2}\ \log(r_\ell^2)+\frac{r_E(6 - 3 r_E - 20 r_\ell^2)}{2(1-r_\ell^2)^2}\label{eq:REE}
\; ,\end{eqnarray}
where $r_\ell=m_\ell/m_\pi$ and $r_E=2\Delta E/m_\pi$. The contribution to the total rate is obtained by setting $\Delta E$ to its maximum value of $m_\pi/2\,(1-r_\ell^2)$. 
Here we introduce the familiar photon energy cutoff $\Delta E$ in the pion's rest frame in order to write a simple explicit formula.  As is described in Sec.\,\ref{sec:final} below, this simple cutoff can be replaced as needed by energy or angle cuts dictated by a particular experimental setup.

The infrared divergences explicitly cancel in the sum $\Gamma_{0C}$ and $\Gamma_{EE}$ and the remaining infrared finite terms are given in Eqs.\,(\ref{eq:D0C})\,-\,(\ref{eq:REE}).

\section{Final Result}\label{sec:final}

The final result for the $O(\alpha_\mathrm{em})$ contributions to $\Gamma(\pi^+\to\ell^+\nu_\ell)+\Gamma(\pi^+\to\ell^+\nu_\ell\gamma)$ consists of a large number of terms presented in different sections and subsections of this paper and we now collect them all together here. We start by writing 
\begin{equation}
\Gamma(\pi^+\to\ell^+\nu_\ell)+\Gamma(\pi^+\to\ell^+\nu_\ell\gamma)=\Gamma_0+(\Gamma_{0A}+\Gamma_{DD})+(\Gamma_{0B}
+\Gamma_{DE})+(\Gamma_{0C}+\Gamma_{EE})\,,
\end{equation}
where $\Gamma_0$, given in Eq.\,(\ref{eq:Gamma0}), is the width without electromagnetic corrections, and the remaining six terms represent the interference of the amplitudes indicated in the subscripts; thus for example, $\Gamma_{0A}$ is the contribution from the interference of the $O(\alpha_\mathrm{em}^0)$ diagram D0 and the $O(\alpha_\mathrm{em})$ diagram A, integrated over phase space. The six contributions at $O(\alpha_\mathrm{em})$ have been grouped into three pairs, each of which is infrared convergent. We now present the results for each of these three pairs in turn, without rewriting all the expressions, but pointing instead to the equations in the text where they can be found. 

For the two-body decay $\pi^+\to\ell^+\nu_\ell$ the integration over the two-body final-state phase space is fixed. For the three-body decay 
$\pi^+\to\ell^+\nu_\ell\gamma$,  it may be appropriate to compute a partial width by introducing 
cuts on the kinematical variables, such as the energy of the photon or the angle between the photon and the charged lepton, in order to match the  
theoretical prediction to experimental measurements. The cancellation of infrared divergences is unaffected, but the remaining finite terms depend on the cuts.  
Below we do not specify whether any such cuts have been imposed and simply write the three-body phase space integral as $\int d\Phi_3(p_\pi;p_\ell,p_\nu,k)$.

\subsection{$\Gamma_{0A}+\Gamma_{DD}$}
Using the notation of this paper, the result for $\Gamma_{0A}$ can be written in the form:
\begin{equation}
\Gamma_{0A}=\Gamma_0\left(\frac{\delta m_\pi}{m_\pi^0}
+\frac{2m_\pi^0}{Z_0e^{-m_\pi^0t_\pi}H_0^4}\left(2
\tilde{C}_{J_W\phi;R_1}^{A}(t_\pi)+\tilde{C}^{A\,S}_{J_W\phi;R_2}(t_\pi)+\tilde{C}^{A\,L\,\mathrm{div}}_{J_W\phi;R_2}(t_\pi)
+\tilde{C}^{A\,L\,\mathrm{con}}_{J_W\phi;R_2}(t_\pi)
\right)\right)\label{eq:Gamma0Atotal}
\end{equation}
where $\tilde{C}_{J_W\phi;R_1}^{A}(t_\pi)=C_{J_W\phi;R_1}^{A}(t_\pi)$ is given in Eqs.\,(\ref{eq:CJWphiR1tot})\,-\,(\ref{eq:CJWphiR1d}), $C_{J_W\phi;R_2}^{A\,S}(t_\pi)$ in Eq.\,(\ref{eq:CAJWR2S}) (or equivalently Eq.\,(\ref{eq:CAJphi2con}) after subtraction of the term proportional to $t_\pi$), 
$\tilde{C}^{A\,L\,\mathrm{div}}_{J_W\phi;R_2}(t_\pi)$ in Eq.\,(\ref{eq:tildeCAJphi2div}) and 
$\tilde{C}^{A\,L\,\mathrm{con}}_{J_W\phi;R_2}(t_\pi)$ in Eq.\,(\ref{eq:tildeCAJphi2con}). The infrared divergence is contained in the term proportional to 
$\tilde{C}^{A\,L\,\mathrm{div}}_{J_W\phi;R_2}(t_\pi)$, i.e. $\Gamma_{0A}^\mathrm{div}$ given in Eqs.\,(\ref{eq:Gamma0Adiv0}) and (\ref{eq:Gamma0Adiv}), and this is treated separately in Appendix\,\ref{subsec:0ADDdiv}. The remaining terms are infrared finite. The new non-perturbative input into these calculations are the $H_{2s}^{\mu\nu}(\vec{x},t)$ for values of $|t|\le t_s$.

$\Gamma_{DD}$ is given by
\begin{eqnarray}
\Gamma_{DD}&=&\frac1{2m_\pi}\int d\Phi_3(p_\pi;p_\ell,p_\nu,k) |M_D|^2=
\frac1{2m_\pi}\int d\Phi_3(p_\pi;p_\ell,p_\nu,k)\, \big|M_D^S+M_D^{L\,\mathrm{div}}+M_D^{L\,\mathrm{con}}\big|^2
\nonumber\\
&=&\frac1{2m_\pi}\int d\Phi_3(p_\pi;p_\ell,p_\nu,k) 
\bigg(|M_D^S|^2+|M_D^{L\,\mathrm{div}}|^2+|M_D^{L\,\mathrm{con}}|^2
+2\,\mathrm{Re}\big[M_D^\mathrm{S}(M_D^{L\,\mathrm{div}})^\dagger\big]\nonumber\\
&&\hspace{1.2in}+2\,\mathrm{Re}\big[M_D^\mathrm{S}(M_D^{L\,\mathrm{con}})^\dagger\big]
+2\,\mathrm{Re}\big[M_D^\mathrm{L\,\mathrm{con}}(M_D^{L\,\mathrm{div}})^\dagger\big]\bigg)\,,
\end{eqnarray}
where $M_D^S$ is given in Eq.\,(\ref{eq:MDS}), $M_D^{L\,\mathrm{div}}$ in Eq.\,(\ref{eq:MDLdiv}) and $M_D^{L\,\mathrm{con}}$ in Eq.\,(\ref{eq:MDLcon}). The infrared divergence is contained in the term with $|M_D^{L\,\mathrm{div}}|^2$ in the integrand and is treated separately in Appendix\,\ref{subsec:0ADDdiv}. The remaining 5 terms 
are all infrared convergent. Note that the only non-perturbative ingredient, which needs ultimately to be computed using lattice QCD, is $H_1^{\mu\nu}(\vec{x},t)$ for time separations between the weak and electromagnetic current which are smaller than or equal to $t_s$, $|t|\le t_s$.

The finite terms which remain after the cancellation of the infrared divergences in $\Gamma_{0A}^\mathrm{div}+\Gamma_{DD}^\mathrm{div}$ depend on the 
three-body phase space over which $|M_D^{L\,\mathrm{div}}|^2$ is integrated. In Appendix\,\ref{subsec:0ADDdiv} we evaluate the finite terms obtained after integrating over the full three-body phase space.

\subsection{$\Gamma_{0B}+\Gamma_{DE}$}

The procedure for $\Gamma_{0B}+\Gamma_{DE}$ is very similar to the above. We write
\begin{eqnarray}
\Gamma_{0B}&=&\frac1{2m_\pi}\int d\Phi_2(p_\pi;p_\ell,p_\nu)~ 2\,\mathrm{Re}[M_BM_0^\dagger]\nonumber\\
&=&\frac1{2m_\pi}\int d\Phi_2(p_\pi;p_\ell,p_\nu)\, 2\,\mathrm{Re}\,\Big[\big(M_B^S+M_B^{L\,\mathrm{div}}+M_B^{L\,\mathrm{con}}\big)\,M_0^\dagger\Big]
\,,\label{eq:Gamma0B}
\end{eqnarray}
where $M_B^S$, $M_B^{L\,\mathrm{div}}$ and $M_B^{L\,\mathrm{con}}$ are given in Eqs.\,(\ref{eq:MBS}), (\ref{eq:MBLdiv}) and (\ref{eq:MBLcon}) respectively
and $M_0$ is given in Eq.\,(\ref{eq:M0Euclidean}}).
The infrared divergence is contained in the term with $M_B^{L\,\mathrm{div}}$ in the integrand of Eq.\,(\ref{eq:Gamma0B}) and is treated separately in Appendix\,\ref{subsec:0BDEdiv}. The remaining 2 terms 
are both infrared convergent. Again the only non-perturbative ingredient, which needs ultimately to be computed using lattice QCD, is $H_1^{\mu\nu}(\vec{x},t)$ for time separations between the weak and electromagnetic current which are smaller than or equal to $t_s$.

The expression for $\Gamma_{DE}$ is 
\begin{eqnarray}
\Gamma_{DE}&=&\frac1{2m_\pi}\int d\Phi_3(p_\pi;p_\ell,p_\nu,k) ~2\,\mathrm{Re}\,[M_DM_E^\dagger]\nonumber\\
&=&\frac1{2m_\pi}\int d\Phi_3(p_\pi;p_\ell,p_\nu,k) ~2\,\mathrm{Re}\,\Big[(M_D^S+M_D^{L\,\mathrm{div}}+M_D^{L\,\mathrm{con}})(M_E^\mathrm{div}+M_E^\mathrm{con})^\dagger\Big]\,,
\end{eqnarray} 
where $M_D^S$ is given in Eq.\,(\ref{eq:MDS}), $M_D^{L\,\mathrm{div}}$ in Eq.\,(\ref{eq:MDLdiv}), $M_D^{L\,\mathrm{con}}$ in Eq.\,(\ref{eq:MDLcon})
$M_E^{\mathrm{div}}$ in Eq.\,(\ref{eq:MEdiv}) and $M_E^{\mathrm{con}}$ in Eq.\,(\ref{eq:MEcon})\,.
The infrared divergence is contained in the term with $\mathrm{Re }[M_D^{L\,\mathrm{div}}M_E^{\mathrm{div}\,\dagger}]$ in the integrand and is treated separately in Appendix\,\ref{subsec:0BDEdiv}. The remaining 5 terms 
are all infrared convergent. There are no non-perturbative QCD ingredients in $M_E^{\mathrm{div}}$ and $M_E^{\mathrm{con}}$ (beyond the lowest order $H_0^4$), and we repeat that the only non-perturbative ingredient in $M_D$ , which needs ultimately to be computed using lattice QCD, is $H_1^{\mu\nu}(\vec{x},t)$ for time separations between the weak and electromagnetic current which are smaller than or equal to $t_s$, $|t|\le t_s$.

The finite terms which remain after the cancellation of the infrared divergences in $\Gamma_{0B}^\mathrm{div}+\Gamma_{DE}^\mathrm{div}$ depend on the 
three-body phase space over which $2\,\mathrm{Re}[M_D^{L\,\mathrm{div}}M_E^{\mathrm{div}\,\dagger}]$ is integrated. In Appendix\,\ref{subsec:0BDEdiv} we evaluate the finite terms obtained after integrating over the full three-body phase space.

\subsection{$\Gamma_{0C}+\Gamma_{EE}$}

For $\Gamma_{0C}+\Gamma_{EE}$ there is no hadronic input beyond the lowest order $H_0^4$. As above, the result depends on the three-body phase-space over which $|M_E|^2$ is integrated. In Eqs.\,(\ref{eq:D0C})\,-\,(\ref{eq:REE}), we present the results corresponding to an upper cut-off $\Delta E$ on the energy of the final state photon in the rest frame of the pion but integrating over the remaining variables\,\cite{Carrasco:2015xwa}. 
The total rate is obtained by setting $\Delta E$ to its maximum value $\Delta E^\mathrm{max}=m_\pi/2(1-m_\ell^2/m_\pi^2)$.

\section{Summary and Conclusions}\label{sec:concs}\,
In this paper we have presented a framework, based on Infinite Volume Reconstruction (IVR) for the evaluation of electromagnetic corrections, at $O(\alpha_\mathrm{em})$, to the leptonic decay widths of pseudoscalar mesons. Although we have used the decays of a $\pi^+$ to illustrate the method, it 
can be applied identically to the decays of other pseudoscalar mesons. The IVR technique is based on the observation for sufficiently large time separations ($t>t_s\lesssim L$ say) between the electromagnetic currents or between an electromagnetic and hadronic weak current, the only significant contribution comes from the propagation of a single pion and a photon. As has been explained in detail in Secs.\,\ref{sec:diagrams} and \ref{sec:ircancellation} this allows the computations of the hadronic matrix elements to be limited to time separations $\le t_s$. We
underline two important consequences of the IVR method:
\begin{enumerate}
\item[i)] The computation of hadronic effects in the electromagnetic corrections to the leptonic decay widths is organised in such a way that infrared divergences are not present. The cancellation of infrared divergences between the contributions to $\Gamma(\pi^+\to\ell^+\nu_\ell)$ and $\Gamma(\pi^+\to\ell^+\nu_\ell\gamma)$ at $O(\alpha_\mathrm{em})$\,\cite{Bloch:1937pw} is demonstrated analytically (see Sec.\,\ref{sec:ircancellation}), so that all the terms which need to be computed to determine $\Gamma(\pi^+\to\ell^+\nu_\ell)+\Gamma(\pi^+\to\ell^+\nu_\ell\gamma)$ are individually infrared finite. 

By contrast, in the QED$_\mathrm{L}$ method of Ref.\,\cite{Carrasco:2015xwa} the cancellation of infrared divergences is achieved by subtracting the contribution to the decay amplitudes obtained perturbatively by treating the pseudoscalar meson as a point-like particle from the non-perturbatively computed infrared divergent amplitudes (the divergences appear in the form $\log(m_\pi L)$). The method has been successfully implemented in Refs.\,\cite{Giusti:2017dwk,DiCarlo:2019thl}. It remains to be seen whether, and by how much, the uncertainties will be reduced by avoiding the subtraction of an analytic perturbative expression (the pointlike contribution to the amplitude) from a non-perturbatively computed term (the finite-volume infrared-divergent amplitude in QCD).
\item[ii)]  The implementation of this method, as described in this paper, results in finite-volume effects which are exponentially small in $L$ as compared to the  $O(1/(m_\pi L)^2)$ structure-dependent finite-volume corrections present with the QED$_\mathrm{L}$ approach\,\cite{Lubicz:2016xro}. The impact on the numerical precision of this remains to be investigated.

Finally, in Appendix\,\ref{sec:Kl3} we have outlined an additional application of IVR to the calculation of the $O(\alpha_\mathrm{em})$ corrections to the more complex decay process $K\to\pi\ell\nu_\ell$.  Here also a complete lattice calculation is possible with all infrared singularities treated analytically and all finite volume effects vanishing exponentially in $L$.
\end{enumerate}

\subsection*{Acknowledgements}
 We warmly thank our colleagues from the RBC-UKQCD collaboration for helpful discussions. CTS thanks his colleagues from RM123 for many fruitful discussions
 on QED corrections to decay amplitudes.

N.H.C. and T.W. acknowledge support from U.S. DOE grant \#DE-32SC0011941. X.F. was supported in part by NSFC of China under Grants No. 12125501, No. 12070131001, and No. 12141501, and National Key Research and Development Program of China under No. 2020YFA0406400. 
L.J. acknowledge support under US DOE grant DE-SC0010339 and US DOE
Office of Science Early Career Award DE-SC0021147. C.T.S. was partially supported by an Emeritus Fellowship from the Leverhulme Trust and by STFC (UK) grant ST/T000775/1.
\appendix

\section{Notation and Conventions}\label{sec:conventions}

We begin our discussion of the diagrams contributing to the physical amplitudes and decay rates in Minkowski space-time, before demonstrating that they can be determined in finite-volume lattice computations in Euclidean space.
In this appendix we briefly summarise our notation and the conventions which we use in the main text of this paper, in both Minkowski and Euclidean space-times. 

\subsection{Minkowski space-time}\label{subsec:conventionsM}
In Minkowski space-time we use the metric $g^{\mu\nu}=\mathrm{diag}(-1,1,1,1)$ and Dirac matrices which satisfy the anti-commutation relations $\{\gamma^\mu,\gamma^\nu\}=-2g^{\mu\nu}$. The electromagnetic current is given by
\begin{equation}
J^\mu_{\mathrm{em}}=\sum_f Q_f\,\bar{q}_f\gamma^\mu q_f-\sum_{\ell}\bar{\ell}\gamma^\mu\ell~,
\label{eq:Jemdef2}
\end{equation}
where the charges $Q_f=+\frac23$ for up-like quarks, $-\frac13$ for down-like ones and $-1$ for the leptons $\ell$. 

The discussion in this paper is presented for the decay $\pi^+\to\nu\ell^+(\gamma)$ but its generalisation to the decays of other pseudoscalar mesons, including those containing bottom and/or charm quarks, is totally straightforward with the natural replacement of the quark fields and CKM matrix elements. The hadronic weak current for the decay of a $\pi^+$ meson is $J_W^\mu=\bar{d}\gamma^\mu(1-\gamma^5)u$ and the corresponding Lagrangian density is 
\begin{equation}
{\cal L}_W(x)=\frac{G_F}{\sqrt{2}}\,V_{ud}^\ast\,g_{\mu\nu}\,J_W^\mu(x)\,\big[\bar{\nu}_\ell(x)\gamma^\nu(1-\gamma^5)\ell(x)\big]\,.
\end{equation}
The photon and lepton propagators are given respectively by
\begin{eqnarray}
S_\gamma^{\mu\nu}(x,y)&=&\int\frac{\dfour k}{(2\pi)^4}\,\frac{-ig^{\mu\nu}}{k^2-i\epsilon}e^{-ik\cdot(x-y)}\qquad\mathrm{(Feynman~gauge)}
\label{eq:photonprop}\\
S_\ell(x,y) &=& i\int\frac{\dfour p}{(2\pi)^4}\,\frac{p_\mu\gamma^\mu-m_\ell}{p^2+m_\ell^2-i\epsilon}e^{ip\cdot(x-y)}\,.
\end{eqnarray}
For decays with a real photon in the final state coupled to the conserved electromagnetic current, we take for the sum over polarisations $\lambda$: 
\begin{equation}\label{eq:polsum}
\sum_\lambda\epsilon_\lambda^\mu\big(\vec{k}\hspace{1pt}\big)\epsilon^{\ast\,\nu}_\lambda\big(\vec{k}\hspace{1pt}\big)=g^{\mu\nu}\,.
\end{equation}

\subsection{Euclidean space}\label{subsec:conventionsE}

Following the continuation to Euclidean space, $t_M\to -it_E$, $\vec{x}_M\to\vec{x}_E$ we relate the Dirac matrices in Euclidean and Minkowski spaces by\footnote{The suffices $E$ and $M$ denote Euclidean and Minkowski spaces respectively.}: 
\begin{equation}\label{eq:gammaME}
\gamma_{4\hspace{0.25pt}E}=\gamma^0_M\,,\qquad \vec{\gamma}_E=-i\vec{\gamma}_M\,.
\end{equation}
The photon propagator in the Feynman gauge is given by 
\begin{equation}\label{eq:SgammaE}
S_\gamma^{\mu\nu}(x,y)=\int\frac{\dfour k}{(2\pi)^4}\,\frac{\delta^{\mu\nu}}{k^2}\,e^{-ik\cdot(x-y)}=
\int\frac{\dthree k}{(2\pi)^3}\,\frac{\delta^{\mu\nu}}{2|\vec{k}|}e^{-|\vec{k}|\,|t_x-t_y|}\,e^{-i\vec{k}\cdot(\vec{x}-\vec{y})}=\frac{\delta^{\mu\nu}}{4\pi^2 |x-y|^2}
\end{equation}
and the lepton propagator is 
\begin{equation}
S_\ell(x,y) = \int\frac{\dfour p}{(2\pi)^4}\,\frac{-ip_\mu\gamma_\mu+m_\ell}{p^2+m_\ell^2}\,e^{ip\cdot(x-y)}
\equiv\int\frac{\dfour p}{(2\pi)^4}~\tilde{S}_\ell(p)\,e^{ip\cdot(x-y)}\,.
\end{equation}
For the polarisation vector in Euclidean space it is convenient to take
$\epsilon^{\ast 0}_{\lambda M}=-i\epsilon^{\ast 4}_{\lambda E}$ and $\epsilon^{\ast i}_{\lambda M}=\epsilon^{\ast i}_{\lambda E}$ ($i=1,2,3$) so that 
\begin{equation}\label{eq:epsilonE}
g_{\mu\nu}\epsilon_{\lambda M}^{\ast \mu}\,\gamma^\nu_M=i \epsilon^{\ast \mu}_{\lambda E}\gamma^\mu_E\,.
\end{equation}

\section{Cancellation of infrared divergences - the finite terms}\label{sec:finiteterms}

In Sec.\,\ref{sec:ircancellation} we have shown that the infrared divergences cancelled separately in $\Gamma_{0A}^\mathrm{div}+\Gamma_{DD}^\mathrm{div}$ (see Eqs.\,(\ref{eq:Gamma0Adiv}) and (\ref{eq:GammadivDD})), in $\Gamma_{0B}^\mathrm{div}+\Gamma_{DE}^\mathrm{div}$ (see Eqs.\,(\ref{eq:Gammadiv0B}) and (\ref{eq:GammadivDE}) and subsequent discussion) and in $\Gamma_{0C}^\mathrm{div}+\Gamma_{EE}^\mathrm{div}$ (see Eqs.\,(\ref{eq:D0C})\,-\,(\ref{eq:REE})). Although the infrared divergences cancel seperately in these three pairs, there remain finite-terms. For $\Gamma_{0C}^\mathrm{div}+\Gamma_{EE}^\mathrm{div}$ the finite terms are presented in Eqs.\,(\ref{eq:D0C})\,-\,(\ref{eq:REE})) as a function of the maximum photon energy $\Delta E$. The results depend on the three-body phase-space over which the widths for the decay $\pi^+\to\ell^+\nu_\ell\gamma$ are integrated. 
In this section we calculate the residual finite terms in $\Gamma_{0A}^\mathrm{div}+\Gamma_{DD}^\mathrm{div}$ and $\Gamma_{0B}^\mathrm{div}+\Gamma_{DE}^\mathrm{div}$ obtained by integrating over the full three-body phase-space, in which the photon energy is integrated up to its maximum value $k_\mathrm{max}=m_\pi/2\,(1-m_\ell^2/m_\pi^2)$. If instead the maximum photon energy is to be taken to be $\Delta E$, then the expressions below should be modified by replacing $k_\mathrm{max}$ by $\Delta E$. If partial widths are studied by imposing kinematical cuts on the lepton momenta then the derivation below should be modified accordingly.

In this appendix we simplify the notation in two ways. Firstly since the diagrams are of $O(\alpha_\mathrm{em})$ we can replace $m_\pi^0$ by $m_\pi$ and secondly, since we have shown that the infrared divergences cancel explicitly we replace $E_\gamma$ by $k=|\vecp{k}|$ where $\vec{k}$ is the three-momentum of the photon.

\subsection{$\Gamma_{0A}^{\mathrm{div}}$+$\Gamma_{DD}^{\mathrm{div}}$}\label{subsec:0ADDdiv}
In this section we evaluate $\Gamma_{0A}^{\mathrm{div}}$+$\Gamma_{DD}^{\mathrm{div}}$. For convenience
we rewrite Eqs.\,(\ref{eq:Gamma0Adiv}) and (\ref{eq:GammadivDD}) here:
\begin{eqnarray}\label{eq:Gamma0Adiv2}
\Gamma_{0A}^\mathrm{div}&=&\frac{e^2}{2m_\pi}\left(\frac{G_F^2|V_{ud}|^2}{2}\right)\big|H_0^4\big|^2\,\int\frac{\dthree k}{(2\pi)^3}\,\frac1{2k} \,
\frac{e^{-k t_s}}{(E_\pi(\vecp{k})+k-m_\pi)^2}\,\int d\Phi_2(p_\pi;p_\ell,p_{\nu_\ell})\,\big|L_0^4(p_\ell,p_\nu)\big|^2
\nonumber\\ 
&\equiv&\frac{e^2}{2m_\pi}\left(\frac{G_F^2|V_{ud}|^2}{2}\right)\big|H_0^4\big|^2\,I_{0A}\label{eq:Gamma0Adiv2}\\
\Gamma_{DD}^\mathrm{div}&=&-\frac{e^2}{2m_\pi}\left(\frac{G_F^2|V_{ud}|^2}{2}\right)\big|H_0^4\big|^2
\int\frac{\dthree k}{(2\pi)^3}\,\frac1{2k} \,
\frac{e^{-2k t_s}}{(E_\pi(\vecp{k})+k-m_\pi)^2}
\int d\Phi_2(p_\pi-k_\gamma;p_\ell,p_{\nu_\ell})\,\big|L_0^4(p_\ell,p_\nu)\big|^2\nonumber\\
&\equiv&\frac{e^2}{2m_\pi}\left(\frac{G_F^2|V_{ud}|^2}{2}\right)\big|H_0^4\big|^2\,I_{DD}\,.
\label{eq:GammadivDD2}
\end{eqnarray}
We recall that $E_\pi(\vecp{k})=\sqrt{k^2+m_\pi^2}$ where $k\equiv|\vecp{k}|$ and below we denote the four momentum of the photon by $k_\gamma=(k,\vecp{k})$. Since we will show explicitly that the infrared divergences cancel in $I_{0A}+I_{DD}$ we denote
the energy of the photon by $E_\gamma$ by $k$.

While the cancellation of the infrared divergences in $I_{0A}+I_{DD}$ is manifest, there are a number of sources of residual finite terms, the evaluation of which is the subject of this section: 
\begin{enumerate}
\item[i)] There is a factor of $e^{-k t_s}$ in the integrand of $I_{0A}$ and $e^{-2k t_s}$ in $I_{DD}$.
\item[ii)] The sum over the lepton polarisations $\big|L_0^4(p_\ell,p_\nu)\big|^2$ is different depending on whether $p_\ell+p_\nu=p_\pi$ as in 
$I_{0A}$ or $p_\ell+p_\nu=p_\pi-k_\gamma$ as in $I_{DD}$, where $k_\gamma$ is the four momentum of the photon\,.
\item[iii)] Similarly the leptonic two-body phase space is different in the two cases.
\item[iv)] Finally the integral over $|\vec{k}|$ runs from 0 to $\infty$ in $I_{0A}$ and from 0 to 
$k_\mathrm{max}=m_\pi/2\,(1-m_\ell^2/m_\pi^2)$ in $I_{DD}$.
\end{enumerate}
Our result is written in the form
\begin{equation}\label{eq:I0AplusIDD}
I_{0A}+I_{DD}=F_1+F_2+F_3+F_4\,,
\end{equation}
where the $F_i$ are simple one or two dimensional integrals which can readily be evaluated for any choice of masses and $t_s$. For the reader's convenience we collect all the results here and then proceed to derive Eq:\,(\ref{eq:I0AplusIDD}):
\begin{eqnarray}
F_1&=&\frac{m_\ell^2}{8\pi^3}\left(1-\frac{m_\ell^2}{m_\pi^2}\right)^{\!\!2}
\int_0^\infty k\, dk\,
\frac{e^{-k t_s}-e^{-2k t_s}}{(\sqrt{k^2+m_\pi^2}+k-m_\pi)^2}\,.\label{eq:F1}\\
F_2&=&-\frac{1}{32\pi^3}\int_0^{k_{\mathrm{max}}} \hspace{-10pt}dk
\frac{e^{-2k t_s}}{(\sqrt{k^2+m_\pi^2}+k-m_\pi)^2}\int_{p_\nu^\mathrm{min}}^{p_\nu^\mathrm{max}}\!\! dp_\nu\times\nonumber\\
&&\hspace{0.5in}\int_{-1}^1dz\,f_{DD}(k,p_\nu,z)\,\delta\bigg(z-\frac{m_\pi^2-2m_\pi k-2(m_\pi-k)p_\nu-m_\ell^2}
{2k p_\nu}\bigg)\label{eq:F2}\\ 
F_3&=&\frac{m_\ell^4}{4\pi^3 m_\pi^2}\left(1-\frac{m_\ell^2}{m_\pi^2}\right)
\int_0^{k_\mathrm{max}}\!\!k^2\,dk~
\frac{e^{-2k t_s}}{(\sqrt{k^2+m^2}+k-m_\pi)^2}
\,\frac{1}{m_\pi-2k}\label{eq:F3}\\
F_4&=&\frac{m_\ell^2}{8\pi^3}\left(1-\frac{m_\ell^2}{m_\pi^2}\right)^{\!\!2}\,\int_{k_\mathrm{max}}^\infty k\,dk\,\frac{e^{-2k t_s}}{(\sqrt{k^2+m^2}+k-m_\pi)^2}\,,
\label{eq:F4}
\end{eqnarray} 
where $z$ is the cosine of the angle between $\vec{p}_\nu$ and $\vec{k}$ ($\vec{p}_\nu\cdot\vec{k}=|\vec{p}_\nu| k z$)
\begin{equation}\label{eq:pnurange}
k_\mathrm{max}=\frac{m_\pi}2\left(1-\frac{m_\ell^2}{m_\pi^2}\right)\,,\qquad
p_\nu^\mathrm{min}=\frac{m_\pi^2-2m_\pi k-m_\ell^2}{2m_\pi}\,,\qquad
p_\nu^\mathrm{max}=\frac{m_\pi^2-2m_\pi k-m_\ell^2}{2(m_\pi-2k)}\,.
\end{equation}
and 
\begin{equation}
f_{DD}(k,|\vec{p}_\nu|,z)=4k^2 -\frac{4k(2m_\pi-k)m_\ell^4}{m_\pi^2(m_\pi-k)^2} 
-\frac{4(2m_\ell^2+2\vec{p}_\nu\cdot\vec{k}+k^2)(2\vec{p}_\nu\cdot\vec{k}+k^2)}{(m_\pi-k)^2}\,.
\label{eq:fDD}
\end{equation}
The quantities $k_\mathrm{max}$ and $p_\nu^{\pm}$ are the kinematical limits on the final state photon's energy and $|\vec{p}_\nu|$ respectively. In order to simplify the notation in Eqs.\,(\ref{eq:F2}) and (\ref{eq:fDD}), we have replaced $|\vecp{p}_\nu|$ by $p_\nu$.
In deriving these equations we have used 
\begin{equation}\label{eq:L040A}
\big|L_0^4(p_\ell,p_\nu)\big|^2=4m_\ell^2\left(1-\frac{m_\ell^2}{m_\pi^2}\right)
\end{equation}
when $p_\ell+p_\nu=p_\pi$ as in $I_{0A}$ and
\begin{equation}\label{eq:L04DD}
\big|L_0^4(p_\ell,p_\nu)\big|^2=4m_\ell^2\left(1-\frac{m_\ell^2}{m_\pi^2}\right)+f_{DD}(k,p_\nu,z)
\end{equation}
when $p_\ell+p_\nu=p_\pi-k_\gamma$ as in $\Gamma_{DD}^\mathrm{div}$.
Since $f_{DD}(0,p_\nu,z)=0$, $F_2$ is infrared convergent. 

Another ingredient in the derivation of Eqs.\,(\ref{eq:F1})\,-\,(\ref{eq:F4}) is the integral over the phase space of the leptons. For $I_{0A}$ the integrand is independent of the integration variables (see Eq.\,(\ref{eq:L040A})) and 
\begin{equation}
\int d\Phi_2(p_\pi;p_\ell,p_\nu)=\frac1{8\pi}\,\left(1-\frac{m_\ell^2}{m_\pi^2}\right)\,.
\end{equation}
For $I_{DD}$ the integrand in general does depend on the integration variables (see Eqs.\,(\ref{eq:L04DD}) and (\ref{eq:fDD})) so that 
\begin{equation}\label{eq:IDDphasespace1}
\int d\Phi_2(p_\pi-k_\gamma;p_\ell,p_\nu)\,f_{DD}(k,p_\nu,z)=\frac{1}{8\pi k}\int_{p_\nu^\mathrm{min}}^{p_\nu^\mathrm{max}}\!\! dp_\nu
\int_{-1}^1dz\,f_{DD}(k,p_\nu,z)\,\delta\bigg(z-\frac{m_\pi^2-2m_\pi k-2(m_\pi-k)p_\nu-m_\ell^2}
{2k p_\nu}\bigg)\,.
\end{equation}
When the integrand is independent of the integration variables:
\begin{equation}\label{eq:IDDphasespace2}
\int d\Phi_2(p_\pi-k_\gamma;p_\ell,p_\nu)=\frac1{8\pi}\frac{m_\pi^2-2m_\pi k-m_\ell^2}{m_\pi^2-2m_\pi k}\,.
\end{equation}

We now ekxplain the origin of the $F_i,~(i=1$\,-\,$4)$. $F_1$ arises because it is convenient to have the same factor $e^{-2k t_s}$ in the numerator of the integrands in the infrared divergent terms of both $I_{0A}$ and $I_{DD}$ and we therefore write
\begin{equation}
I_{0A}=\frac{m_\ell^2}{2\pi}\left(1-\frac{m_\ell^2}{m_\pi^2}\right)^{\!\!2}
\int\frac{\dthree k}{(2\pi)^3}\,\frac1{2k} \,
\frac{e^{-2k t_s}}{(E_\pi(\vecp{k})+k-m_\pi)^2}+F_1\,.\label{eq:I0Aaux1}
\end{equation}
Using Eqs.\,(\ref{eq:L04DD}), (\ref{eq:IDDphasespace1}) and (\ref{eq:IDDphasespace2}) we rewrite $I_{DD}$ in the form
\begin{eqnarray}
I_{DD}&=&-\frac{m_\ell^2}{2\pi m_\pi}\left(1-\frac{m_\ell^2}{m_\pi^2}\right)
\int \frac{\dthree k}{(2\pi)^3}\frac{1}{2k}
\frac{e^{-2k t_s}}{(E_\pi(\vecp{k})+k-m_\pi)^2}
\,\frac{m_\pi^2-2m_\pi k-m_\ell^2}{m_\pi-2k}+F_2\\ 
&=&-\frac{m_\ell^2}{2\pi}\left(1-\frac{m_\ell^2}{m_\pi^2}\right)^{\!\!2}
\int \frac{\dthree k}{(2\pi)^3}\frac{1}{2k}
\frac{e^{-2k t_s}}{(E_\pi(\vecp{k})+k-m_\pi)^2}+F_2+F_3\,.\label{eq:IDDaux1}
\end{eqnarray}

Finally, we recall that the range of the $k=|\vecp{k}|$ integration in $I_{0A}$ is $(0,\infty)$ and in $I_{DD}$ it is $(0,k_\mathrm{max})$ so whilst the integrands in the first terms on the right-hand sides of Eqs.\,(\ref{eq:I0Aaux1}) and (\ref{eq:IDDaux1}) are equal and opposite, the integrals do not cancel exactly and the sum of the two integrals is $F_4$. 

We have therefore shown that 
\begin{equation}
\Gamma_{0A}^\mathrm{div}+\Gamma_{DD}^\mathrm{div}=\frac{e^2}{2m_\pi}\left(\frac{G_F^2|V_{ud}|^2}{2}\right)\big|H_0^4\big|^2\,\big(
F_1+F_2+F_3+F_4)\,,
\end{equation}
where the $F_i$ are simple finite one or two-dimensional integrals which can readily be evaluated numerically for any choice of masses and $t_s$.

\subsection{$\Gamma_{0B}^{\mathrm{div}}$+$\Gamma_{DE}^{\mathrm{div}}$}\label{subsec:0BDEdiv}
We now repeat the evaluation of the finite-terms remaining after the cancellation of infrared divergences in $\Gamma_{0B}^{\mathrm{div}}$+$\Gamma_{DE}^{\mathrm{div}}$. Again the cancellation of infrared divergences is manifest, but there are a number of finite terms remaining which are the subject of this section.

We start be rewriting the integral expressions for these two terms, i.e. Eqs.(\ref{eq:Gammadiv0B}) and (\ref{eq:GammadivDE})
\begin{eqnarray}
\Gamma_{0B}^{\mathrm{div}}&=&\frac{e^2G_F^2}{2m_\pi}\,|V_{ud}|^2\,\big|H_0^4\big|^2\,\int d\Phi_2(p_\ell,p_{\nu_\ell})L_0^{4\,\dagger}\,\times\nonumber\\
&&\int\frac{\dthree k}{(2\pi)^3}\,
\frac{e^{-k t_s}}{2k}
\left\{\frac{\bar{u}(p_{\nu_\ell})\gamma^4(1-\gamma^5)(-\Elpk\gamma_4+i(\vec{p}_\ell+\vec{k})\cdot\vec{\gamma}+m_\ell)\gamma^4 v(p_\ell)}{2\Elpk(E_\ell^\prime(\vec{k})+k-E_\ell)(E_\pi(\vec{k})+k-m_\pi)}\right\}\nonumber\\
&\equiv&\frac{e^2G_F^2}{2m_\pi}\,|V_{ud}|^2\,\big|H_0^4\big|^2\,I_{0B}\,\label{eq:Gamma0B2}
\end{eqnarray}
and
\begin{eqnarray}
\Gamma_{DE}^{\mathrm{div}}&=&\frac{e^2G_F^2}{2m_\pi}\big|H_0^4\big|^2\,|V_{ud}|^2
\,\int d\Phi_3(p_\ell,p_{\nu_\ell},k)\,L_0^{4\,\dagger}\,e^{-k t_s}\times\nonumber\\
&&\bar{u}(p_{\nu_\ell})\gamma^{4}(1-\gamma^5)\bigg\{\frac{\big(-\Elpk\gamma^4+i(\vec{p}_\ell+\vec{k})\cdot\vec{\gamma}+m_\ell\big)}{2\Elpk\,
(E_\ell^\prime(\vec{k}\,)-k-E_\ell)\,(\Epik+k-m_\pi)}\bigg\}\gamma^{4} v(p_\ell)\nonumber\\
&\equiv&\frac{e^2G_F^2}{2m_\pi}\,\big|H_0^4\big|^2\,|V_{ud}|^2\,I_{DE}\,.
\end{eqnarray}
We recall that $E_\ell=\sqrt{\vecpsq{p}_\ell+m_\ell^2}$, 
$E_\ell^\prime(\vecp{k})=\sqrt{(\vec{p}_\ell+\vecp{k})^2+m_\ell^2}$, $k=|\vecp{k}|$ and 
$E_\pi(\vecp{k})=\sqrt{k^2+m_\pi^2}$.
In Eq.\,(\ref{eq:Gamma0B2}) we have changed variables $\vec{k}\to -\vec{k}$ so that the lepton trace, written in terms of the momenta and energies, is the same in both cases. We write the lepton trace as $L_{0B}=L_{DE}=L_{0B}^\mathrm{div}+L_{0B}^\mathrm{con}$ where 
\begin{eqnarray}
L_{0B}^\mathrm{div}=L_{DE}^\mathrm{div}&=&-16E_\nu E_\ell^2-16E_\ell\, \vec{p}_\ell\cdot \vec{p}_\nu\\ 
L_{0B}^\mathrm{con}=L_{DE}^\mathrm{con}&=&-8(\Delta E_\ell(\vecp{k})) E_\ell E_\nu -8(\Delta E_\ell(\vecp{k}))\vec{p}_\ell\cdot\vec{p}_\nu - 8 E_\ell (\vec{p}_\nu\cdot\vec{k})
-8E_\nu(\vec{p}_\ell\cdot\vec{k})\,,
\end{eqnarray}
where $\Delta E_\ell(\vecp{k})=E_\ell^\prime(\vecp{k})-E_\ell$.
At small photon momenta $L_{0B}^\mathrm{con}=O(k)$ and there is then no infrared divergence. The integrals can readily be performed numerically as we explain towards the end of this section.

We now start by considering the contributions from the divergent terms. For the two-body decay for which $\Gamma_{0B}$ contributes to the width, $E_\ell$ and $|\vecp{p}_\ell|$ are fixed, $E_\ell=(m_\pi^2+m_\ell^2)/2m_\pi$ and $|\vecp{p}_\ell|=(m_\pi^2-m_\ell^2)/2m_\pi$ and 
\begin{eqnarray}
I_{0B}^\mathrm{div}&\equiv&\frac1{16\pi}\left(1-\frac{m_\ell^2}{m_\pi^2}\right)\int\frac{\dthree k}{(2\pi)^3}\,
\frac{e^{-k t_s}}{2k}\,\frac{1}{(E_\pi(\vec{k})+k-m_\pi)}\,\int_{-1}^1 dz_\ell~
\frac{-16E_\nu E_\ell^2-16E_\ell\, \vec{p}_\ell\cdot \vec{p}_\nu}{2\Elpk(E_\ell^\prime(\vec{k})+k-E_\ell)}\nonumber\\
&=&-\frac{m_\ell^2}{4\pi}\left(1-\frac{m_\ell^2}{m_\pi^2}\right)^{\!\!\!2}\int\frac{\dthree k}{(2\pi)^3}\,\frac{e^{-k t_s}}{2k}\,\frac{1}{(E_\pi(\vec{k})+k-m_\pi)}\,
\int_{-1}^1 dz_\ell~\frac{E_\ell}{\Elpk(E_\ell^\prime(\vec{k})+k-E_\ell)}\nonumber\\
&=&-\frac{m_\ell^2}{4\pi}\left(1-\frac{m_\ell^2}{m_\pi^2}\right)^{\!\!\!2}\int\frac{\dthree k}{(2\pi)^3}\,\frac{e^{-k t_s}}{2k}\,\frac{1}{(E_\pi(\vec{k})+k-m_\pi)}\,
\int_{-1}^1 dz_\ell~\frac{1}{(E_\ell^\prime(\vec{k})+k-E_\ell)}+F_{1;0B}\nonumber\\
&=&-\frac{m_\ell^2}{4\pi}\left(1-\frac{m_\ell^2}{m_\pi^2}\right)^{\!\!\!2}\int\frac{\dthree k}{(2\pi)^3}\,\frac{e^{-k t_s}}{2k}\,\frac{1}{(E_\pi(\vec{k})+k-m_\pi)}\,
\int_{-1}^1 dz_\ell~\frac{1}{k+\frac{\vec{p}_\ell\cdot\vec{k}}{E_\ell}}+F_{1;0B}+F_{2;0B}\,,\label{eq:I0B3}
\end{eqnarray}
where $z_\ell$ is the cosine of the angle between $\vec{k}$ and $\vec{p}_\ell$ so that $\vec{p}_\ell\cdot\vec{k}=|\vec{p}_\ell| k z_\ell$ and
\begin{eqnarray}
F_{1;0B}&=&\frac{m_\ell^2}{4\pi}\left(1-\frac{m_\ell^2}{m_\pi^2}\right)^{\!\!\!2}\int\frac{\dthree k}{(2\pi)^3}\,\frac{e^{-k t_s}}{2k}\,\frac{1}{(E_\pi(\vec{k})+k-m_\pi)}\,
\int_{-1}^1 dz_\ell~\frac{\Delta E_\ell(\vecp{k})}{\Elpk(E_\ell^\prime(\vec{k})+k-E_\ell)}\\
F_{2;0B}&=&-\frac{m_\ell^2}{4\pi}\left(1-\frac{m_\ell^2}{m_\pi^2}\right)^{\!\!\!2}\int\frac{\dthree k}{(2\pi)^3}\,\frac{e^{-k t_s}}{2k}\,\frac{1}{(E_\pi(\vec{k})+k-m_\pi)}\,
\int_{-1}^1 dz_\ell~
\frac{\frac{\vec{p}_\ell\cdot\vec{k}}{E_\ell}-\Delta E_\ell(\vecp{k})}{(\Delta E_\ell(\vecp{k})+k)\left(k+\frac{\vec{p}_\ell\cdot\vec{k}}{E_\ell}\right)}
\end{eqnarray}
are finite integrals which can readily be evaluated numerically. 

In the first term on the right-hand side of Eq.\,(\ref{eq:I0B3}) the $z_\ell$ integration can be performed to give
\begin{equation}
\int_{-1}^1 dz_\ell~\frac{1}{k+\frac{\vec{p}_\ell\cdot\vec{k}}{E_\ell}}=\frac{E_\ell}{p_\ell k}\,\log\frac{E_\ell+p_\ell}{E_\ell-p_\ell}=\frac{1}{k}\frac{m_\pi^2+m_\ell^2}{m_\pi^2-m_\ell^2}
\,\log\frac{m_\pi^2}{m_\ell^2}\,,
\end{equation}
where $p_\ell=|\vec{p}_\ell|$, so that 
\begin{eqnarray}
I_{0B}^\mathrm{div}=-
\frac{m_\ell^2}{16\pi^3}\left(1-\frac{m_\ell^2}{m_\pi^2}\right)\left(1+\frac{m_\ell^2}{m_\pi^2}\right)\,\log\frac{m_\pi^2}{m_\ell^2}\int_0^\infty 
\hspace{-0.05in}dk~ \frac{e^{-k t_s}}{E_\pi(\vec{k})+k-m_\pi}\,
+F_{1;0B}+F_{2;0B}\,.\label{eq:I0B5}
\end{eqnarray}
The two finite terms, $F_{1;0B}$ and $F_{2;0B}$ are simple two-dimensional integrals (over $k$ and $z_\ell$) which can readily be evaluated numerically for any values of the masses and $t_s$. The two-body phase-space integral of a general function $f(k,p_\ell,z_\ell)$, where $p_\ell=|\vec{p}_\ell|$ can be reduced to 
\begin{equation}
\int d\Phi_2(p_\pi;p_\ell,p_{\nu_\ell})f(k,p_\ell,z_\ell)=
\frac1{16\pi}\left(1-\frac{m_\ell^2}{m_\pi^2}\right)
\int_{-1}^{1}dz_\ell\,f(k,p_\ell,z_\ell)\,.
\end{equation}
The finite contributions corresponding to $L_{0B}^\mathrm{con}$ in the numerator are evaluated similarly.

We now consider $I_{DE}^\mathrm{div}$. Following the corresponding steps to those in Eq.\,(\ref{eq:I0B3}) we have
\begin{eqnarray}
I_{DE}^\mathrm{div}&=&\int\frac{\dthree k}{(2\pi)^3}\,\frac{e^{-k t_s}}{2k}\int d\Phi_2(p_\pi-k;p_\ell,p_{\nu_\ell})
\frac{-16E_\nu E_\ell^2-16E_\ell \,\vec{p}_\nu\cdot \vec{p}_\ell}{2E_\ell^\prime(\vecp{k})(E_\ell^\prime(\vec{k})-k-E_\ell)(E_\pi(\vec{k})+k-m_\pi)}\nonumber\\
&&\hspace{-0.75in}=-4m_\ell^2\left(1-\frac{m_\ell^2}{m_\pi^2}\right)\int\frac{\dthree k}{(2\pi)^3}\,\frac{e^{-k t_s}}{2k(E_\pi(\vec{k})+k-m_\pi)}\int d\Phi_2(p_\pi-k;p_\ell,p_{\nu_\ell})
\frac{E_\ell}{E_\ell^\prime(\vecp{k})(E_\ell^\prime(\vec{k})-k-E_\ell)}
+F_{1;DE}\nonumber\\
&&\hspace{-0.75in}=-4m_\ell^2\left(1-\frac{m_\ell^2}{m_\pi^2}\right)\int\frac{\dthree k}{(2\pi)^3}\,\frac{e^{-k t_s}}{2k(E_\pi(\vec{k})+k-m_\pi)}\int d\Phi_2(p_\pi-k;p_\ell,p_{\nu_\ell})
\frac{1}{(E_\ell^\prime(\vec{k})-k-E_\ell)}
+F_{1;DE}+F_{2;DE}\nonumber\\
&&\hspace{-0.75in}=4m_\ell^2\left(1-\frac{m_\ell^2}{m_\pi^2}\right)\int\frac{\dthree k}{(2\pi)^3}\,\frac{e^{-k t_s}}
{2k(E_\pi(\vec{k})+k-m_\pi)}\int d\Phi_2(p_\pi-k;p_\ell,p_{\nu_\ell})\frac1{k-\frac{\vec{p}_\ell\cdot\vec{k}}{E_\ell}}+F_{1;DE}+F_{2;DE}+F_{3;DE}\,,\label{eq:IDEaux3}
\end{eqnarray}
where $F_{1;DE}$, $F_{2;DE}$ and $F_{3;DE}$ are infrared finite:
\begin{eqnarray}
F_{1;DE}&=&\int\frac{\dthree k}{(2\pi)^3}\,\frac{e^{-k t_s}}{2k(E_\pi(\vec{k})+k-m_\pi)}\int d\Phi_2(p_\pi-k;p_\ell,p_{\nu_\ell})\times
\nonumber\\
&&\hspace{1in}\frac{8E_\ell\left(\left(1-\frac{m_\ell^2}{m_\pi^2}\right)m_\ell^2-4(m_\pi-k-E_\ell)E_\ell+m_\pi(m_\pi-2k) -m_\ell^2\right)}
{2E_\ell^\prime(\vecp{k})(E_\ell^\prime(\vec{k})-k-E_\ell)}
\nonumber\\
F_{2;DE}&=&4m_\ell^2\left(1-\frac{m_\ell^2}{m_\pi^2}\right)\int\frac{\dthree k}{(2\pi)^3}\,\frac{e^{-k t_s}}
{2k(E_\pi(\vec{k})+k-m_\pi)}\int d\Phi_2(p_\pi-k;p_\ell,p_{\nu_\ell})
\frac{\Delta E_\ell(\vecp{k})}{E_\ell^\prime(\vecp{k})(E_\ell^\prime(\vec{k})-k-E_\ell)}
\nonumber\\ 
F_{3;DE}&=&-4m_\ell^2\left(1-\frac{m_\ell^2}{m_\pi^2}\right)\int\frac{\dthree k}{(2\pi)^3}\,\frac{e^{-k t_s}}
{2k(E_\pi(\vec{k})+k-m_\pi)}\int d\Phi_2(p_\pi-k;p_\ell,p_{\nu_\ell})\frac{\Delta E_\ell(\vecp{k})-\frac{\vec{p}_\ell\cdot\vec{k}}
{E_\ell(\vec{k})}}
{(k-\frac{\vec{p}_\ell\cdot\vec{k}}{E_\ell})(\Delta E_\ell-k)}\,.\label{eq:F123DE}
\end{eqnarray}

The infrared divergence is contained in the first term on the right-hand side of Eq.\,(\ref{eq:IDEaux3}) and we now evaluate
\begin{equation}
\int d\Phi_2(p_\pi-k;p_\ell,p_{\nu_\ell})\frac{E_\ell}{E_\ell k-\vec{p}_\ell\cdot\vec{k}}=-
\frac{m_\pi^2+m_\ell^2}{2m_\pi}\int d\Phi_2(p_\pi-k;p_\ell,p_{\nu_\ell})\frac{1}{p_\ell\cdot k}
+\int d\Phi_2(p_\pi-k;p_\ell,p_{\nu_\ell})\frac{E_\ell-\frac{m_\pi^2+m_\ell^2}{2m_\pi}}{E_\ell k-\vec{p}_\ell\cdot\vec{k}}\,.\label{eq:IDEaux4}
\end{equation}
The second term on the right-hand side of Eq.\,(\ref{eq:IDEaux4}) is infrared convergent and we now focus on the first term. The integrand is Lorentz invariant and so we can evaluate the integral in the rest-frame of the lepton system
\begin{eqnarray}
-\frac{m_\pi^2+m_\ell^2}{2m_\pi}\int d\Phi_2(p_\pi-k;p_\ell,p_{\nu_\ell})\frac{1}{p_\ell\cdot k}
&=&\frac1{8\pi m_\pi k}\frac{m_\pi^2+m_\ell^2}{2m_\pi}
\,\log\frac{E_\ell^\ast+p_\ell^\ast}{E_\ell^\ast-p_\ell^\ast}
\end{eqnarray}
where $E_\ell^\ast, p_\ell^\ast$ are the variables in the lepton rest frame:
\begin{equation}
E_\ell^\ast=\frac{m_\pi^2+m_\ell^2-2m_\pi k}{2\sqrt{m_\pi^2-2m_\pi k}}\,,\qquad
p_\ell^\ast=\frac{m_\pi^2-m_\ell^2-2m_\pi k}{2\sqrt{m_\pi^2-2m_\pi k}}\,.
\end{equation}
Thus we can write
\begin{equation}
I_{DE}^\mathrm{div}=\frac{m_\ell^2}{16\pi^3}\left(1+\frac{m_\ell^2}{m_\pi^2}\right)
\left(1-\frac{m_\ell^2}{m_\pi^2}\right)\log\frac{m_\pi^2}{m_\ell^2}\int_0^{k_\mathrm{max}} \hspace{-0.1in}dk~\frac{e^{-k t_s}}
{E_\pi(\vec{k})+k-m_\pi}+F_{DE}\,,
\end{equation}
where $k_\mathrm{max}=m_\pi/2(1-m_\ell^2/m_\pi^2)$ is the maximum value of $k$ in the three-body decay and the finite term $F_{DE}$ is given by
\begin{equation}
F_{DE}=\sum_{i=1}^5F_{i;DE}\,,
\end{equation}
$F_{1;DE},\,F_{2;DE}$ and $F_{3;DE}$ are given in Eqs.\,(\ref{eq:F123DE}) and 
\begin{eqnarray}
F_{4;DE}&=&
4m_\ell^2\left(1-\frac{m_\ell^2}{m_\pi^2}\right)\int\frac{\dthree k}{(2\pi)^3}\,\frac{e^{-k t_s}}
{2k(E_\pi(\vec{k})+k-m_\pi)}\int d\Phi_2(p_\pi-k;p_\ell,p_{\nu_\ell})
\frac{E_\ell-\frac{m_\pi^2+m_\ell^2}{2m_\pi}}{E_\ell k-\vec{p}_\ell\cdot\vec{k}}
\nonumber\\
F_{5;DE}&=&\frac{m_\ell^2}{8\pi}\left(1-\frac{m_\ell^2}{m_\pi^2}\right)\!\!
\left(1+\frac{m_\ell^2}{m_\pi^2}\right)\!\!
\int\frac{\dthree k}{(2\pi)^3}\,\frac{e^{-k t_s}}
{k^2(E_\pi(\vec{k})+k-m_\pi)}
\,\log\frac{m_\pi(m_\pi-2k)}{m_\pi^2}
\,.\label{eq:F45DE}
\end{eqnarray}

Thus finally we have 
\begin{equation}
I_{0B}^\mathrm{div}+I_{DE}^\mathrm{div}=
-\frac{m_\ell^2}{16\pi^3}\left(1-\frac{m_\ell^2}{m_\pi^2}\right)\left(1+\frac{m_\ell^2}{m_\pi^2}\right)\,\log\frac{m_\pi^2}{m_\ell^2}\int_{k_\mathrm{max}}^\infty \hspace{-5pt}dk\,\frac{e^{-k t_s}}{E_\pi(\vec{k})+k-m_\pi}\,
+F_{1;0B}+F_{2;0B}+F_{DE}\,.\label{eq:I0BplusIDEdiv}
\end{equation}
All the terms on the right-hand side of Eq.\,(\ref{eq:I0BplusIDEdiv}) are infrared finite.

The finite terms $F_{i;DE}$ can also be readily evaluated numerically for any values of the masses and $t_s$. $F_{5;DE}$ is a one-dimensional integral whereas $F_{i;DE}$,
$i=(1\,$-$\,4)$, are two-dimensional integrals. In evaluating these it is
natural to use the mass-shell condition for the neutrino to determine $z_\ell$, the cosine of the angle between $\vec{p}$ and $\vec{k}$:
\begin{equation}\label{eq:zlexpression}
z_\ell=\frac{m_\pi^2+m_\ell^2-2m_\pi k-2E_\ell(m_\pi-k)}{2p_\ell k}\,,
\end{equation}where $p_\ell=|\vec{p}_\ell|$.
The range of integration over $p_\ell=|\vec{p}_\ell|$, or equivalently $E_\ell$, can then be determined from Eq.\,(\ref{eq:pnurange}):
\begin{eqnarray}
E_\ell\le E_\ell^\mathrm{max}=m_\pi-k-p_\nu^\mathrm{min}=\frac{m_\pi^2+m_\ell^2}{2m_\pi}\qquad&\mathrm{and}&\qquad
p_\ell^\mathrm{max}=\frac{m_\pi^2-m_\ell^2}{2m_\pi}\\
E_\ell\ge E_\ell^\mathrm{min}=m_\pi-k-p_\nu^\mathrm{max}=\frac{(m_\pi-2k)^2+m_\ell^2}{2(m_\pi-2k)}\qquad&\mathrm{and}&\qquad
p_\ell^\mathrm{min}=\frac{(m_\pi-2k)^2-m_\ell^2}{2(m_\pi-2k)}
\end{eqnarray}
so that for a general function $f(k,p_\ell,z_\ell)$
\begin{eqnarray}
\int d\Phi_2(p_\pi-k;p_\ell,p_{\nu_\ell})f(k,p_\ell,z_\ell)&=&
\frac1{8\pi k}\int^{p_\ell^\mathrm{max}}_{p_\ell^\mathrm{min}}
\frac{p_\ell\,dp_\ell}{E_\ell}\,f(k,p_\ell,z_\ell)\nonumber\\
&=&\frac1{8\pi k}\int^{E_\ell^\mathrm{max}}_{E_\ell^\mathrm{min}}
dE_\ell\,f(k,p_\ell,z_\ell)\,,\label{eq:pellrange}
\end{eqnarray}
where $z_\ell$ is given in terms of the integration variable $p_\ell$ by Eq.\,(\ref{eq:zlexpression}).
Note that the factor of $1/k$ in front of the integrals in Eq.\,(\ref{eq:pellrange}) is compensated at small $k$ by the ranges of integration being of $O(k)$.

All the finite terms contributing to $\Gamma_{DE}$ listed above (with the exception of $F_{5;DE}$ which is a simple one-dimensional integral), including those with 
$L_{DE}^\mathrm{con}$ in the numerator, can readily be evaluated using Eq,\,(\ref{eq:pellrange})
for any specified values of the masses and $t_s$.

\section{Electromagnetic corrections to $\mathbf{K_{\ell3}}$ decays}\label{sec:Kl3}

In the preceding sections of this paper we have developed a method using infinite-volume reconstruction to calculate the radiative corrections to leptonic decays, such as $\pi_{\ell2}$ and $K_{\ell2}$, that promises greater precision than approaches in which the amplitude is fully computed in a finite volume, such as that based on the QED$_\mathrm{L}$ treatment of electromagnetism.
This refinement replaces power-law finite-volume corrections with corrections which are exponentially suppressed in the linear size of the finite volume.  The additional analytic control provided by the IVR method also allows the analytic cancellation of infrared divergences so that all expressions which are evaluated numerically require no infrared regulator.  

In this appendix we generalize this approach to treat the electromagnetic corrections to the $K_{\ell3}$ decay, a process where there is no alternative approach currently known that permits a first-principles lattice calculation.  The fundamental diagrams for the $K_{\ell3}$ process are similar to those in Fig.\,\ref{fig:diagrams} except that the initial meson is a kaon and a pion emerges from the hadronic weak vertex.  In this case diagrams A and B in Fig.\,\ref{fig:diagrams}, where the photon is attached to one or two quark lines, will now include the case where one or more of these electromagnetic vertices appear on quarks of the final-state meson. We show in Fig.\,\ref{fig:Coulomb} an example of such a new type-B diagram in which the photon propagator connects the lepton with a quark appearing in the final-state pion.

\begin{figure}[t]
\centering
\includegraphics[width=0.45\textwidth]{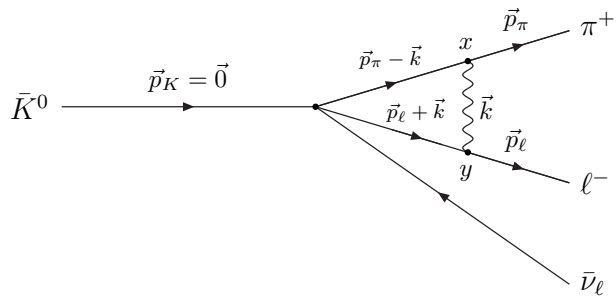}
\caption{A new representative of a type B diagram identified in Fig.\,\ref{fig:diagrams}  that can appear when computing the electromagnetic corrections to $K_{\ell 3}$ decay.  Here the three-momenta carried by the propagators have been labeled in a fashion consistent with Eq.\,\eqref{eq:IVR2A} below.}
\label{fig:Coulomb}
\end{figure}

Calculating the electromagnetic corrections to $K_{\ell3}$ decays involves two new difficulties not present in leptonic decays, such as $K_{\ell 2}$\,\cite{Sachrajda:2019uhh,Sachrajda:2021enz}.
Both difficulties are associated with the exchange of a photon between the two charged final-state particles, the pion and the lepton as shown in Fig.\,\ref{fig:Coulomb}~\footnote{The difficulties are also present in the semileptonic decay of a charged kaon, e.g. $K^-\to\pi^0\ell^-\bar{\nu}_\ell$. In this case however, the imaginary part of the amplitude in Minkowski space is not infrared divergent.}.  
In the final state, these two particles are non-interacting and each carry the three momentum that is determined by the lattice interpolating operator which annihilates them. Thus, the total energy carried by the pion and lepton, $E_{\pi\ell}$, is also determined. However, the intermediate state can also consist of a pion and a lepton, where the individual particles have spatial momenta that are different from those of the final-state particles.  This is illustrated in Fig.\,\ref{fig:Coulomb}, by the pion with momentum $p-k$ and the lepton with momentum $l+k$.
The $\pi$-$\ell$ intermediate state can therefore have an energy $E_{\pi\ell}^\prime$ which is lower than $E_{\pi\ell}$. 
This creates a familiar difficulty in a Euclidean-space lattice calculation since such a lower-energy pion-lepton intermediate state will result in an exponentially growing term in the $\bra{0}\phi_\pi(t)\phi_\ell(t) J^\mu_\mathrm{em}(t_x) J^\nu_\mathrm{em}(t_y) H_W(0) \phi_K^\dagger(t_K)\ket{0}$ correlation function, where $\phi_K^\dagger$, $\phi_{\pi}$ and $\phi_\ell$
are interpolating operators used to create or annihilate the corresponding particles and we have only exhibited the time-dependences.
In this case the lower-energy pion-lepton intermediate state will be favored because of its less-rapid exponentially-falling Euclidean-space time dependence, leading to an exponentially growing relative factor proportional to $e^{(E_{\pi\ell}-E_{\pi\ell}^\prime)t}$.  In a conventional Euclidean-space lattice calculation such an unphysical term must be carefully identified and subtracted.  

In the infinite-volume integration over the photon's momentum $k$, as $E_{\pi\ell}^\prime$ approaches $E_{\pi\ell}$ a singular energy denominator appears.  Applying the Feynman prescription of introducing the usual $-i\epsilon$ in the denominators of the lepton and photon propagators results in a complex amplitude.  The real part of this amplitude is obtained by a principal-part recipe while the imaginary part comes from a delta function, giving a result dictated by the standard optical theorem.  A finite-volume Euclidean calculation would miss this imaginary contribution and the approximation of the principal part by a discrete sum would introduce potentially large finite-volume corrections\,\cite{Christ:2015pwa}.  

These difficulties arise from the space-time region in which the photon is exchanged between the pion and the lepton at increasingly late times in the decay, i.e. with $t_x$ and $t_y$ close to $t$.  This is precisely the region that can be treated analytically using IVR.  In fact, using IVR a lattice QCD calculation can treat the final state pion directly in Minkowski space avoiding the difficulties described above.   If we assume that other possible $X\ell$ intermediate states with energy $E_{X\ell}^\prime$ smaller than $E_{\pi\ell}$ are unimportant, then this provides a complete treatment of the radiative corrections to $K_{\ell3}$ decays.  Here the most important hadronic state $X$ is the two pion state whose lack of significance is suggested by the ratio of partial widths: $\Gamma(K_L\to\pi^+\pi^0\ell^-\overline{\nu}_\ell)/\Gamma(K_L\to\pi^+\ell^-\overline{\nu}_\ell) \approx 10^{-4}$.  In addition, if desired such two-pion intermediate states can be further suppressed or avoided altogether by considering $K_{\ell3}$ decays in the kinematic region in which the neutrino carries substantial energy. 

A discussion of the radiative corrections to $K_{\ell3}$ decays which is as detailed as that presented here for leptonic decays is beyond the scope of this paper.  However, given the absence of other lattice approaches to the calculation of these corrections and its value as a further example of the methods developed in this paper, we present a broad outline of this approach in this appendix.   The critical step is the use of IVR to determine the contribution of an intermediate pion carrying a known spatial momentum.  We begin with the relevant hadronic matrix element expressed as a sum over intermediate states:
\begin{eqnarray}
\langle \pi(\vec p_\pi)|J_\mu(x) J_\nu^W(0)|K(\vecp 0)\rangle_E
                                  &=& \sum_n \langle \pi(\vec p_\pi)|J^\mu_\mathrm{em}(x)|n\rangle\langle n|J^\nu_W(0)|K(\vecp 0)\rangle_E
                                  \label{eq:IVR1a} \\
                             &\simeq& \int \frac{d^3 p^{\,\prime}}{2E_\pi(\vec p\,^\prime)} e^{-x_4\bigl(E_\pi(\vec p\,^\prime) -E_\pi\bigr)} \langle \pi(\vec{p}_\pi)|J^\mu_\mathrm{em}(\vec{x},0)|\pi(\vec p\,')\rangle\langle \pi(\vec p\,')|J^\nu_W(0)|K(\vecp 0)\rangle,
 \label{eq:IVR1b}
\end{eqnarray}
where $x_4>0$ and the subscript $E$ has been introduced when necessary to indicate a Euclidean-space amplitude.   For a generic momentum $\vec{q}$ we define $E_\pi(\vec{q})=\sqrt{\vecp{q}^2+m_\pi^2}$, and in order to simplify the notation we define $E_\pi$ to be the energy of the external pion, $E_\pi=\sqrt{\vec{p}^2+m_\pi^2}$. As has already been extensively discussed, by taking the Euclidean time $x_4$ to be sufficiently large we can insure that only the intermediate pion state contributes as described by Eq.\,\eqref{eq:IVR1b}.

Following the now familiar steps taken earlier, we can Fourier transform Eq.~\eqref{eq:IVR1b} to determine the pion contribution $\mathcal{A}_\pi(\vec p,\vec p\,',x_0)$ to this amplitude at an arbitrary time $x_0$ in Minkowski space from our Euclidean lattice result:.
\begin{eqnarray}
\mathcal{A}^{\mu\nu}_\pi(\vec p_\pi,\,\vecp{p}^{\hspace{0.5pt}\prime}\!,x_0) &\equiv& \langle \pi(\vec{p}_\pi)|J^\mu_\mathrm{em}(0)|\pi(\vec p\,')\rangle
                                               \langle \pi(\vec p\,')|J^\nu_W(0)|K(\vecp 0)\rangle  e^{-ix_0(E_\pi(\vec p\,') -E_\pi)}
\label{eq:A_pi}\\
&=& h^{\mu\rho} h^{\nu\sigma}
\int d^3 x~ e^{i(\vec p -\vec p\,')\cdot \vec x} \,e^{(t_s -ix_0)(E_\pi(\vec p\,') -E_\pi)}
                                                             \langle \pi(\vecp p)|J^\rho_\mathrm{em}(\vec x, t_s) J^\sigma_W(0)|K(\vecp 0)\rangle_E\,,\label{eq:A_pi2}
\end{eqnarray}
provided the real Euclidean time $t_s$ is sufficiently large and positive.  Here $h=\mathrm{diag}(1,i,i,i)$ is introduced to take into account that the currents in Eq.\,(\ref{eq:A_pi}) are defined in Minkowski space whereas those in Eq.\,(\ref{eq:A_pi2}) are defined in Euclidean space using the conventions in Appendix\,\ref{sec:conventions}.
We have also replaced the variable $x_4$ by $t_s$ to follow more closely the conventions used earlier.  Thus, the Minkowski-space amplitude $\mathcal{A}^{\mu\nu}_\pi(\vec p_\pi,\,\vecp{p}^{\hspace{0.5pt}\prime}\!,x_0)$ 
can be determined directly from a Euclidean lattice calculation.   All finite-volume errors will remain exponentially small in the size of the spatial volume provided we keep $t_s \lesssim L$.

We can now use the amplitude $\mathcal{A}^{\mu\nu}_\pi(\vec p,\,\vecp{p}^{\hspace{0.5pt}\prime}\!,x_0)$ to avoid both of the difficulties described above that are involved in the calculation of the radiative correction to $K_{\ell3}$ decays.  Firstly, the amplitude $\mathcal{A}^{\mu\nu}_\pi(\vec p,\,\vecp{p}^{\hspace{0.5pt}\prime}\!,x_0)$ can be substituted directly into the Minkowski-space calculation of the contribution of the $\pi\ell$ intermediate state to the $K_{\ell3}$ decay.  There will be no terms with exponentially growing time dependence since the calculation is performed in Minkowski space and the unwanted term that oscillates 
at large times can be isolated in this analytic calculation and dropped as was done in Ref.\,\cite{Christ:2020hwe}.  The resulting complex amplitude will obey the optical theorem~\footnote{Note, this result will contain a physical infrared divergence that results from the logarithmic radial dependence of the Coulomb wave functions.  This divergence can be regulated by adding a photon mass and removed by including screening effects or evaluating a ratio in which these effects cancel. This divergence contributes to the imaginary part of the amplitude and hence does not enter the $O(\alpha_\mathrm{em})$ correction to the decay rate being considered here.}.

An explicit expression for this Minkowski-space amplitude coming from a single pion intermediate state can be readily written down directly in terms of the underlying lattice QCD amplitude:
\begin{eqnarray}
  \int_{-\infty}^\infty dk_0 \int \dthree{k}\, e^{t_s(E_\pi(\vec{p}_\pi-\vecp{k}) -E_\pi)} \int \dthree{x}~ e^{i\vec x \cdot \vec k}
                                             ~  \frac{h^{\mu\rho}h^{\nu\sigma}\langle \pi(\vec p_\pi)|J^\rho_\mathrm{em}(\vec x, t_s) J^\sigma_W(0)|K(\vecp{0})\rangle_E}
                                                              {E_\pi - E_\pi(\vec p_\pi -\vecp{k})-k_0 +i\epsilon}\,\times &&
\nonumber \\
   && \hspace{-2.8in} \frac{1}{k^2-i\epsilon}\frac{\bar{u}_\ell(\vec{p}_\ell) \gamma_\mu(\gamma\cdot(p_\ell+k) +m_\ell)\gamma_\nu(1-\gamma^5)v_{\overline{\nu}}(\vec p_{\bar{\nu}})}{(p_\ell+k)^2 +m_\ell^2 -i\epsilon}\,. \label{eq:IVR2A}
\end{eqnarray} 
The four-vector $k$ is the Minkowski-space momentum carried by the photon propagator, $p_\ell$ is the four-momentum of the final-state lepton and $p_{\overline{\nu}}$ the four-momentum of the final-state anti-neutrino.  The routing of momenta adopted in Eq.\,\eqref{eq:IVR2A} is shown in Fig.\,\ref{fig:Coulomb}.  This expression is independent of the parameter $t_s$ when $t_s$ is sufficiently large that intermediate states more massive than the pion can be neglected.  

As in our earlier derivations, the analytic integrals over $k_0$ and $\vec k$ can be performed at fixed $\vec x$ allowing the quantity in Eq.\,\eqref{eq:IVR2A} to be expressed as the product of a Euclidean-space, finite-volume lattice amplitude and an analytic kernel which in this case has both real and imaginary parts.  Equation\,\eqref{eq:IVR2A} isolates the contribution of the pion intermediate state inserted between the currents $J^\mu_\mathrm{em}(\vec x, x_4) J^\nu_W(0)$ for the case that $x_4>0$.  We should recognize that all features of the electromagnetic interaction of a physical pion are captured by this Euclidean-space matrix element, including the pion's electromagnetic form factor.

The second step of the calculation targets the remaining terms in the sum over intermediate states that appear in Eq.\,\eqref{eq:IVR1a}.  These terms can be written in Minkowski space using the notation introduced in Eqs.\,\eqref{eq:IVR1a}, \eqref{eq:A_pi} and \eqref{eq:IVR2A}:
\begin{eqnarray}
\tilde{A}&\equiv&  \int_{-\infty}^\infty \frac{dk_0}{2\pi} \int \frac{\dthree k}{(2\pi)^3}\, \int_0^\infty\!\! dx_0 ~e^{-ix_0k_0}\left\{
                          \int \dthree x ~e^{i\vec x \cdot \vec k}   \langle \pi(\vec p_\pi)|J^\mu_\mathrm{em}(\vec x, x_0) J^\nu_W(0)|K(\vecp 0)\rangle
                                       - \mathcal{A}_\pi^{\mu\nu}(\vec p_\pi,\vec p_\pi -\vec k, x_0)\right\} ~\times\nonumber \\
   && \hspace{1.5in}\frac{1}{k^2-i\epsilon}~\frac{\overline{u}_\ell(\vec{p}_\ell) \gamma_\mu(\gamma\cdot(p_\ell+k) -m_\ell)\gamma_\nu(1-\gamma^5)v_{\bar{\nu}}(\vec p_{\bar{\nu}})}{(p_\ell +k)^2 +m_\ell^2 -i\epsilon}\,. 
\label{eq:IVR2B}
\end{eqnarray}
This Minkowski amplitude is written as a product of a time-ordered QCD matrix element multiplied by covariant Feynman propagators for the photon and lepton.  This can of course be re-expressed as a conventional time-ordered matrix element of one weak current and two electromagnetic currents in which the photon and leptons as well as the QCD degrees of freedom all appear as intermediate states.  By removing the contribution of the single-pion state to these intermediate states and neglecting the small contributions of two- and three-pion states, we guarantee that no intermediate states appear in the right-hand side of Eq.\,(\ref{eq:IVR2B}) with lower energy than $m_K$.
 
Under these circumstances, the same result for this subtracted decay amplitude will be obtained in either Minkowski or Euclidean space.  Thus, Eq.\,(\ref{eq:IVR2B}) can be re-expressed as the product of a time-ordered QCD matrix element multiplied by covariant Feynman propagators for the photon and lepton, all expressed in Euclidean space.   The resulting amplitude will fall exponentially as the separation between the hadronic weak and electromagnetic currents increases, allowing the hadronic matrix element to be computed in lattice QCD with only exponentially suppressed finite volume errors.   

In this appendix we have focussed on the region $x_0>0$, since this is where the difficulties discussed above, associated with intermediate states with energies lower than $m_K$, appear. For $x_0<0$ there are no such difficulties; the Minkowski and Euclidean integrals over negative $x_0$ and $x_4$ respectively are equal, so that the corresponding contribution to the physical hadronic matrix element can also be computed in lattice QCD with only exponentially suppressed finite volume errors.
 
Finally we summarize the results of this appendix.  We have provided a further application of the IVR method to treat the electromagnetic corrections to $K_{\ell3}$ decays.  The resulting approach has the same important features as the treatment of the electromagnetic corrections to the decay of a pseudoscalar meson into a lepton and neutrino which was the main topic of this paper: i)  All errors resulting from the finite volume in which the lattice QCD portions of the calculation are performed fall exponentially with increasing lattice volume.  ii) Infrared divergences appear only in the analytic parts of the calculation, leaving the amplitudes to be computed using lattice QCD infrared finite.  

The electromagnetic corrections to $K_{\ell3}$ decays are more complex than those needed for the leptonic decays of a pseudo-scalar meson and have until now alluded a treatment in lattice QCD.  The most significant obstacle to such a lattice calculation is the photon exchange between the final-state pion and lepton.  The difficulties associated with this photon exchange contribution are the infrared-singular imaginary part and the appearance of intermediate pion-lepton states that are less energetic than the final pion-lepton.   Both difficulties can be resolved analytically in the IVR approach allowing this pion-lepton scattering contribution to be computed in a finite-volume lattice calculation of the $\phi_K^\dagger$-$J_W$-$J_{\mathrm{em}}$-$\phi_\pi$ four-point function as summarized in Eq.\,\eqref{eq:IVR2A}.  After the contribution of this single-pion intermediate state has been evaluated,  the remaining contribution can be directly evaluated from the lattice QCD calculation of a finite-volume Euclidean-space amplitude as indicated in Eq.\,\eqref{eq:IVR2B}.

\end{document}